\documentclass[10pt,journal,cspaper,compsoc]{IEEEtran}
\usepackage{mathptmx}
\usepackage{graphicx}
\usepackage{times}
\usepackage{caption}
\usepackage{morefloats}
\usepackage[tight,normalsize,sf,SF]{subfigure}

\usepackage{amsmath,amscd}
\usepackage{amsfonts}
\usepackage[ruled,vlined]{algorithm2e}
\usepackage{epsfig}
\usepackage{enumerate}
\usepackage{multirow}
\usepackage{array}
\usepackage{diagrams}
\newtheorem{theorem}{Theorem}
\newtheorem{definition}{Definition}

\ifCLASSOPTIONcompsoc
\else
\fi
\ifCLASSINFOpdf
\else
\fi
\begin{document}
%
\title{Spherical Parameterization Balancing Angle and Area Distortions}
%
%
%
%

\author{Saad~Nadeem, Zhengyu~Su, Wei~Zeng,
        Arie~Kaufman,~\IEEEmembership{Fellow,~IEEE,}
        and~Xianfeng~Gu
\IEEEcompsocitemizethanks{\IEEEcompsocthanksitem Saad Nadeem, Zhengyu Su, Arie Kaufman and Xianfeng Gu are with Computer Science Department, Stony Brook University, Stony Brook, NY 11794-2424.\protect\\
E-mail: \{sanadeem, zhsu, ari, gu\}@cs.stonybrook.edu.
\IEEEcompsocthanksitem Wei Zeng is with Florida International University.\protect\\
 E-mail: wzeng@cs.fiu.edu}
\thanks{}}

\IEEEcompsoctitleabstractindextext{%
\begin{abstract}
This work presents a novel framework for spherical mesh parameterization. An efficient angle-preserving spherical parameterization algorithm is introduced, which is based on dynamic Yamabe flow and the conformal welding method with solid theoretic foundation. An area-preserving spherical parameterization is also discussed, which is based on discrete optimal mass transport theory. Furthermore, a spherical parameterization algorithm, which is based on the polar decomposition method, balancing angle distortion and area distortion is presented. The algorithms are tested on 3D geometric data and the experiments demonstrate the efficiency and efficacy of the proposed methods.
\end{abstract}

}

\maketitle

\IEEEdisplaynotcompsoctitleabstractindextext

%
\IEEEpeerreviewmaketitle

\section{Introduction}

\vspace{-0.5mm}
\subsection{Motivation}

Mesh parameterization refers to the process of bijectively mapping a mesh onto a domain in a canonical space, generally the plane, the sphere or the hyperbolic disk. It plays a fundamental role in computer graphics, visualization, computer vision and medical imaging. The main criterion for mesh parameterization quality is the induced distortion. In general, the mapping distortions can be classified into angle distortion and area distortion. A mapping preserving both angle structure and the area element must be isometric, thus preserving Gaussian curvature. Therefore, in general cases, it is possible for a parameterization algorithm to be either angle-preserving or area-preserving, but not both.

\begin{figure*}[th]
\begin{center}
\begin{tabular*}{1\textwidth}{@{\extracolsep{\fill}}cccc}
\includegraphics[width=0.27\textwidth]{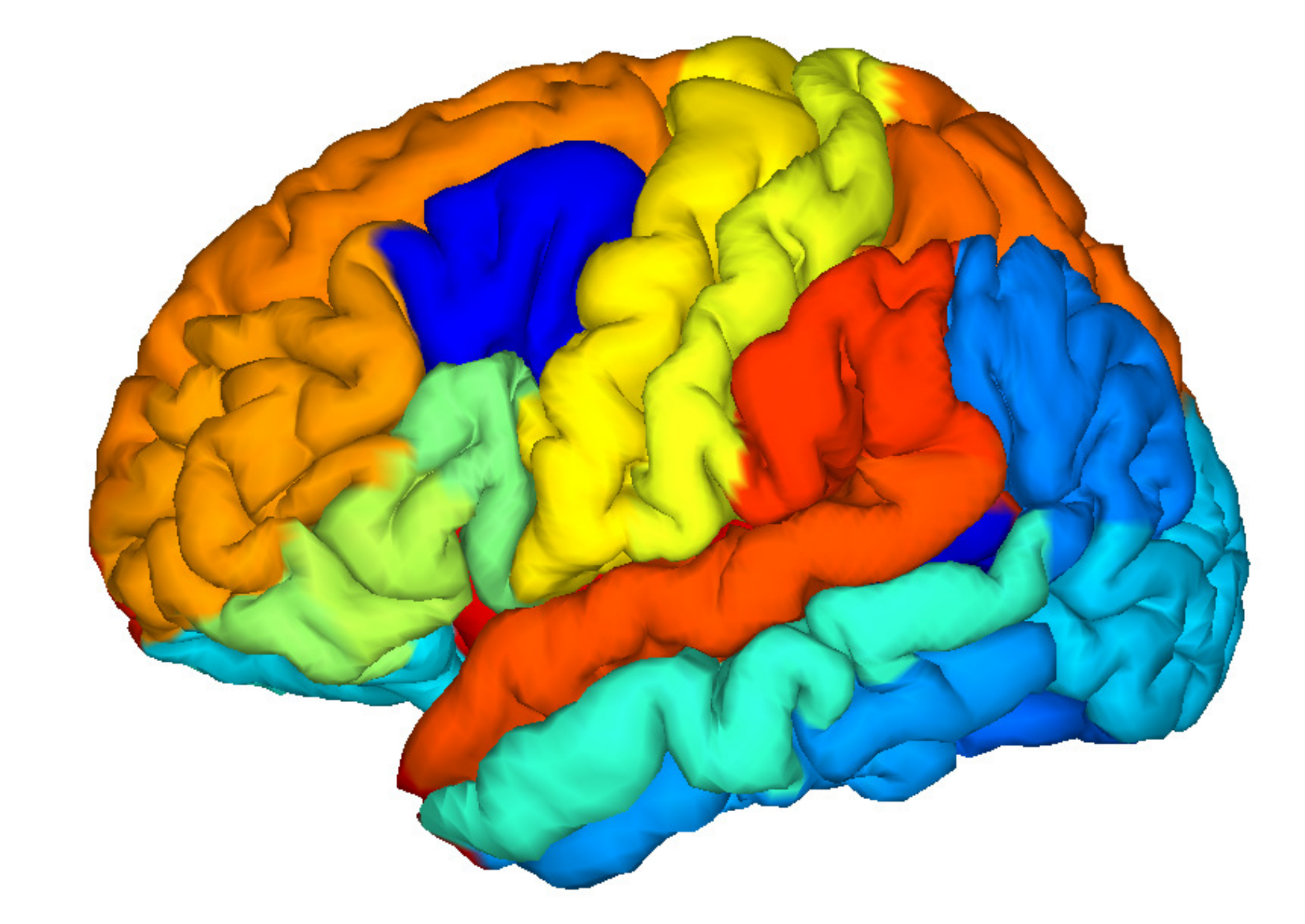}&
\includegraphics[width=0.27\textwidth]{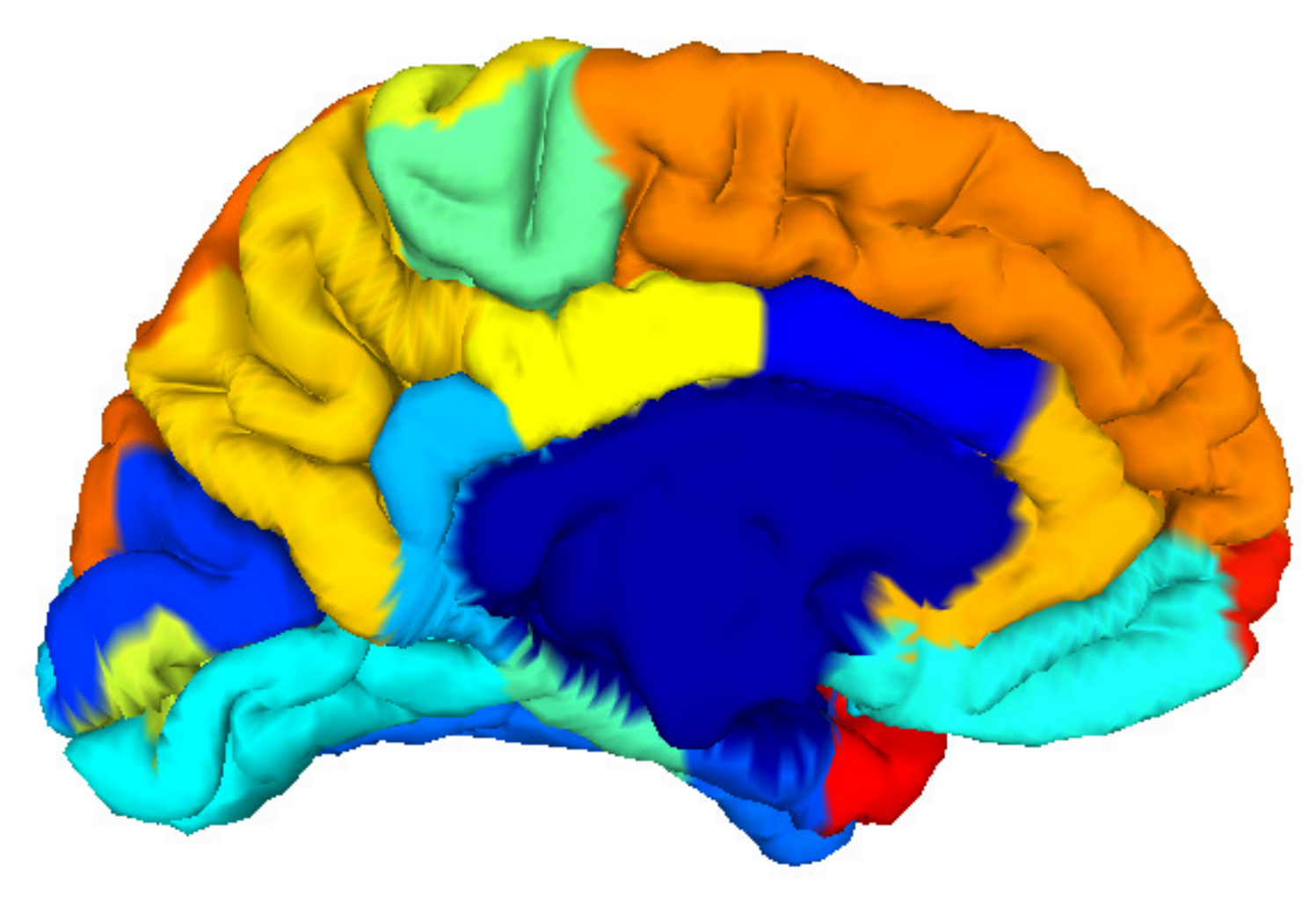}&
\includegraphics[width=0.20\textwidth]{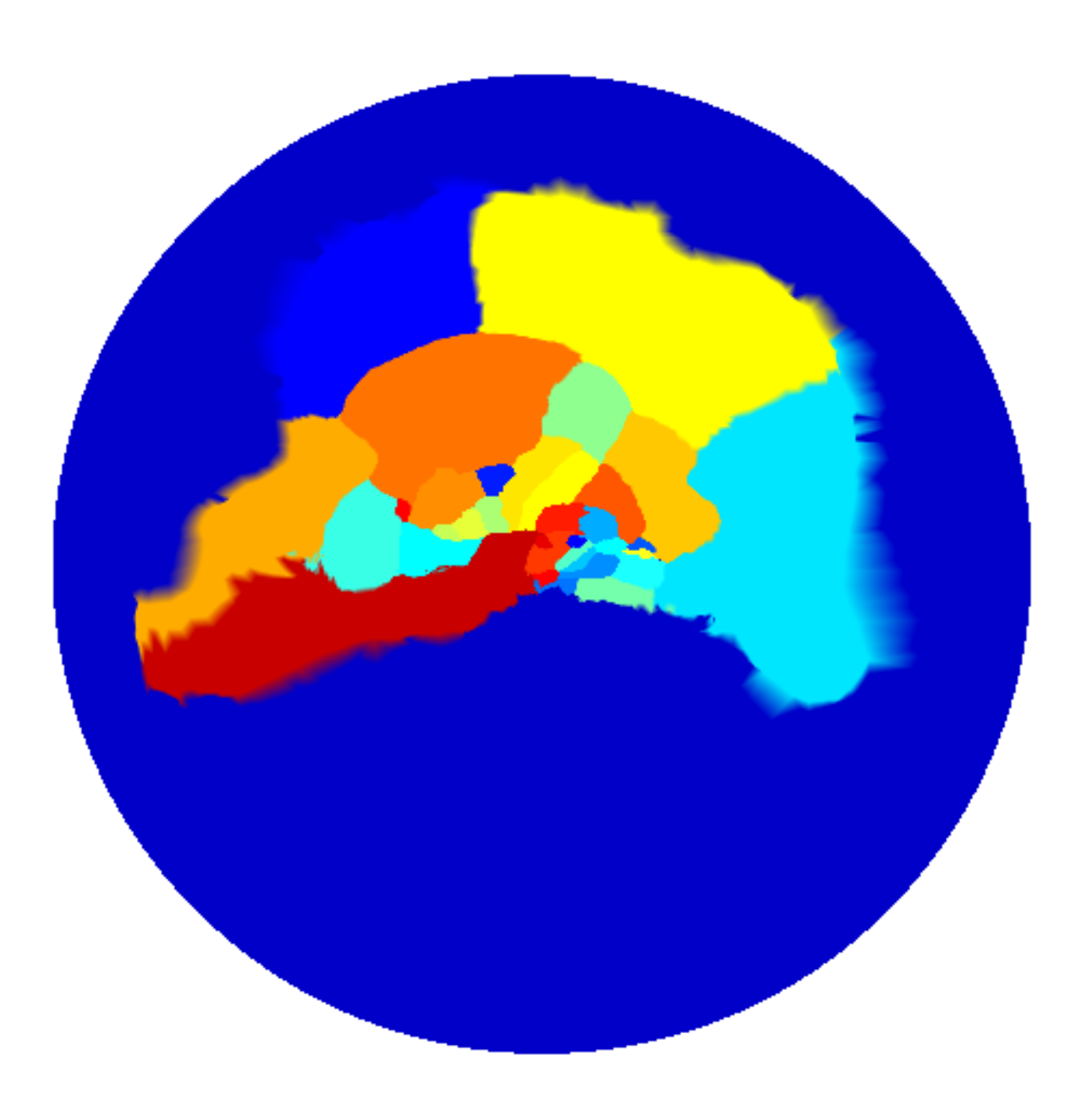}&
\includegraphics[width=0.19\textwidth]{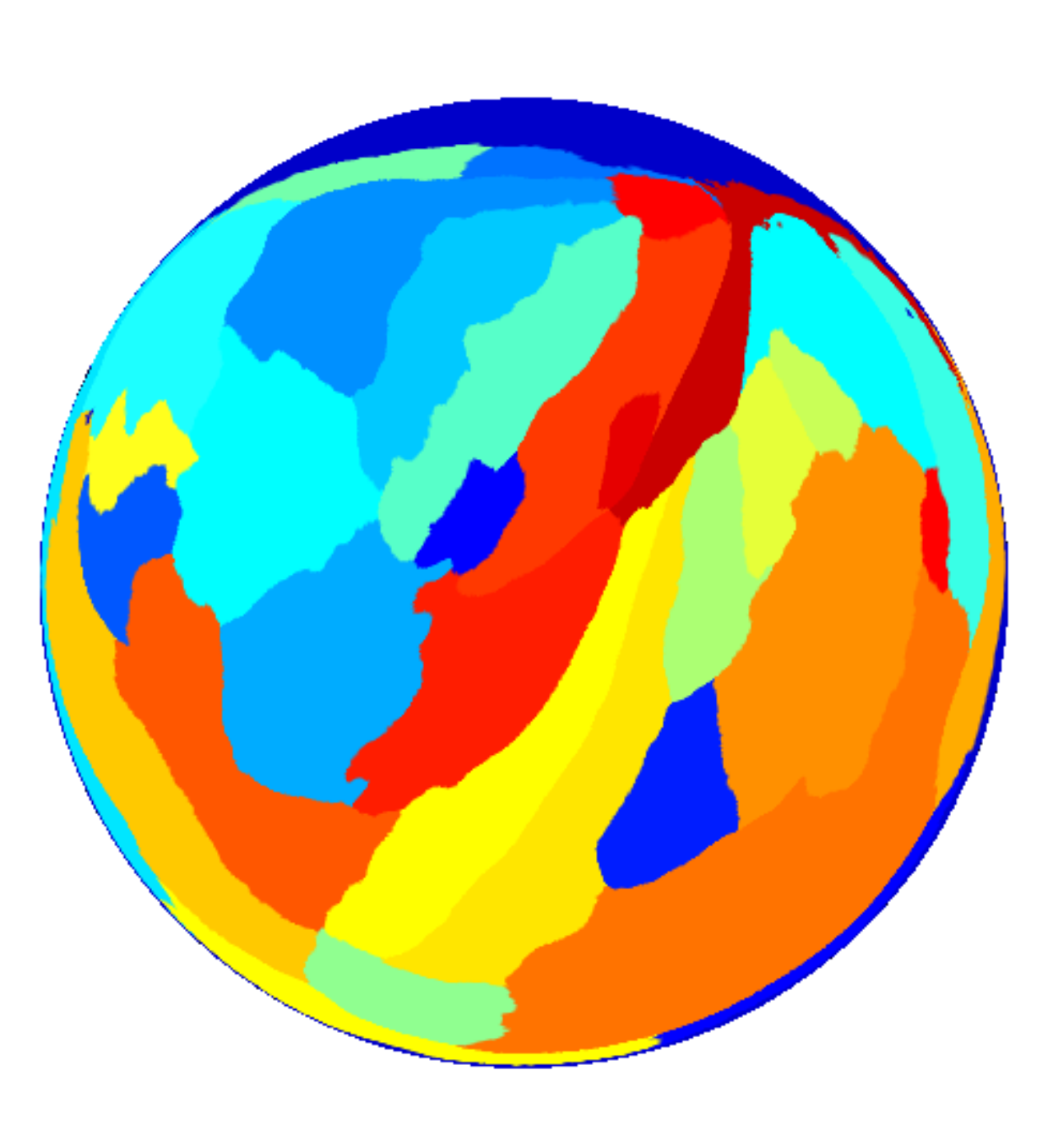}\\
(a) Superior view & (b) Inferior view & (c) Conformal map & (d) Area-preserving\\
& & & map
\end{tabular*}
\end{center}
\vspace{-3mm}
\caption{ Cortical surface mappings: Brain cortical surface with (a) superior and (b) inferior views, and the corresponding (c) conformal (angle-preserving) and (d) area-preserving parameterizations.\label{fig:brain_ap_ap}}
\vspace{-5mm}
\end{figure*}

A parameterization is angle-preserving, or conformal, if it preserves the intersection angles between arbitrary curves; or equivalently, the mapping is locally a scaling transformation. Therefore, a conformal mapping preserves local shapes. However, conformal mapping may induce large area distortions, as shown in Fig.~\ref{fig:brain_ap_ap}(c). On the other hand, a parameterization is area-preserving if it preserves the area element. However, an area-preserving mapping may induce large local shape distortions, as shown in Fig.~\ref{fig:brain_ap_ap}(d). In practice, for some applications, such as Alzheimer's disease diagnosis using brain morphometry, the areas of each functional region are crucial, and therefore the parameterization is required to preserve the area element. For other applications, such as cancer detection, the local shapes are more important, and therefore conformal mapping is preferred. However, in the case of brain mapping, virtual colonoscopy, deformable surface registration, dynamic surface tracking and mesh spline fitting, it is highly desirable to maintain a good balance between angle and area distortion.

\vspace{-3.5mm}
\subsection{Our Approach}

We focus on algorithms for finding angle-preserving, area-preserving and balanced parameterizations for genus zero surfaces without boundaries, namely topological spheres. For conformal mapping, the spherical parameterization method \cite{gotsman:2003} minimizes harmonic energy using a non-linear heat diffusion method. This method is highly non-linear, and sensitive to the choice of the initial condition. Another method \cite{haker:2000} maps one vertex to infinity, which induces a large deformation in that neighborhood. In order to overcome these disadvantages, we propose the following divide-and-conquer method: first we divide the input mesh into two segments with roughly equal areas, and each segment is then conformally mapped onto the planar disk using the discrete Ricci flow method \cite{Zengbook:2013}. The Ricci flow method is equivalent to convex optimization; the existence and the uniqueness of the solution have theoretic guarantees. Then, the two planar disks are glued together to cover the whole complex plane, including the infinity point, using a conformal welding method, such as the zipper algorithm \cite{Marshall2007}. This avoids the singularity issue found in the conventional methods.

More specifically, we use dynamic Yamabe flow \cite{gu:2013_2:arXiv} to conformally map the segments onto the respective planar disks. Yamabe flow is a scheme of Ricci flow, which deforms the Riemannian metric proportional to the curvature, such that the curvature evolves according to a non-linear heat diffusion process and becomes constant everywhere. Dynamic Yamabe flow, on the other hand, keeps the triangulation Delaunay during the flow, which guarantees the convergence, stability and the existence of the solution.

For an area-preserving method, we propose using our recently developed discrete optimal mass transport map theory \cite{gu:2013:arXiv}, which is equivalent to a convex optimization, and ensures the existence and the uniqueness of the solution, and that the mapping is area-distortion free. Algorithmically, this method can be converted to a power Voronoi diagram algorithm. The optimal mass transport map is solely determined by the source and the target area element (measures) on the sphere.

In order to achieve a good balance between angle distortion and area distortion, we propose using the polar decomposition method \cite{Brenier}. Suppose $\varphi: S\to \mathbb{S}^2$ is a conformal parameterization from the surface to the unit sphere, then $\varphi$ can be decomposed as $\varphi=\eta \circ \sigma$, where $\sigma : S \to \mathbb{S}^2$ is an area-preserving map, and $\eta: \mathbb{S}^2 \to \mathbb{S}^2$ is induced by an optimal mass transport map. By varying the area element on the unit sphere, we can change the optimal transport map $\eta$, and then construct a one-parameter family of mappings, connecting the area-preserving mapping $\sigma$ to the angle-preserving mapping $\varphi$. One can choose an intermediate map to achieve a good balance between angle and area distortions.

In essence, mesh parameterization unavoidably introduces distortions. These distortions can be classified into angle distortion and area distortion. It is impossible to achieve both angle distortion-free and area distortion-free parameterization simultaneoulsy. Therefore, the research focus in this paper is to balance between angle and area distortions. Conformal parameterization preserves angles, optimal transportation parameterization preserves area element. By combining them, and manipulating the target measure, optimal transport method is capable of achieving mesh parameterizations with a good balance between angle and area distortions.

\setlength{\tabcolsep}{2pt}

\vspace{-4mm}
\subsection{Contributions}

Our contributions can be summarized as follows:
\begin{enumerate}
\item A novel divide-and-conquer algorithm for conformal spherical parameterization based on dynamic Yamabe flow and conformal welding. Unlike conventional methods, this method is more efficient and has more rigorous theoretical foundations.
\item A novel balanced spherical parametrization based on polar decomposition. The discrete Yamabe flow and the conformal welding have been explored in isolation in previous works. However, the combination of conformal mapping and optimal mass transport is novel, which allows the user to trade off area distortion and angle distortion to best fit the requirements in practice.
\end{enumerate}
\vspace{-2.5mm}
\section{Previous Work}

The literature on mesh parameterization is vast, and a thorough survey is beyond the scope of the current work. Rather, we focus on the most directly related works, and refer the readers to comprehensive surveys \cite{FH05,sheffer:2007,sheffer2006}.

\vspace{-4mm}
\subsection{Optimal Mass Transport}
For optimal mass transport (OMT), approaches based on Monge-Kantorovich theory \cite{Kantorovich48} have been proposed. OMT was applied for flattening blood vessels in an area-preserving way for medical visualization~\cite{zhu:03:APVMS}. Haker et al. \cite{haker:2004} proposed using OMT for image registration and warping; the method is parameter-free and has a unique global optimum. OMT was used for texture mapping \cite{Dominitz10} by starting with an angle-preserving mapping and refining it using the mass transport procedure derived via gradient flow. A method was given for 3D image registration based on the OMT problem \cite{Rehmana09}. They stress that since optimization of OMT is computationally expensive, it is important to find efficient numerical methods to solve this issue, and that it is also crucial to extend the results to 3D surfaces.

There is work based on Monge-Brenier theory \cite{Brenier}. Our prior work~\cite{su:2013:CVPR} proposed an area-preserving brain mapping for brain morphological study, but it can only compute the map with the unit disk parameter domain. M\'{e}rigot \cite{Merigot} proposed a multi-scale approach to solve the optimal transport problem. An optimal-transport driven approach for 2D shape reconstruction and simplification was provided \cite{deGoes:2011:AOTARRSS}, as well as a formulation of capacity-constrained Voronoi tessellation as an optimal transport problem for image processing \cite{deGoes:2012:BNOT}. It produces high-quality blue noise point sets with improved spectral and spatial properties. Excepting our prior work \cite{su:2013:CVPR}, other Monge-Brenier theory-based methods (e.g., \cite{deGoes:2012:BNOT,deGoes:2011:AOTARRSS,Merigot}) are all applied to 2D image matching and registration. Our work applies a Monge-Brenier based OMT method for 3D surfaces with spherical topology. Recent work computing OMT for geometric data processing (\cite{solomon2015,solomon2014}) uses a heat kernel for the approximation. Our method converts the OMT problem to a convex optimization solved by Newton's method.

\vspace{-4.5mm}
\subsection{Spherical Mesh Parameterization}

Several methods have been developed for direct parameterizations on a topological sphere. Based on the type of parametric distortion minimized in each method, they can be classified into three groups: methods that do not explicitly address the issue of distortion, methods minimizing angular distortion, and methods minimizing area distortion.

In practice, most existing parameterization techniques belong to the first group (see \cite{alexa:2002,kobbelt:1999}). For instance, Alexa \cite{alexa:2002} proposed a heuristic iterative procedure that converges to a valid parameterization by applying local improvement (relaxation) rules. In this technique, an initial guess is computed and vertices are moved one at a time by computing a 3D position for the vertex using a barycentric formulation and then projecting the vertex to the unit sphere. An alternative was proposed using a multiresolution technique that involves a simplification of the mesh until it becomes a tetrahedron (or at least, convex) \cite{shapiro:1998}. The simplified model is then embedded in the sphere, and the vertices are inserted back one by one in order to preserve the bijectivity of the mapping. This process is efficient and stable, but optimizing the parameterization is difficult.

Spherical parameterization can be conducted in several ways (e.g., \cite{gotsman:2003,Gu04,saba:2005,sheffer:2005,zayer:2006}).
In the conformal method \cite{haker:2004}, one triangle is first cut out, the remaining surface is conformally mapped to an infinite plane, and the inverse stereo projection is used to map the plane to the sphere. This was applied to texture mapping \cite{haker:2000}. When applied to piecewise linear surfaces (meshes), embedding cannot be guaranteed for maps that are bijective and conformal for smooth surfaces and sometimes produces flipped triangles. In Haker et al. \cite{haker:2004} thin obtuse triangles are flipped by the stereographic projection \cite{sheffer:2007} and the distortion around the punched point is also high.

A similar approach is taken by others \cite{bobenko2010,springborn:2008}, where a polygonal boundary is formed by removing an arbitrary triangle from a closed mesh. The method is based on the introduction of cone singularities \cite{kharevych:2006}. The main idea is that instead of introducing artificial boundaries to absorb the undesired curvature, the entire Gaussian curvature of the mesh is redistributed so that it is concentrated at a few designated places (i.e., cone singularities). The main problem with these methods is that they do not modify the triangulation during curvature flow, such that the triangulation is always Delaunay. This makes it hard to guarantee the existence of a solution and may produce degenerate triangles leading to the collapse of the curvature flow.

Gu et al. \cite{Gu04} gave a nonlinear optimization for computing global conformal parameterization of genus-0 surfaces by minimizing harmonic energy, performing optimization in the tangent spaces of the sphere. With no stereographic projection, the method is more stable than \cite{haker:2004}, though it depends on a chosen initial mapping. The optimization may stay in local minima, instead of a global one.

Gotsman et al. \cite{gotsman:2003} provided a spherical equivalent of the barycentric formulation in the form of a quadratic system of equations, which can generate a bijective conformal mapping using appropriate weights in this scheme. A method was introduced to efficiently solve this system \cite{saba:2005}.

A parameterization method that cuts the mesh along a line connecting user-prescribed poles was given \cite{zayer:2006}. The mesh becomes topologically equivalent to a disk and an initial parameterization is found by solving a Laplace equation in curvilinear coordinates. Parameterization distortion is reduced by a variant of quasiharmonic maps and tangential Laplacian smoothing reduces distortion at the seam.

Taking into account angle distortion, a highly nonlinear optimization procedure that utilizes angles of the spherical triangulation (instead of vertex positions) was proposed \cite{sheffer:2005}. They specify a set of constraints that the angle values need to satisfy to define a planar triangular mesh. Angles as close as possible to the original 3D mesh angles and those that satisfy those constraints are then converted to actual vertex coordinates. In this method, constraints can be defined on the angles and on the triangle areas.

Spherical parameterization was solved using an iterative method \cite{kazhdan2012}, with each step solving a linear system. The method is extrinsic; it modifies the vertex positions to find the mapping and cannot be applied to abstract surfaces without embedding. Moreover, this method is incapable of finding a conformal metric with prescribed curvature, which is more flexible.

Spherical parameterization was computed using Willmore flow \cite{crane2013}. It computes a conformal homotopy using an iterative method. The method is extrinsic and cannot be applied for abstract surfaces without embedding. It cannot find a conformal metric which is not realizable in $\mathbb{R}^3$, such as the mappings from (b) to (e) and (c) to (f) in Figs. \ref{fig:closed_surface_uniformization} and \ref{fig:open_surface_uniformization} of the Appendix.

A major concern with conformal mapping is area distortion. A method was given for minimizing area distortion \cite{degener:2003} which is an extension of the existing MIPS method \cite{hormann:2000}. It attempts to minimize angle distortion by optimizing a nonlinear functional that measures mesh conformality. They added a term measuring area distortion to their energy functional and mediate between angle and area deformations by changing the powers of the components in the functional.

In contrast to the above, our method has solid theoretical foundations and precisely controls angle and area distortion. Given a desired area measure, we achieve the exact solution. Our Yamabe flow method handles surfaces with arbitrary topology and the OMT can be generalized to high genus surfaces. 
\vspace{-1.5mm}
\section{Computational Algorithms}

In this section, we explain the major algorithms in detail. The theoretical foundations necessary for the current work can be found in the Appendix.

\vspace{-2.5mm}
\subsection{Angle-Preserving Mapping}
\label{sec:map_angle}
In this section, we explain the discrete Yamabe flow theory and algorithm. This algorithm is necessary for computing conformal mapping of a topological disk to a planar disk, once we have split the given genus-0 surface into two topological disks (as explained in Section \ref{sec:conf_spherical_mapping}).

\subsubsection{Discrete Dynamic Yamabe Flow} Angle-preserving mappings can be achieved using the discrete Ricci flow method. In the following, we generalize surface Ricci flow to the discrete setting, and focus on the dynamic Yamabe flow method for discrete surface Ricci flow.

On computers, smooth surfaces are approximated by triangulated polyhedral surfaces, namely, a triangle mesh. A mesh is denoted as
$M=(V,E,F)$, where $V$, $E$ and $F$ represent vertex, edge and face sets, respectively. Each face is a Euclidean triangle.

A \emph{discrete Riemannian metric} \cite{Zengbook:2013} is represented as the edge length function $l:E\to\mathbb{R}^+$, satisfying the triangle inequality
on each face. On each face, the three corner angles are determined by the Euclidean cosine law using the edge lengths.

\begin{definition}[Delaunay Triangulation]
The triangulation is \emph{Delaunay} if for each edge $e$, the sum of two corner angles against it is no greater than $\pi$.
\end{definition}

\begin{figure}[h]
\begin{center}
\begin{tabular}{ccc}
\includegraphics[width=0.15\textwidth]{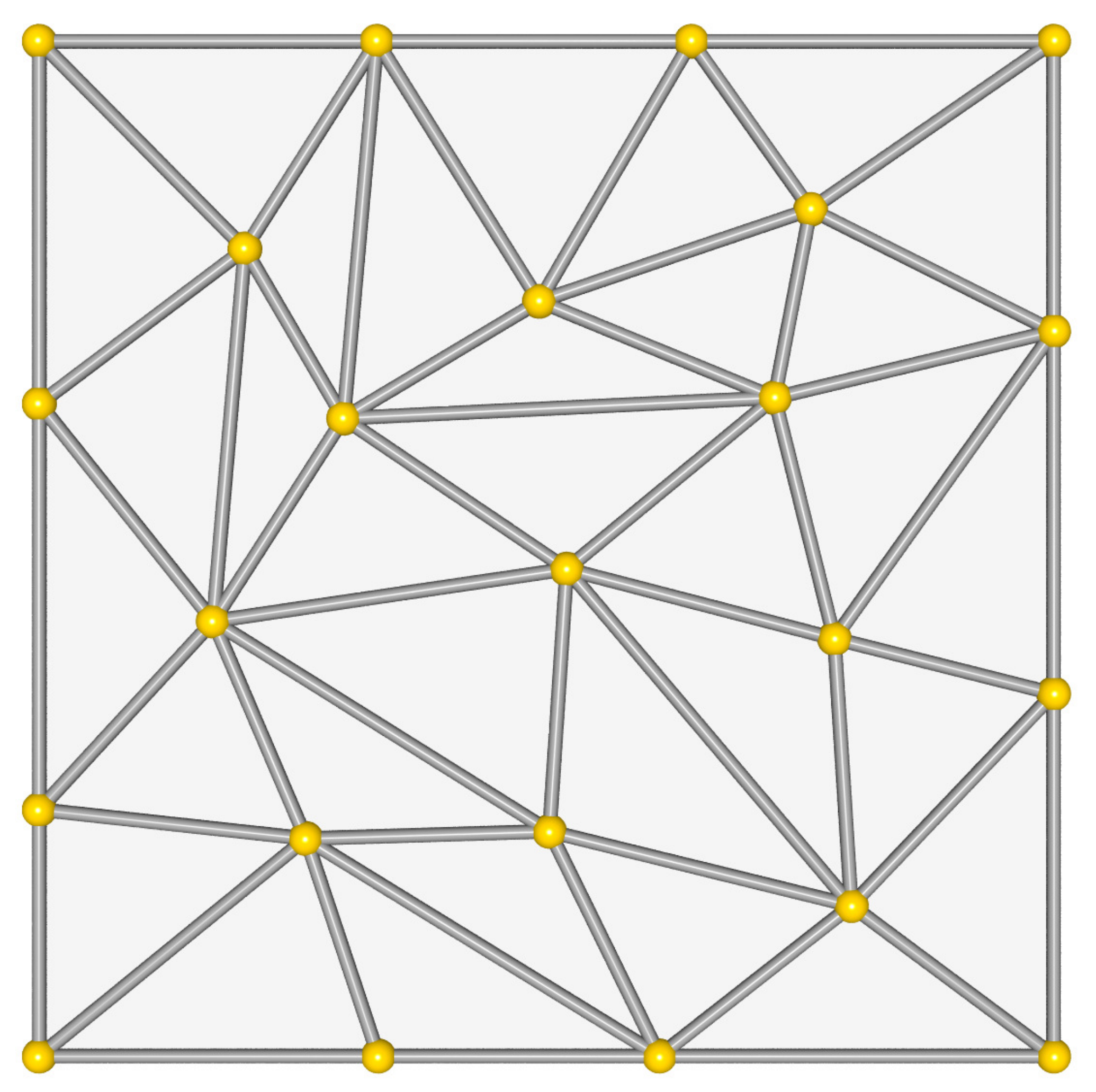}&
\includegraphics[width=0.15\textwidth]{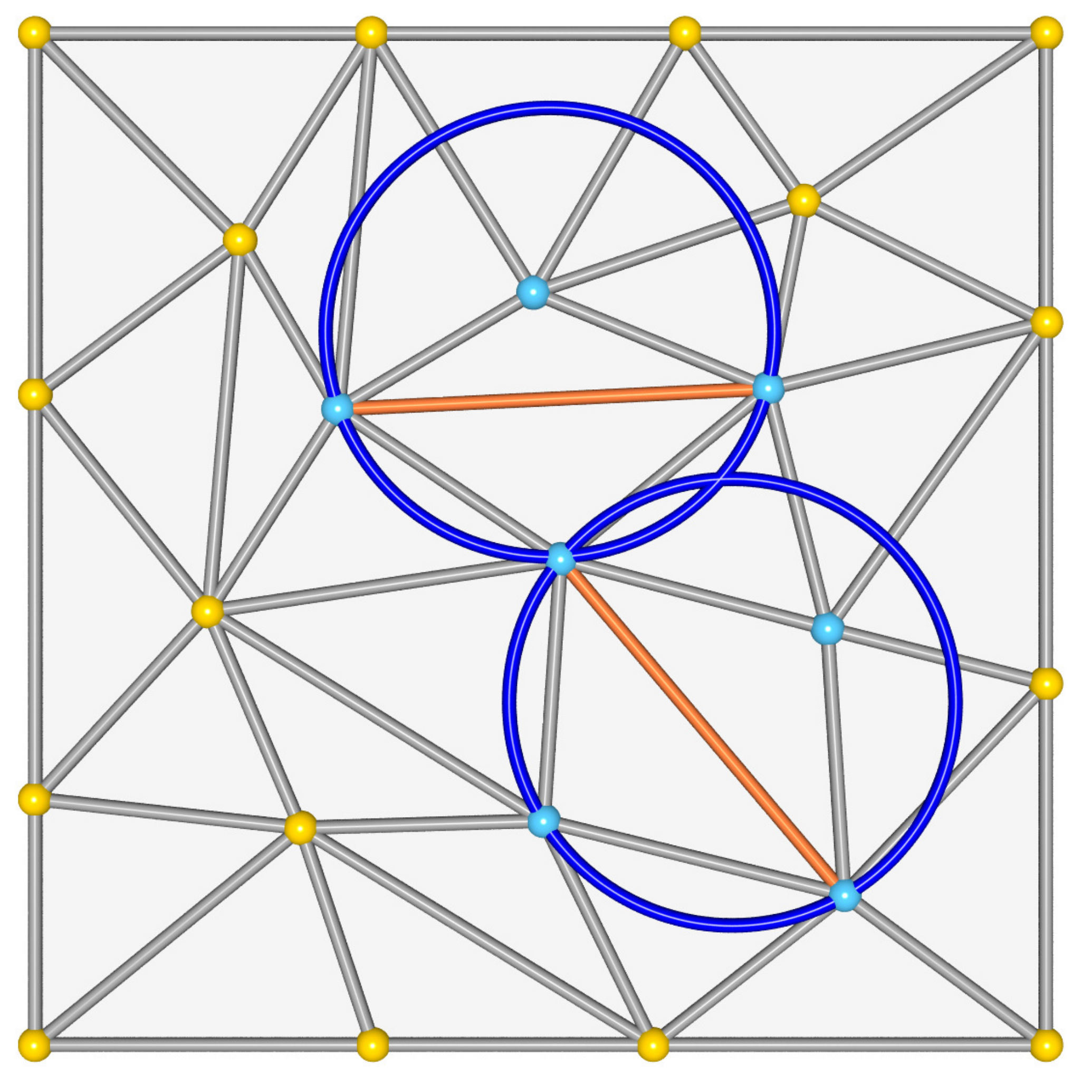}&
\includegraphics[width=0.15\textwidth]{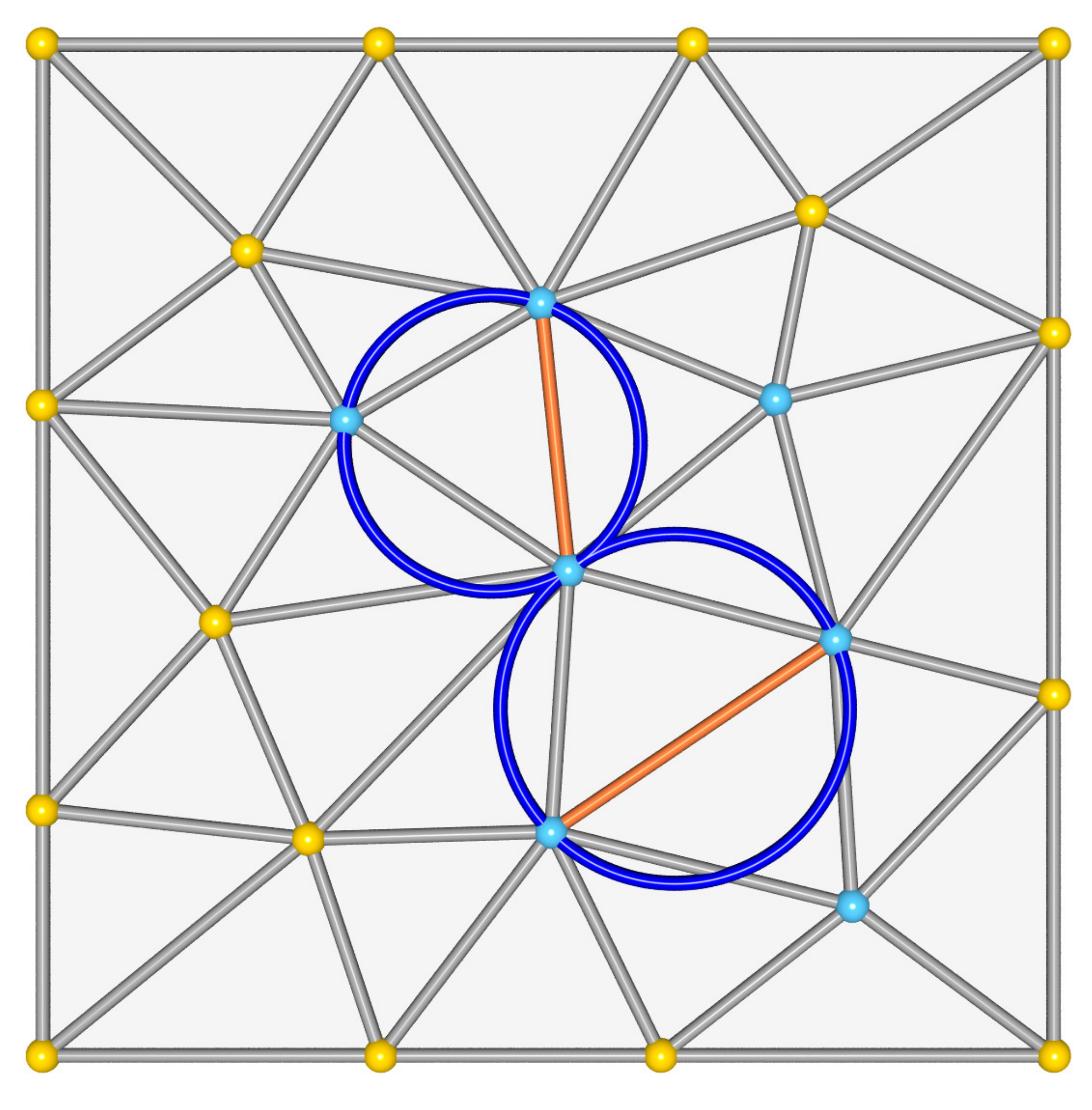}\\
(a) & (b) & (c)
\end{tabular}
\end{center}
\vspace{-3mm}
\caption{ Delaunay triangulation via diagonal switch: (a) A non-Delaunay triangulation with (b) diagonal orange edges switch leads to (c) Delaunay triangulation.
\label{fig:delaunay}}
\vspace{-3mm}
\end{figure}

Given an initial triangulation,
one can achieve Delaunay triangulation by \emph{diagonal switch}: two adjacent triangles are flattened on the plane, the diagonal
is swapped on the plane, and the two new triangles replace the original ones. We illustrate this concept in Figure \ref{fig:delaunay}.

The \emph{discrete Gaussian curvature} \cite{Zengbook:2013} is defined as angle deficit: for an interior vertex, its Gaussian curvature is $2\pi$ minus
the surrounding corner angles; for a boundary vertex, its geodesic curvature is $\pi$ minus the surrounding corner angles.
\begin{definition}[Discrete Gaussian Curvature] Given a triangle mesh $M$ with a discrete Riemannian metric, the curvature for a vertex $v_i\in V$ is defined as:
\[
    K(v_i) = \left\{
    \begin{array}{rr}
    2\pi - \sum_{jk} \theta_i^{jk} & v_i \not \in \partial M\\
    \pi - \sum_{jk} \theta_i^{jk} & v_i \in \partial M\\
    \end{array}
    \right.
\]
\end{definition}
It can be easily shown that the total curvature is a topological invariant.

\begin{theorem}[Gauss-Bonnet] The total curvature equals $2\pi$ multiplied by
the Euler characteristic number of the mesh,
\[
    \sum_{v} K(v) = 2\pi \chi(M)
\]
where $\chi(M)$ is the Euler charactieristic number of $M$ and is defined according to the formula, $\chi(M)=V-E+F$.
\end{theorem}

A \emph{discrete conformal factor} is a function defined on vertices
$u:V\to\mathbb{R}$.

\begin{definition}[Discrete Conformal Metric Deformation]
Given a triangle mesh $M$, with a discrete Riemannian metric, and conformal factor $u:V\to \mathbb{R}$,
suppose an edge $e\in E$ has end vertices $v_i$ and $v_j$, and its original length is $l_{ij}$, then the deformation is:
\[
    l_{ij} \mapsto e^{u_i} l_{ij} e^{u_j}.
\]
\end{definition}
The discrete Ricci flow is defined in the same way as its smooth counterpart.

\begin{definition}[Dynamic Discrete Yamabe Flow \cite{gu:2013_2:arXiv}] Given a triangle mesh $M$ with an initial discrete metric and
the target curvature $\bar{K}$, the discrete Yamabe flow is given by:
\[
    \frac{du_i}{dt} =\bar{K}_i - K_i.
\]
Furthermore, during the flow, the
triangulation is maintained to be Delaunay by diagonal switches.
\end{definition}

The following fundamental theorem has been recently proved \cite{gu:2013_2:arXiv}, which guarantees the existence of solutions.

\begin{theorem}[Dynamic Discrete Yamabe Flow] If the target
curvature satisfies the Gauss-Bonnet condition, and for each vertex
$v_i\in V$, $\bar{K}_i <2\pi$, then the solution to dynamic discrete
Yamabe flow exists, and is uniquely updated to a constant, which is the
unique optimal point of the convex \emph{discrete Ricci energy}:
\begin{equation}
    E(\mathbf{u}) = \int^{\mathbf{u}}_{\mathbf{0}} \sum_{i=1}^n (\bar{K}_i - K_i) du_i,\hspace{8mm} \mathbf{u}=(u_1,u_2,\cdots,u_n).
\label{eqn:ricci_energy}
\end{equation}
\end{theorem}
Therefore, one computational algorithm is to optimize the convex
Ricci energy using Newton's method. The gradient of the energy is
the curvature difference, $\nabla
E(\mathbf{u})=(\mathbf{\bar{K}}-\mathbf{K})^T$, the Hessian matrix
of the energy is the conventional Laplace-Beltrami matrix of the
mesh,
\begin{equation}
    \frac{\partial^2 E(\mathbf{u})}{\partial u_i \partial u_j } = \left\{
    \begin{array}{cc}
    \cot \theta_k^{ij} + \cot \theta_l^{jl} & v_i \sim v_j \\
    0 & v_i \not\sim v_j
    \end{array}
    \right.
    \label{eqn:Laplace}
\end{equation}
where the angles $\theta_k^{ij}$ and $\theta_l^{ji}$ are the two
corner angles against the edge connecting $v_i$ and $v_j$. All the conformal uniformization can be directly carried out by
discrete Yamabe flow. The algorithm for discrete Yamabe flow \cite{gu:2013_2:arXiv} is given in Alg.~\ref{alg:yamabe_flow} as pseudo-code.

\begin{algorithm}[!h]
\caption{Discrete Yamabe Flow}
    \textbf{(1)} The user determines the target curvature $\bar{K}:V\rightarrow \mathbb{R}$, such that for each vertex $v_i \in V$, $\bar{K}(v_i) < 2\pi$ and the total curvature satisfies the Gauss-Bonnet condition, $\sum_{v_i \in V} \bar{K}(v_i) = 2\pi\chi(M)$.

    \textbf{(2)} Initialize the conformal factor as zeros $u_t = 0$, for all vertices.

    \textbf{(3)} Compute the current edge length using equation
    \[
        l_{ij} = e^{u_i}\beta_{ij}e^{u_j},
    \]
    compute the corner angles using Euclidean cosine law and compute the vertex curvatures using equation
    \[
        K(v_i)=2\pi - \sum_{jk}\theta_{i}^{jk}.
    \]

    \textbf{(4)} Compute the gradient of the entropy energy
    \[
        \nabla E(\mathbf{u}) = (\bar{K_1}-K_1,\bar{K_2}-K_2, \cdots, \bar{K_n}-K_n)^T.
    \]

    \textbf{(5)} Compute the Hessian matrix of the entropy energy using Eqn. (4)
    \[
        H = \frac{\partial^2 E(\mathbf{u})}{\partial u_i \partial u_j } = \left\{
    \begin{array}{cc}
    -w_{ij} & v_i \sim v_j \\
    \sum_{k} w_{ik} & i=j\\
    0 & v_i \not\sim v_j
    \end{array}
    \right.
    \]

    \textbf{(6)} Solve linear system $\nabla E=Hx$.

    \textbf{(7)} Update the conformal factor $\mathbf{u} = \mathbf{u}-\delta x$, where $\delta$ is the step length.

    \textbf{(8)} Repeat step 4 through 8 until
    \[
        \max_{v_i \in V} |\bar{K}-K(v_i)| < \epsilon,
    \]
    where $\epsilon$ is a threshold.
\label{alg:yamabe_flow}
\end{algorithm}

\vspace{-2.5mm}
\subsubsection{Discrete Riemann Mapping}

\begin{figure}[h]
\vspace{-5mm}
\begin{center}
\begin{tabular}{cc}
\includegraphics[width=0.22\textwidth]{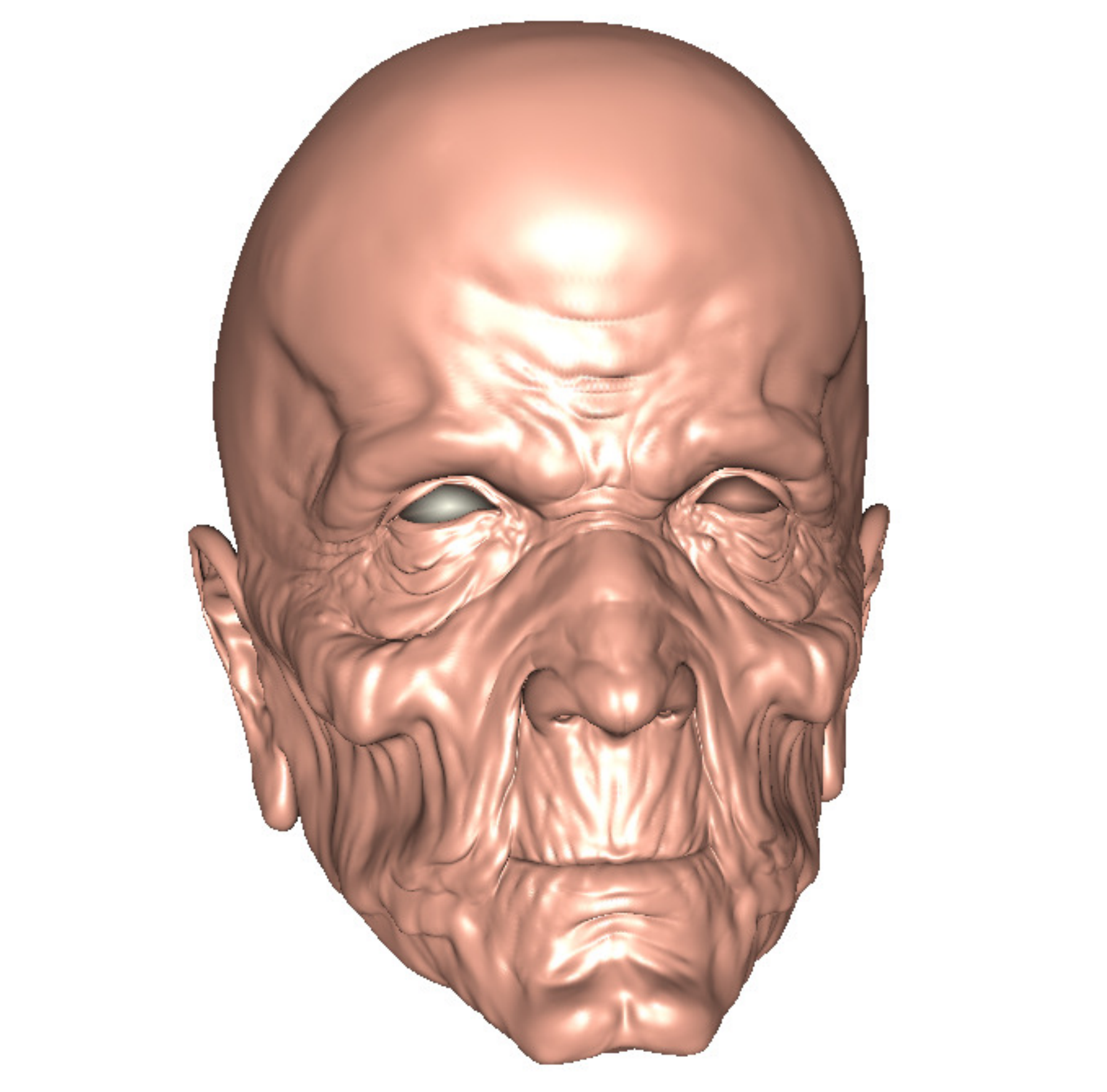}&
\includegraphics[width=0.22\textwidth]{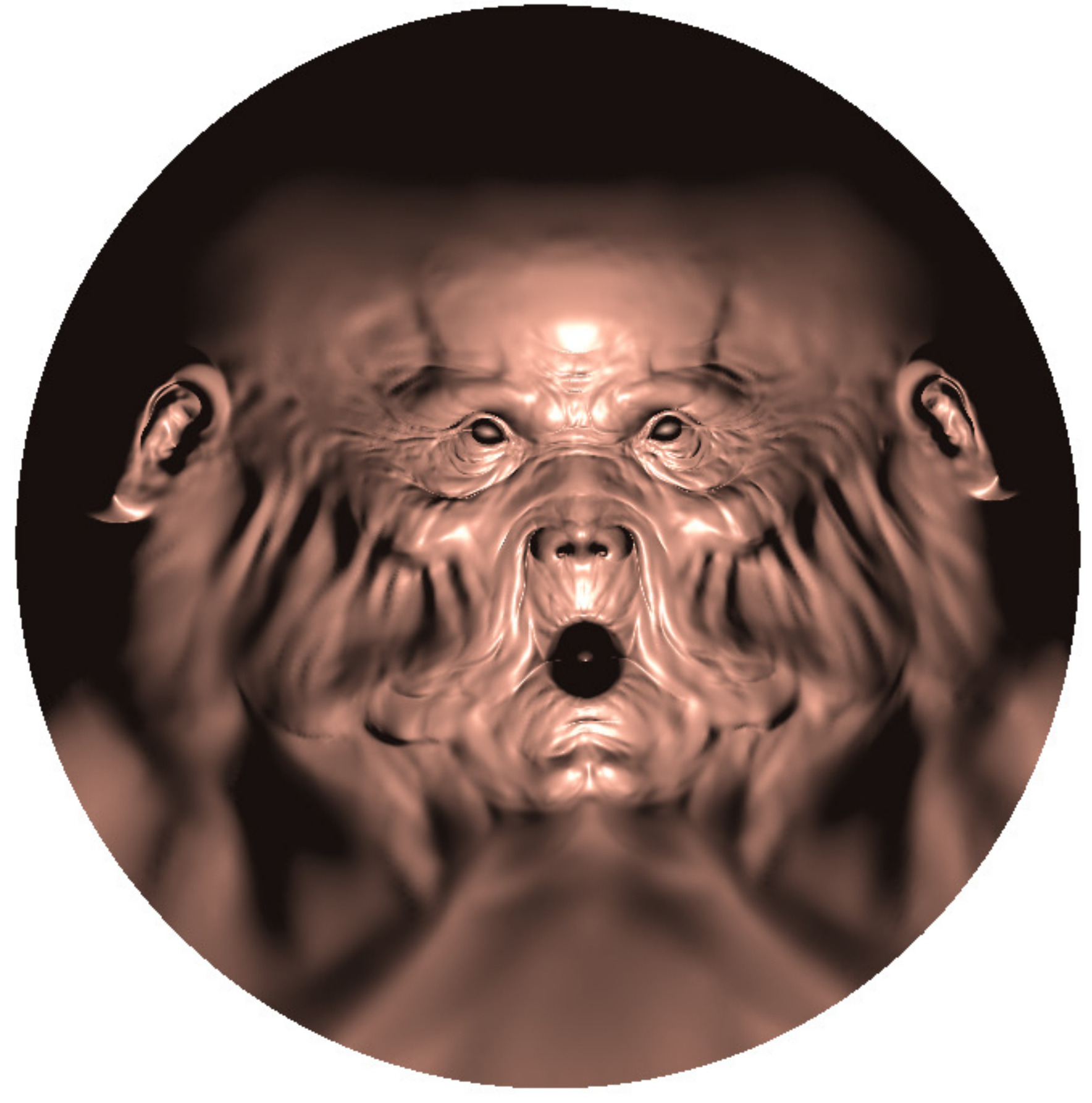}\\
(a) & (b)
\end{tabular}
\end{center}
\vspace{-3mm}
\caption{ Riemann mapping: (a) a genus-0 surface with a single boundary can be mapped onto the (b) unit disk conformally. The Riemann mapping is computed using Ricci flow. \label{fig:oldman}}
\vspace{-6mm}
\end{figure}

A Riemann mapping is a conformal mapping between a metric surface with a disk topology and the unit planar disk. This section focuses on how to compute the discrete approximation of the smooth Riemann mapping, the so-called discrete Riemann mapping.  First, we introduce an algorithm to compute the conformal mapping from a topological annulus onto a planar annulus. Then, for a topological disk, we puncture a small hole in the center, convert it to a topological annulus, and apply the topological annulus method.

\begin{figure}[h]
\begin{center}
\begin{tabular}{ccc}
\includegraphics[width=0.19\textwidth]{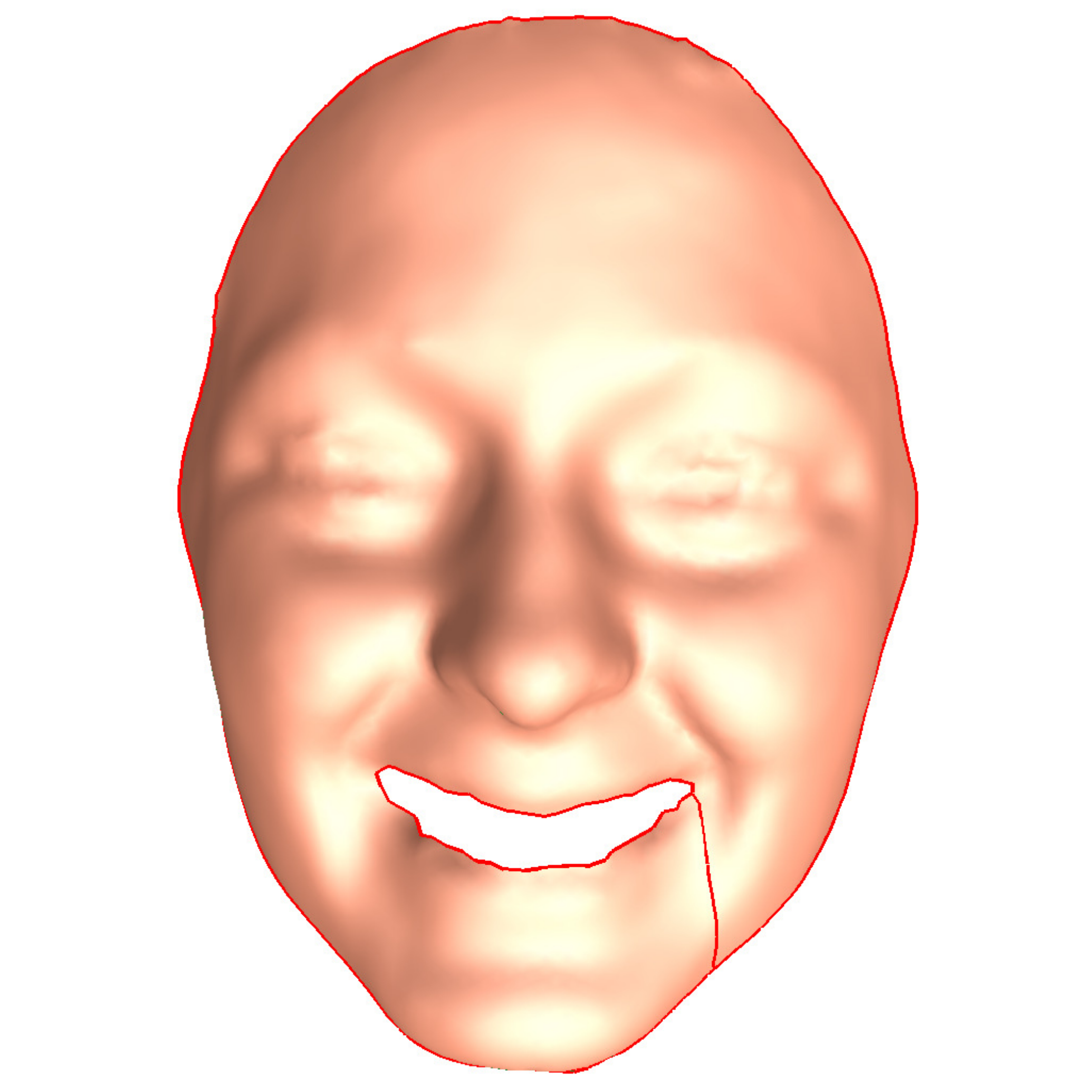}&
\includegraphics[width=0.052\textwidth]{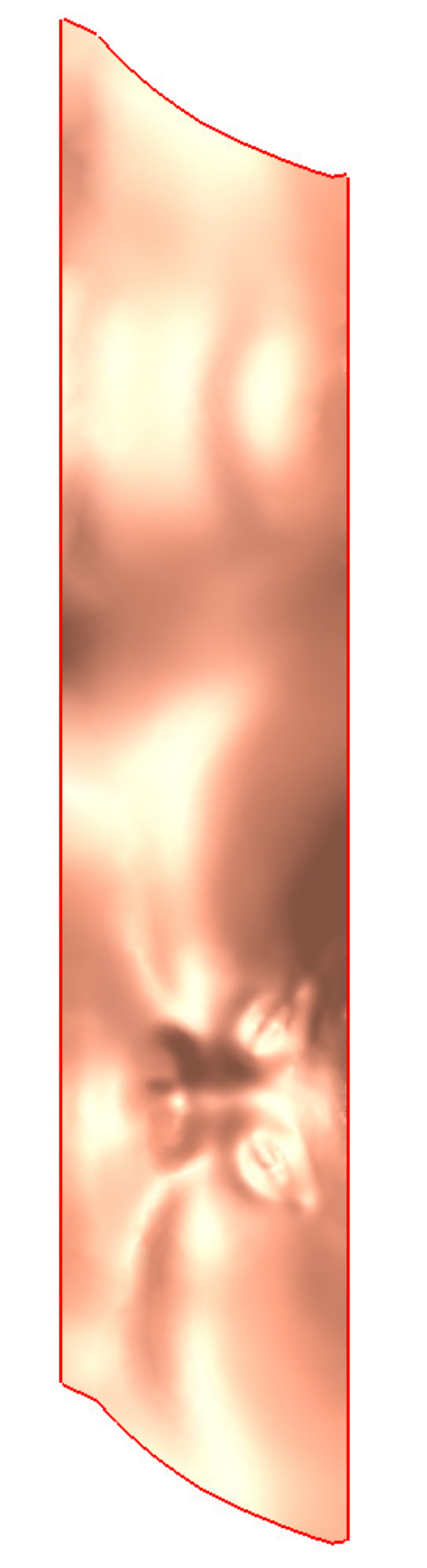}&
\includegraphics[width=0.19\textwidth]{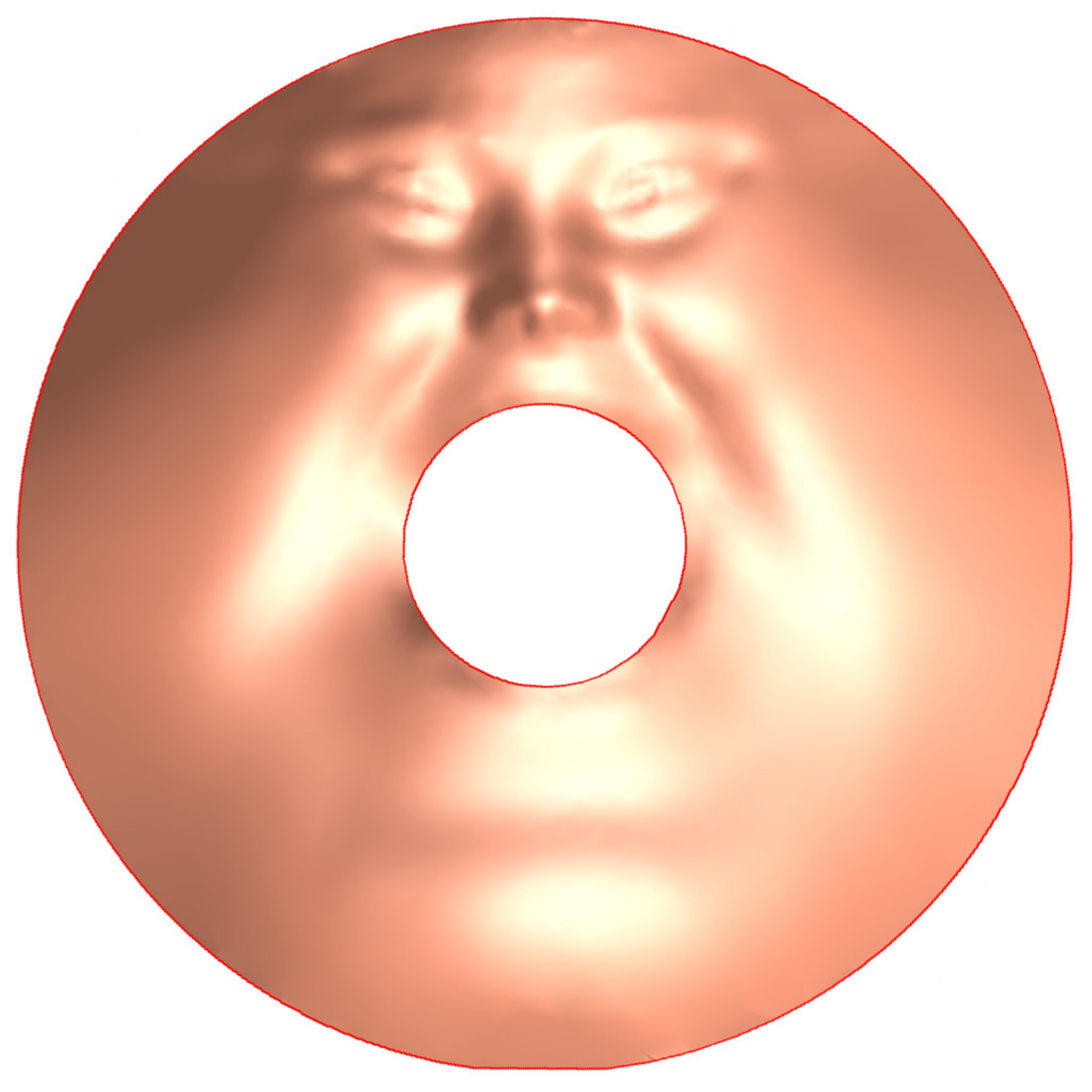}\\
(a) & (b) & (c)
\end{tabular}
\end{center}
\vspace{-3mm}
\caption{ Conformal mapping of a topological annulus: (a) A human facial surface with mouth open; the red curve between the two boundaries will cut the surface open. (b) The sliced surface is mapped to a topological cylinder (periodic rectangle). (c) The rectangle is mapped to the unit disk with a concentric circular hole by exponential map.
\label{fig:HA}}
\vspace{-4mm}
\end{figure}

As shown in Fig.~\ref{fig:HA}, suppose the input mesh $N$ is a topological annulus (a genus-0 mesh with two boundaries) as shown in Fig.~\ref{fig:HA}(a), we set the target curvature to be zero everywhere, including both the interior vertices and boundary vertices, and run the dynamic Yamabe flow to obtain a flat metric of the mesh. Then, we compute a shortest path $\gamma$ connecting the two boundaries and slice the mesh along the path to get a simply connected mesh $\bar{N}$. We flatten the mesh $\bar{N}$ isometrically onto the plane using the flat metric just computed and map it onto a parallelogram. By translation and rotation, we align the parallelogram with the virtual axis, and scale its height to be $2\pi$, as shown in Fig.~\ref{fig:HA}(b). Finally, we use the complex exponential map $z \to exp(z)$ to map the parallelogram to a planar annulus as shown in Fig.~\ref{fig:HA}(c). This procedure maps a topological annulus conformally to a canonical planar annulus.

Suppose we are given a topological disk $M$, we can compute the discrete Riemann mapping to map it onto the unit planar disk by using the above algorithm. We choose an interior face $f_0$, and remove it from $M$ to get a topological annulus. Then, we apply the above algorithm to map the punctured mesh onto a planar annulus using Yamabe flow. Finally, we fill the center hole on the planar annulus by one triangle. This process gives the discrete Riemann mapping.

\vspace{-3mm}
\subsection{Discrete Conformal Welding}
\label{sec:conformal_welding}
In this section, we introduce the conformal welding algorithm. This algorithm is necessary for welding together the two planar disks, seamlessly, that we obtain from conformal mapping of the topological disks using Riemann mapping algorithm (as explained in Section \ref{sec:map_angle}).

\setlength{\tabcolsep}{5pt}
\begin{figure}[h]
\begin{center}
\begin{tabular}{cc}
\includegraphics[width=0.45\textwidth]{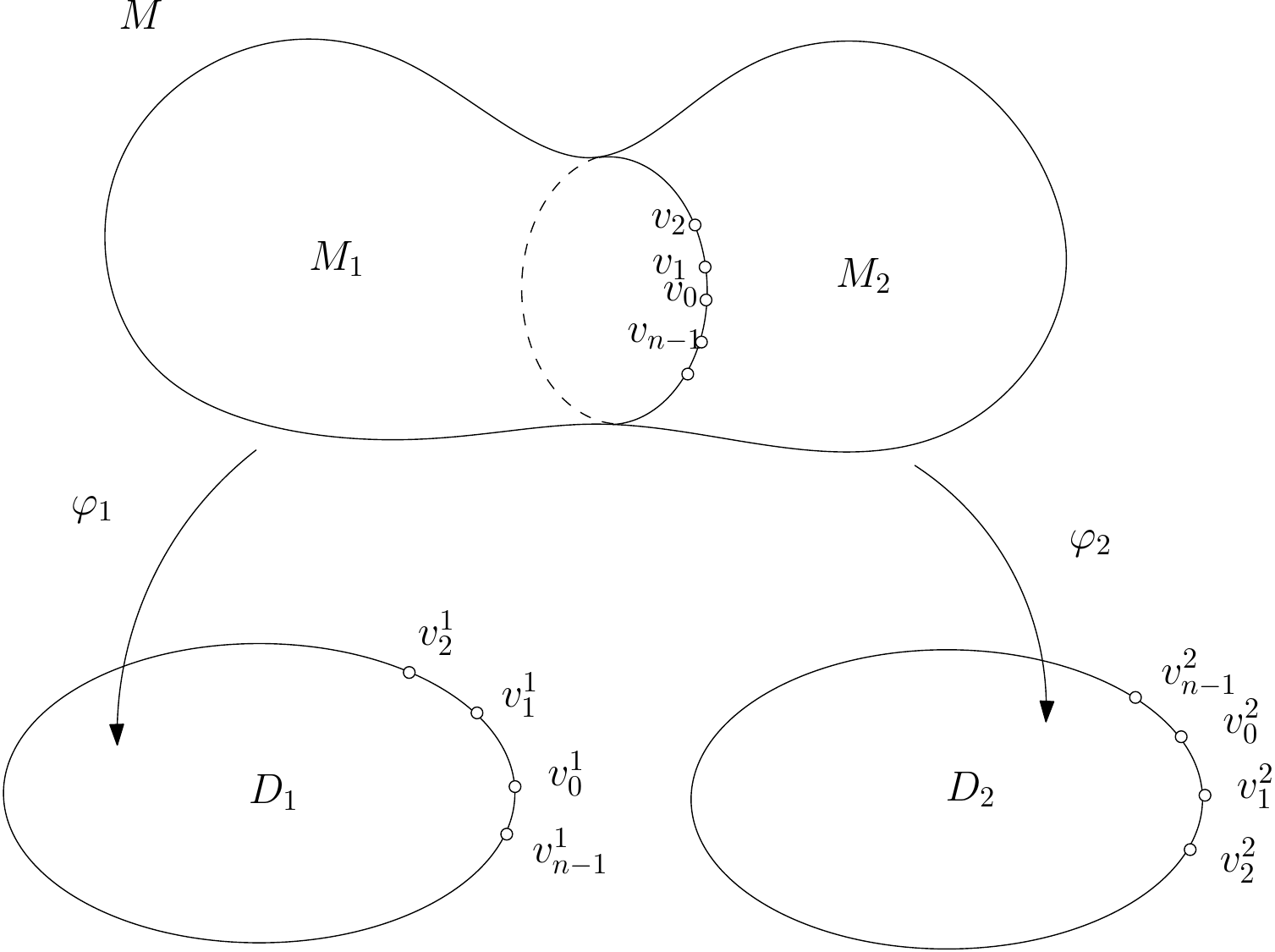}
\end{tabular}
\end{center}
\vspace{-3mm}
\caption{The mesh is divided into two segments and each one is mapped onto the planar disk using conformal mapping.}
\label{fig:initial}
\vspace{-2mm}
\end{figure}

The following algorithm is a variant of the zipper algorithm \cite{Marshall2007}, where the theoretical proof for the convergence can be found.
As shown in Fig.~\ref{fig:initial}, the original mesh $M$ is separated by the cutting loop $\gamma$ into two parts $M_1$ and $M_2$, and each component is mapped to the unit disk by discrete Riemann mapping $\varphi_k: M_k\to \mathbb{D}$. We denote the two images as $D_k = \varphi_k (M_k), k=1,2$. Each vertex $v_i$ on $\gamma$ has a unique corresponding vertex $v_i^1$ on $\partial D_1$ and $v_i^2$ on $\partial D_2$, this induces a mapping, $\eta: \partial D_1 \to \partial D_2$, called the \emph{discrete conformal welding signature}. We sort the vertices on the boundary of $D_k$'s as $\partial D_k = v_0^k,v_1^k,v_2^k,\cdots v_{n-1}^k, k = 1,2$.

For the convenience of visualization, we use the following mapping to map the upper half plane to the interior of the unit circle, and the lower half plane to the exterior of the unit circle, $\sigma: \hat{\mathbb{C}}\to \hat{\mathbb{C}}$, its inverse $\sigma^{-1}$ maps the unit disk to the upper half plane:
\[
    \sigma: w = \frac{z-i}{z+i}, ~~~\sigma^{-1}: z = i\frac{1+w}{1-w}.
\]

As shown in Fig.~\ref{fig:conformal_welding}(a), $D_1$ is the interior of the unit circle and $D_2$ is the exterior of the circle.

\setlength{\tabcolsep}{5pt}
\begin{figure}[h]
\begin{center}
\begin{tabular}{cc}
\includegraphics[width=0.45\textwidth]{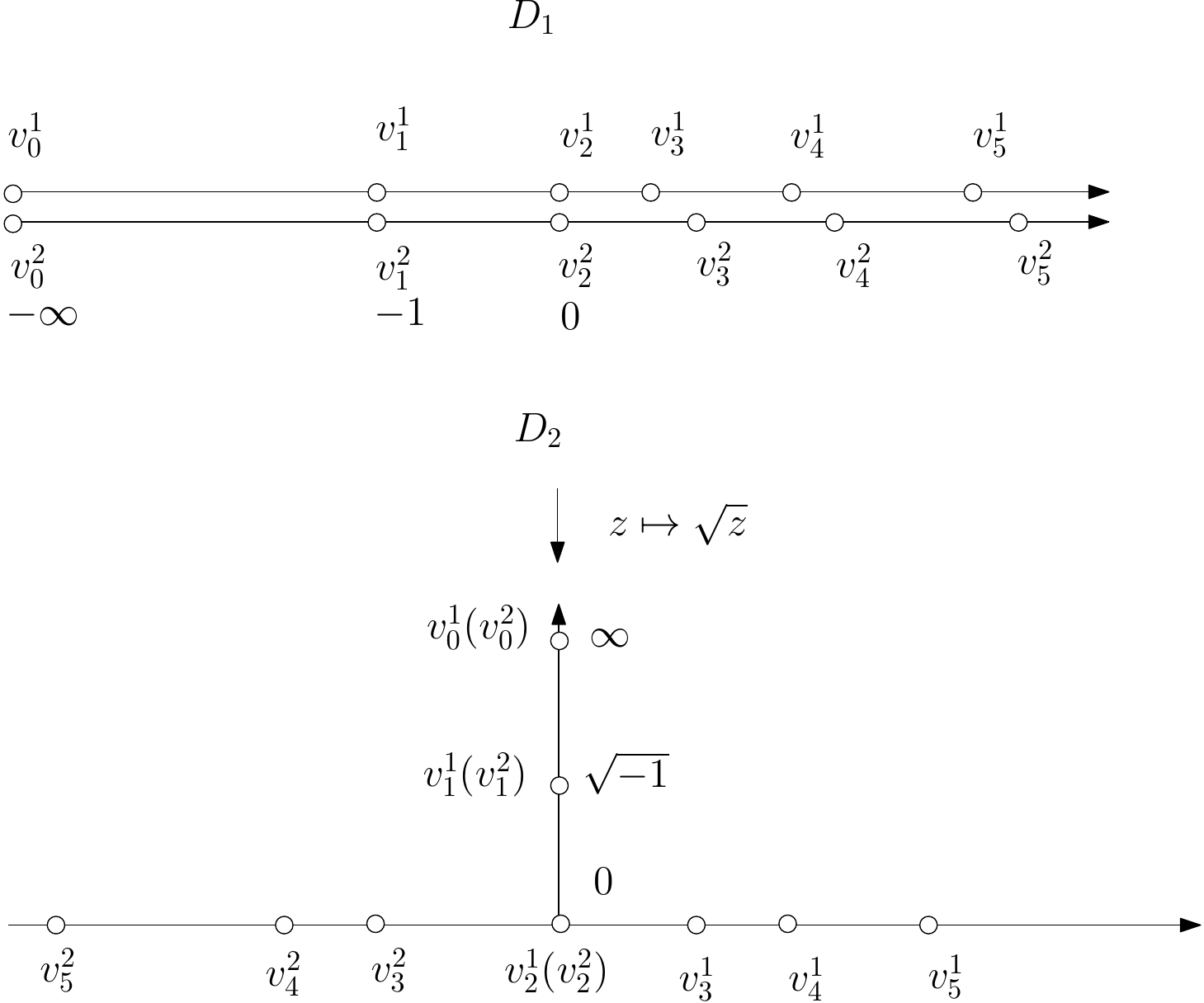}
\end{tabular}
\end{center}
\vspace{-3mm}
\caption{Step 1 glues $\{v_0^1, v_1^1, v_2^1\}$ with $\{v_0^2, v_1^2, v_2^2\}$.}
\label{fig:step_1}
\vspace{-2mm}
\end{figure}

\noindent{\bf Step 1. Glue $v_k^1$ with $v_k^2$, $0\le k \le 2$}: First, we use $\tau$ to map $D_1$ to the upper half plane $y\ge0$. Then, we use a M\"obius transformation to map $\{v_0^1, v_1^1, v_2^1\}$ to $\{-\infty, -1, 0\}$,
\[
    \frac{z-v_2^1}{z-v_0^1} \frac{v_1^1-v_0^1}{v_2^1-v_1^1}.
\]
Similarly, we map $D_2$ to the lower half plane $y\le 0$ and use a M\"obius transformation to map $\{v_0^2, v_1^2, v_2^2\}$ to $\{-\infty,-1,0\}$. Then, we glue $D_1$ and $D_2$ along the negative half real axis, namely, we glue $\{v_0^1, v_1^1, v_2^1\}$ with $\{v_0^2, v_1^2, v_2^2\}$, and take the square root to map the union of $D_1$ and $D_2$ to the upper half plane, $z \to \sqrt{z}$. As shown in Figs.~\ref{fig:step_1} and \ref{fig:conformal_welding}(b), the first $3$ vertices are glued together, and mapped to $\{-\infty,i,0\}$ ($\{+1,0,-1\}$ in the disk view).\\

\setlength{\tabcolsep}{5pt}
\begin{figure}[h]
\begin{center}
\begin{tabular}{cc}
\includegraphics[width=0.45\textwidth]{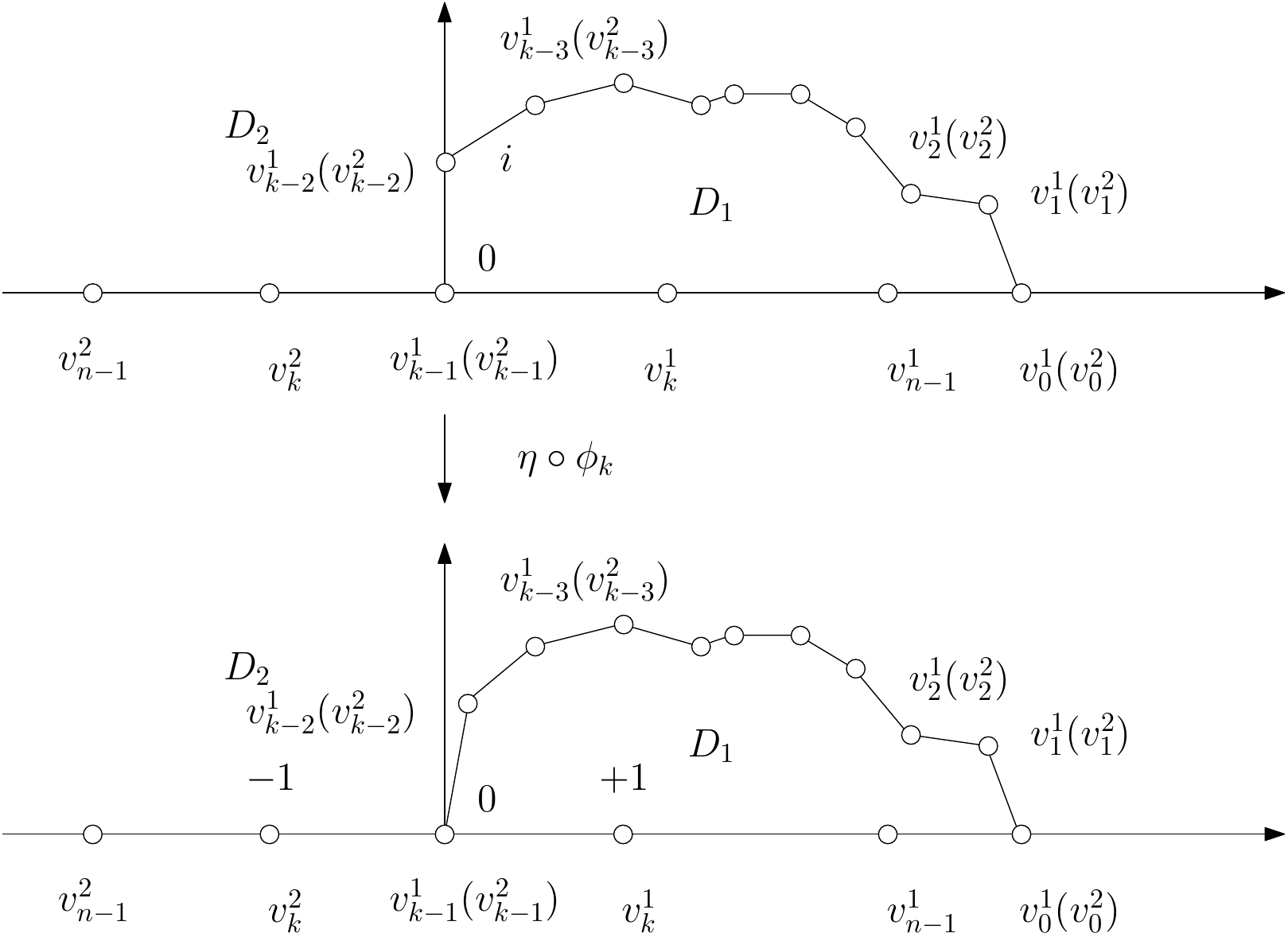}
\end{tabular}
\end{center}
\vspace{-3mm}
\caption{Step 2 maps $\{v_k^2,0,v_k^1\}$ to $\{-1,0,+1\}$.}
\label{fig:step_2}
\vspace{-5mm}
\end{figure}

\noindent{\bf Step 2. Glue $v_k^1$ with $v_k^2$, $3\le k \le n-1$}: We take a M\"obius transformation to map $\{v_k^2,0,v_k^1\}$ to $\{-1,0,+1\}$. First, we use a M\"obius transformation $\phi_1$ to map $\{v_k^2,0,v_k^1\}$ to $\{0,1,\infty\}$
\[
    \phi_k(z)= \frac{z-v_k^2}{z-v_k^1} \frac{v_k^1}{v_k^2}.
\]
Then, we use $\eta$ to map $\{0,1,\infty\}$ to $\{-1,0,+1\}$,
 \[
    \eta(z) = - \frac{z+1}{z-1}.
 \]
The composition $\eta\circ \phi_k$ maps $\{v_k^2,0,v_k^1\}$ to $\{-1,0,+1\}$, as shown in Fig.~\ref{fig:step_2}.

\setlength{\tabcolsep}{5pt}
\begin{figure}[h]
\begin{center}
\begin{tabular}{cc}
\includegraphics[width=0.45\textwidth]{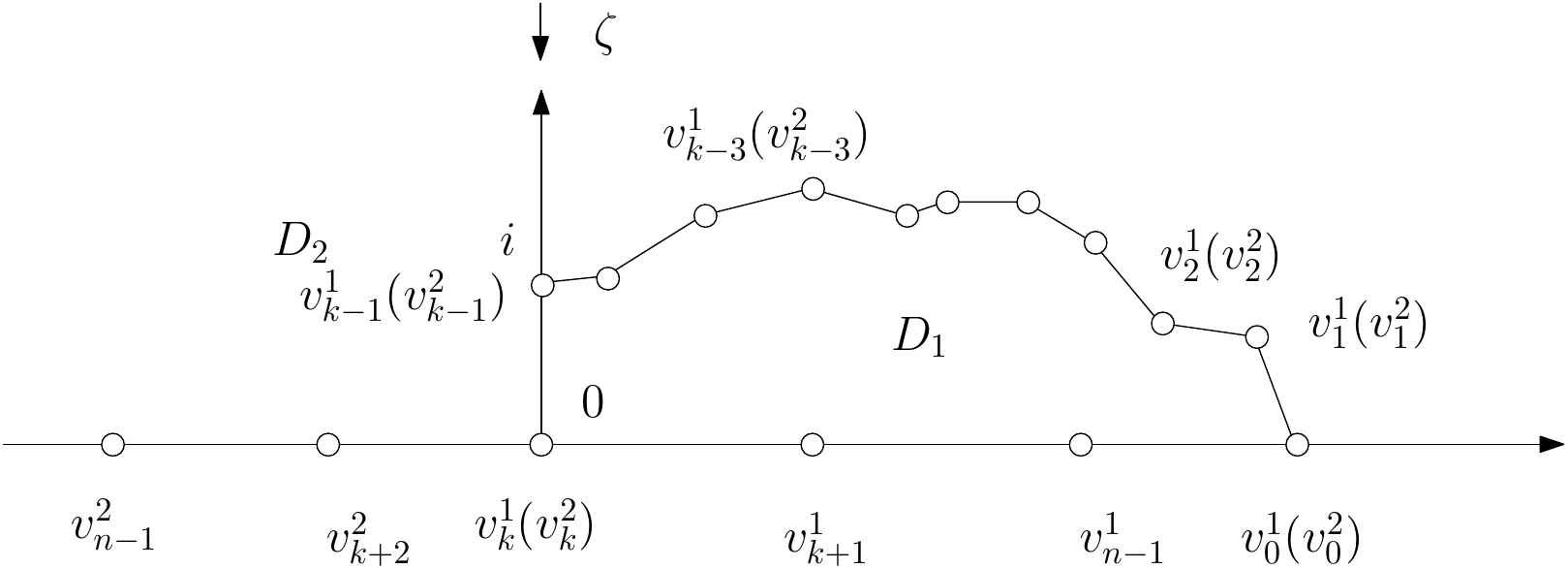}
\end{tabular}
\end{center}
\caption{Step 2 glues the line interval $[v_{k}^2,v_{k-1}^2]$ with $[v_{k-1}^1,v_{k}^1]$, and maps $v_k^2$ and $v_k^1$ to $0$.}
\label{fig:step_3}
\vspace{-2mm}
\end{figure}

We take the map, $\zeta: z \to \sqrt{z^2+1}$, to glue the interval $[-1,0]$ with $[0,+1]$. The composition $\zeta\circ \tau\circ \phi_k$ glues the line interval $[v_{k}^2,v_{k-1}^2]$ with $[v_{k-1}^1,v_{k}^1]$, and maps $v_k^2$ and $v_k^1$ to $0$, as shown in Fig.~\ref{fig:step_3}. We repeat this procedure for $k=3,4,\cdots,n-1$. As shown in Fig.~\ref{fig:conformal_welding}(c)-(e), the glued boundary segment is inside the unit circle, the unglued boundary segments are on the unit circle. \\

\setlength{\tabcolsep}{5pt}
\begin{figure}[h]
\vspace{-4mm}
\begin{center}
\begin{tabular}{cc}
\includegraphics[width=0.25\textwidth]{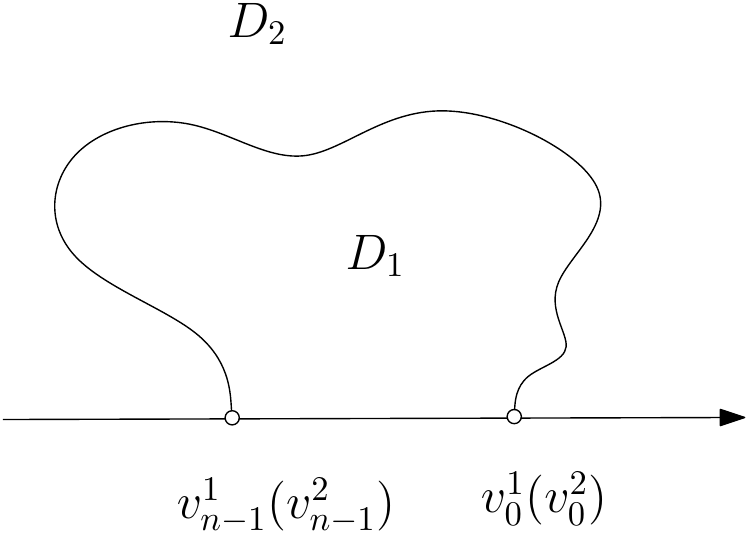}
\end{tabular}
\end{center}
\vspace{-3mm}
\caption{Step 3 glues the line interval  $[v_{n-1}^1,v_0^1]$ with $[v_{0}^2,v_{n-1}^2]$.}
\label{fig:step_4}
\vspace{-2mm}
\end{figure}

\noindent{\bf Step 3. Glue $[v_{n-1}^1,v_0^1]$ with $[v_{0}^2,v_{n-1}^2]$}: At this stage, $v_{n-1}^1$ and $v_{n-1}^2$ are at $0$, $v_0^1$ and $v_0^2$ coincide together on real axis. First, we map $v_0^k$ to $\infty$ and fix $0$,
\[
    z \to \frac{z}{z-v_0^1}.
\]
This maps the union of $D_1$ and $D_2$ to the upper half plane. Then, we use $z \to z^2$ to map the union to the whole extended plane, as shown in Fig.~\ref{fig:step_4}. After this, we apply
$\tau^{-1}$ to map the upper plane to the interior of the unit disk.

The boundaries $\partial D_1$ and $\partial D_2$ are mapped to a
Jordan curve $\gamma$ on the plane. As shown in
Fig.~\ref{fig:conformal_welding}(f), two disks are welded
together and their boundaries are connected to a Jordan curve $\gamma$;
$D_1$ is the domain interior to the curve $\gamma$ and $D_2$ is
exterior to $\gamma$. A Jordan curve here refers to a closed planar polygonal curve; in Fig. \ref{fig:conformal_welding}(a), the curve
separates the two planar domains, $D_{1}$ (white) and $D_{2}$ (yellow).

\vspace{-3mm}
\subsection{Conformal Spherical Mapping}
\label{sec:conf_spherical_mapping}
In this section, we combine the algorithms in Sections \ref{sec:map_angle} and \ref{sec:conformal_welding} to compute the spherical mapping: the topological sphere is split into two topological disks, then each disk is mapped onto the planar disk using the Riemann mapping algorithm in Section \ref{sec:map_angle}, and finally the two planar disks are welded together using the algorithm in Section \ref{sec:conformal_welding} to form the conformal spherical mapping.

Suppose we have a closed genus $0$ mesh $M$, we first cut it into two topological disks. Let $\Delta$ be the discrete Laplace-Beltrami operator (Eq.~\ref{eqn:Laplace}). The first eigenfunction is given by $\Delta f_0 = \lambda f_0$, where $\lambda_0$ is the minimal positive eigenvalue. The zero level set of $f_0$ is a closed curve $\gamma$, which divides $M$ into two segments $M_0$ and $M_1$. We normalized $f_0$, such that it integrates to zero and square-integrates to one. According to the Riemannian manifold spectrum theory \cite{lablee2015spectral}, the areas of the two segments are almost equal.

Then, we use discrete dynamic Yamabe flow to compute the discrete Riemann mappings, $\varphi_k: M_k\to \mathbb{D}_k$, $k=0,1$. Next, we use the conformal welding algorithm to glue the two disks $D_k$ to the extended complex plane $\hat{\mathbb{C}}=\mathbb{C}\cup\{\infty\}$, then further to the unit sphere $\mathbb{S}^2$ by stereographic projection $\tau:\hat{\mathbb{C}}\to \mathbb{S}^2$:
\[
\tau(x,y)= \left( \frac{2x}{1+x^2+y^2}, \frac{2y}{1+x^2+y^2}, \frac{x^2+y^2-1}{x^2+y^2+1}\right).
\]
Unlike some methods based on linear finite elements, the usage of the stereographic projection in our case does not cause imprecision near the poles. This is because the current method maps two connected components onto the planar unit disk and then uses the zipper algorithm to weld the two disks. The zipper algorithm is symbolic, and the numerical computation explicitly involves infinity. Details can be found in Alg.~\ref{alg:spherical_conformal_map}.

\setlength{\tabcolsep}{5pt}
\begin{figure}[th]
\begin{center}
\begin{tabular}{cc}
\includegraphics[width=0.22\textwidth]{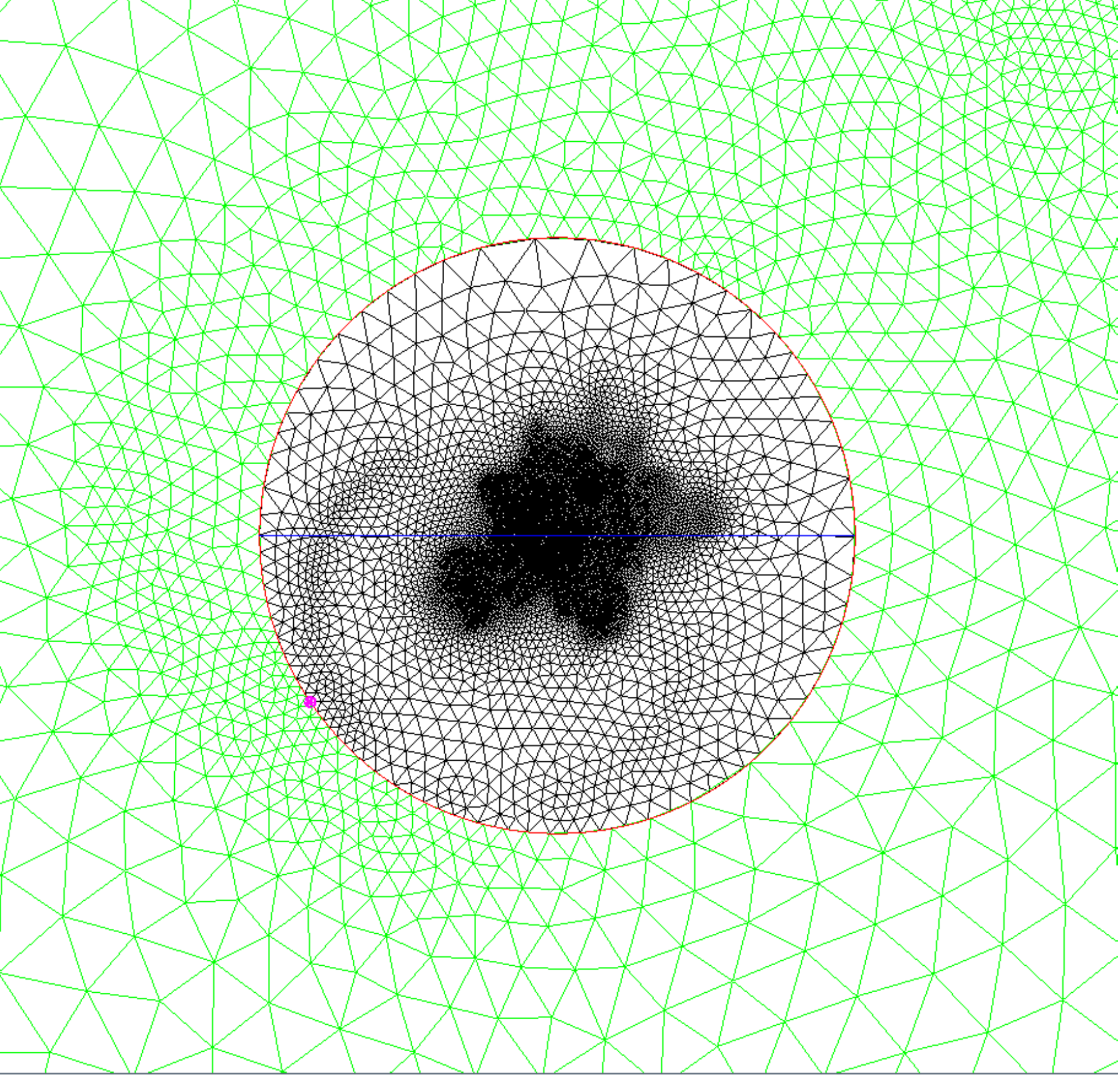}&
\includegraphics[width=0.22\textwidth]{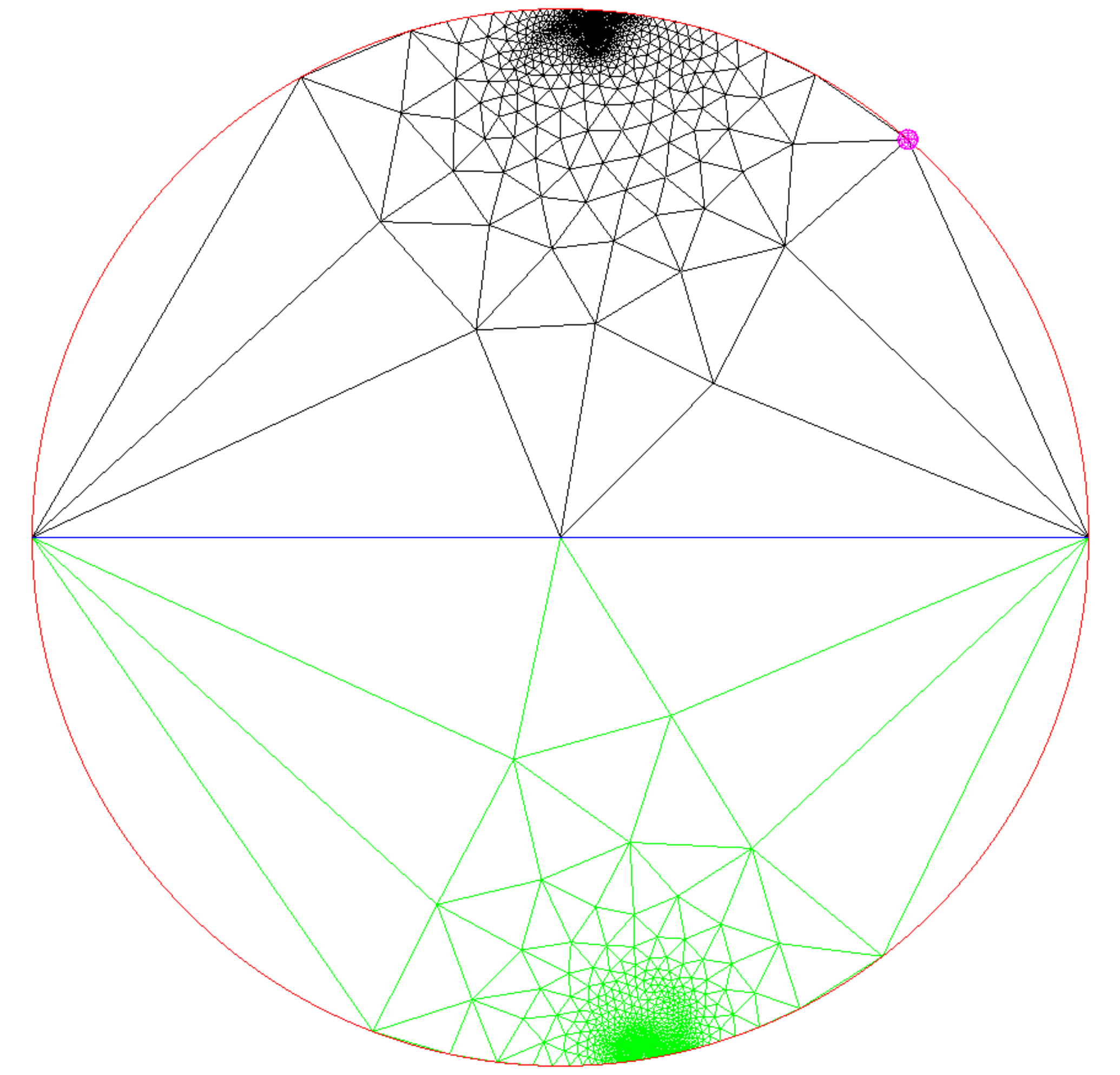}\\
(a) Step 1 & (b) Step 1\\
\includegraphics[width=0.22\textwidth]{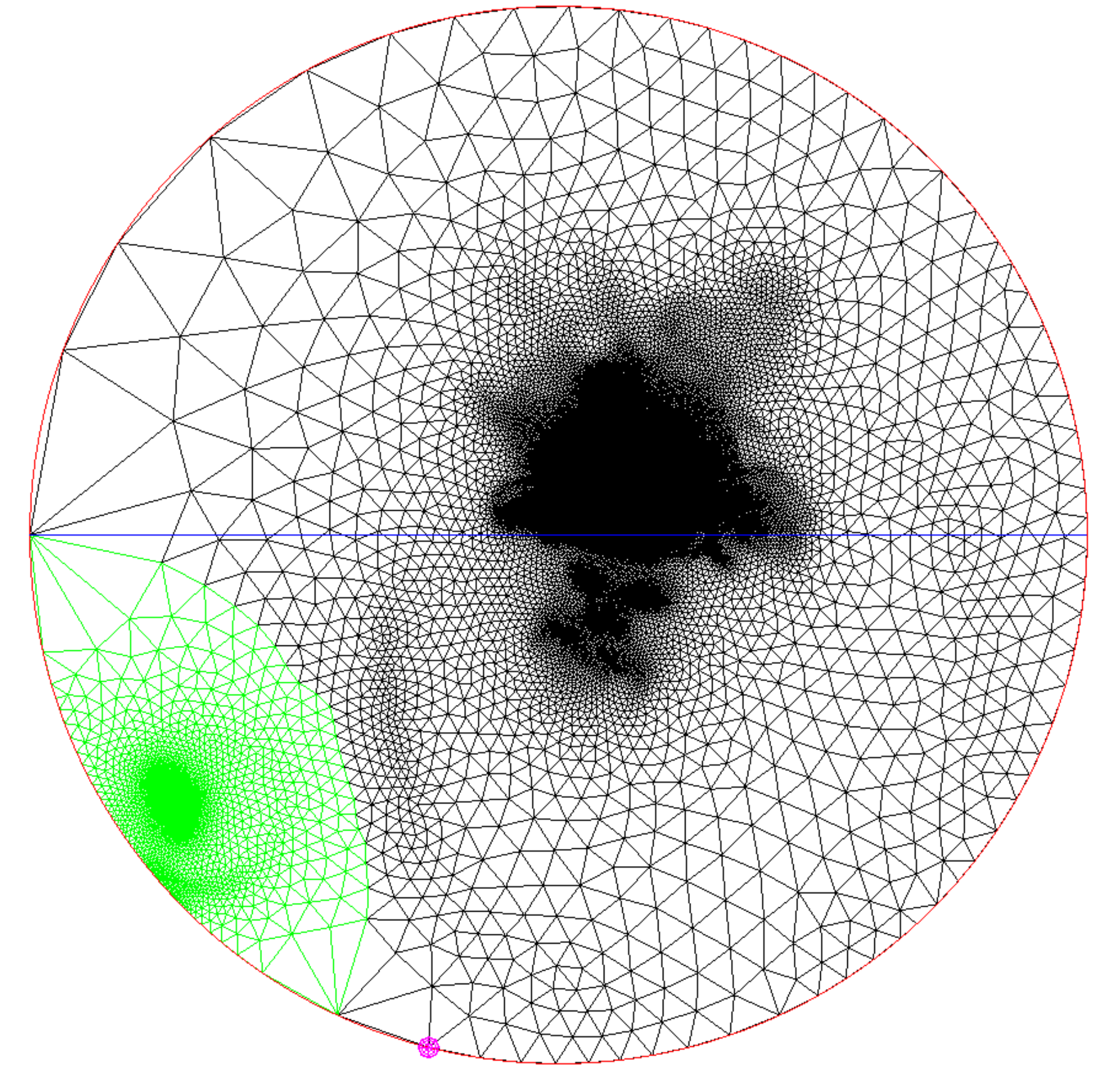}&
\includegraphics[width=0.22\textwidth]{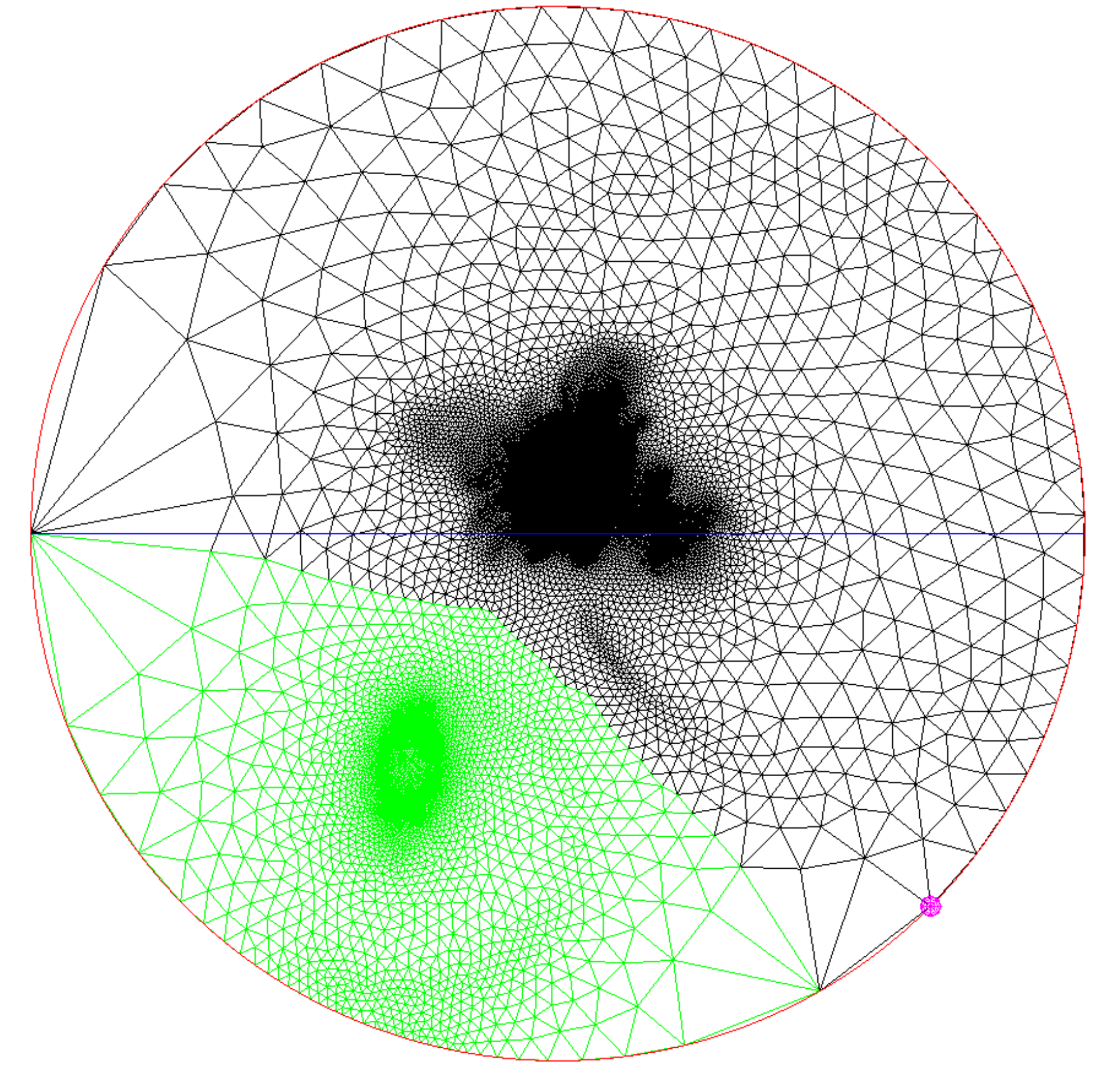}\\
(c) Step 2 & (d) Step 2\\
\includegraphics[width=0.22\textwidth]{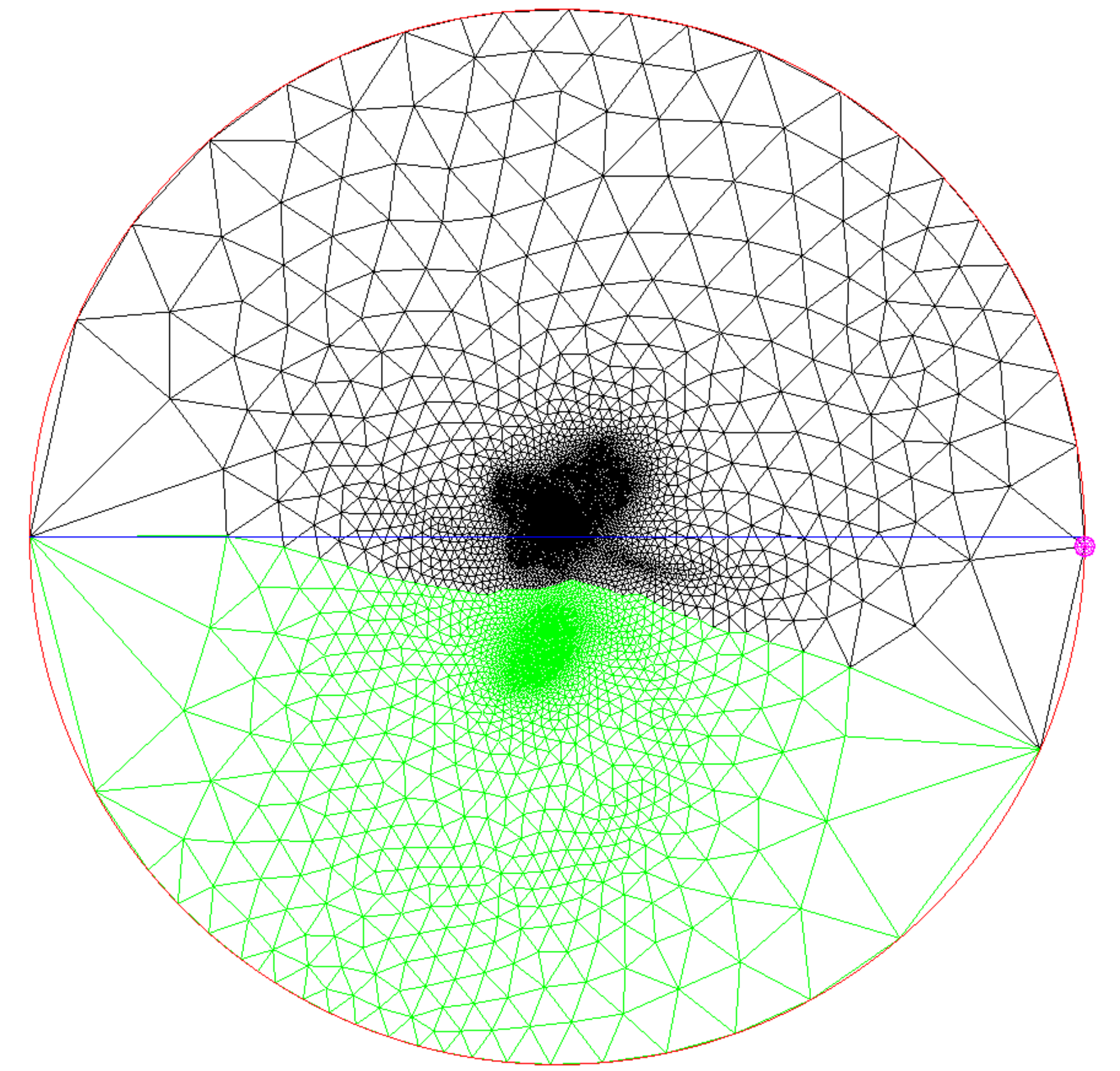}&
\includegraphics[width=0.22\textwidth]{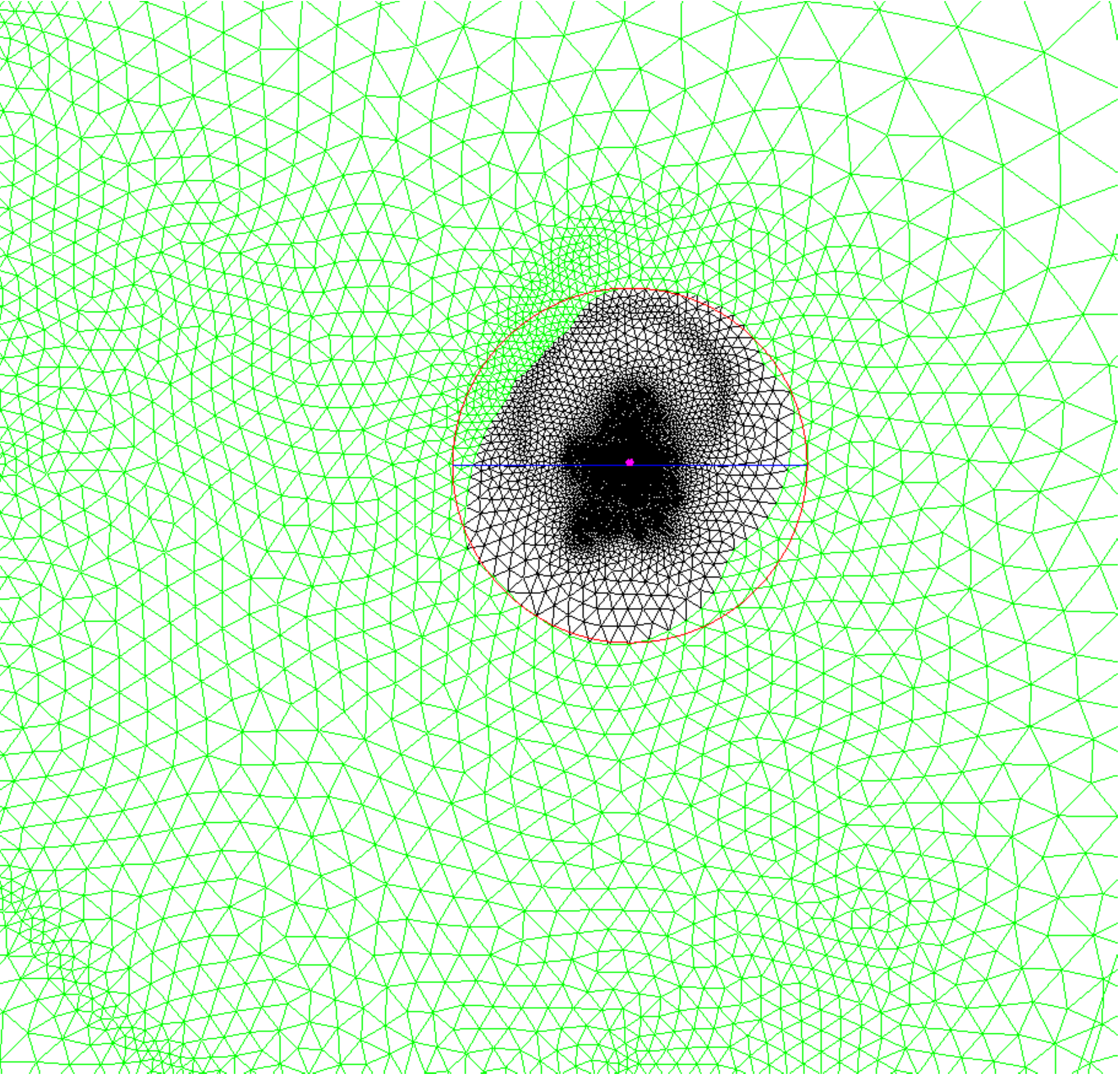}\\
(e) Step 3 & (f) Step 3
\end{tabular}
\end{center}
\vspace{-3mm}
\caption{ Conformal welding. Given the two Riemann mapping results of a given 3D input model (a), the two results are conformally welded by taking one as an interior and the other as an exterior region and following the 3 steps, as highlighted in (b-f).}
\label{fig:conformal_welding}
\vspace{-3mm}
\end{figure}

The conformal mappings between two spherical surface form a 6 dimensional group, the so-called M\"obius transformation group. In our application, we need to add special
normalization conditions to choose a unique one. Our goal is to find the one with balanced mass distribution, namely, the mass center of the image on the unit sphere coincides with the center of the sphere, this removes 3 degrees of freedom. Then we fix the top point and the most frontal point of the surface onto the north hole, and onto the x-axis, this will completely fix the conformal map.

\begin{algorithm}[!h]
\caption{Conformal Spherical Mapping}
    \textbf { Input:}  Closed genus zero surface mesh $M$ with total area $4\pi$.

    \textbf { Output:} A unique diffeomorphic angle preserving mapping $f: M \rightarrow \Omega$, where $\mathbb{D}$ is a unit sphere.

    \textbf{(1)} A cutting loop curve is found as shown in Fig.~\ref{fig:bimba_pipeline}(b), based on the first non-trivial eigenfunction of the Laplace-Beltrami Operator and the surface is segmented into two parts by the cutting loop.

    \textbf{(2)} Each part is conformally mapped onto the planar unit disk using Riemann mapping, as shown in Figs.~\ref{fig:bimba_pipeline}(c) and \ref{fig:bimba_pipeline}(d).\\

    \textbf{(3)} Two planar disks are glued together using the conformal welding method, as shown in Fig.~\ref{fig:bimba_pipeline}(e).\\

    \textbf{(4)} The extended plane is mapped onto the sphere using stereographic projection, as shown in Fig.~\ref{fig:bimba_pipeline}(f).\\

\label{alg:spherical_conformal_map}
\end{algorithm}

\setlength{\tabcolsep}{0pt}
\begin{figure}[h]
\begin{center}
\begin{tabular}{cc}
\includegraphics[height=0.24\textwidth]{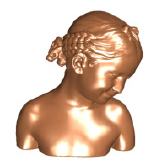}&
\includegraphics[height=0.24\textwidth]{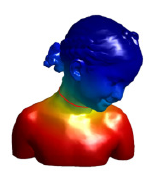}\\
(a) Input surface & (b) Eigenfunction\\
&  and cutting loop\\
\includegraphics[height=0.24\textwidth]{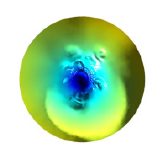}&
\includegraphics[height=0.24\textwidth]{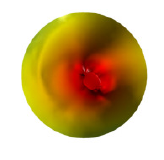}\\
(c) Riemann mapping & (d) Riemann mapping\\
of mesh part & of mesh part\\
above cutting loop in (b) & below cutting loop in (b)\\
\multicolumn{2}{c}
{\includegraphics[height=0.23\textwidth]{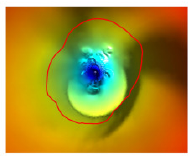}}\\
\multicolumn{2}{c}
{(e) Conformal welding of (c) and (d)}\\
\includegraphics[height=0.23\textwidth]{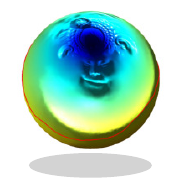}&
\includegraphics[height=0.23\textwidth]{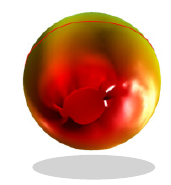}\\
(f) Front view & (g) Back view\\
 spherical mapping &  spherical mapping
\end{tabular}
\end{center}
\vspace{-2.5mm}
\caption{Conformal spherical mapping algorithm pipeline.
\label{fig:bimba_pipeline}}
\vspace{-5mm}
\end{figure}

\setlength{\tabcolsep}{0pt}
\begin{figure}[h]
\begin{center}
\begin{tabular}{cc}
\includegraphics[height=0.24\textwidth]{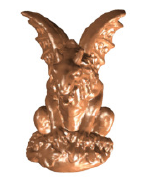}&
\includegraphics[height=0.24\textwidth]{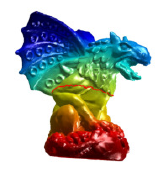}\\
(a) Input surface & (b) Eigenfunction\\
& and cutting loop\\
\multicolumn{2}{c}
{\includegraphics[height=0.23\textwidth]{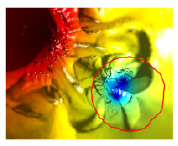}}\\
\multicolumn{2}{c}
{(c) Conformal welding result}\\
\includegraphics[height=0.23\textwidth]{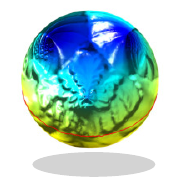}&
\includegraphics[height=0.23\textwidth]{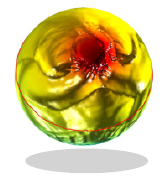}\\
(d) Front view & (e) Back view\\
spherical mapping & spherical mapping
\end{tabular}
\end{center}
\vspace{-2.5mm}
\caption{Conformal spherical mapping for the gargoyle model.
\label{fig:gargoyle2_pipeline}}
\vspace{-5mm}
\end{figure}

\setlength{\tabcolsep}{0pt}
\begin{figure}[h]
\begin{center}
\begin{tabular}{cc}
\includegraphics[height=0.25\textwidth]{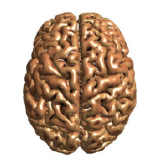}&
\includegraphics[height=0.25\textwidth]{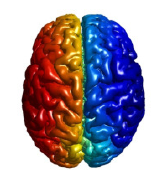}\\
(a) Input surface & (b) Eigenfunction\\
& and cutting loop\\
\multicolumn{2}{c}
{\includegraphics[height=0.23\textwidth]{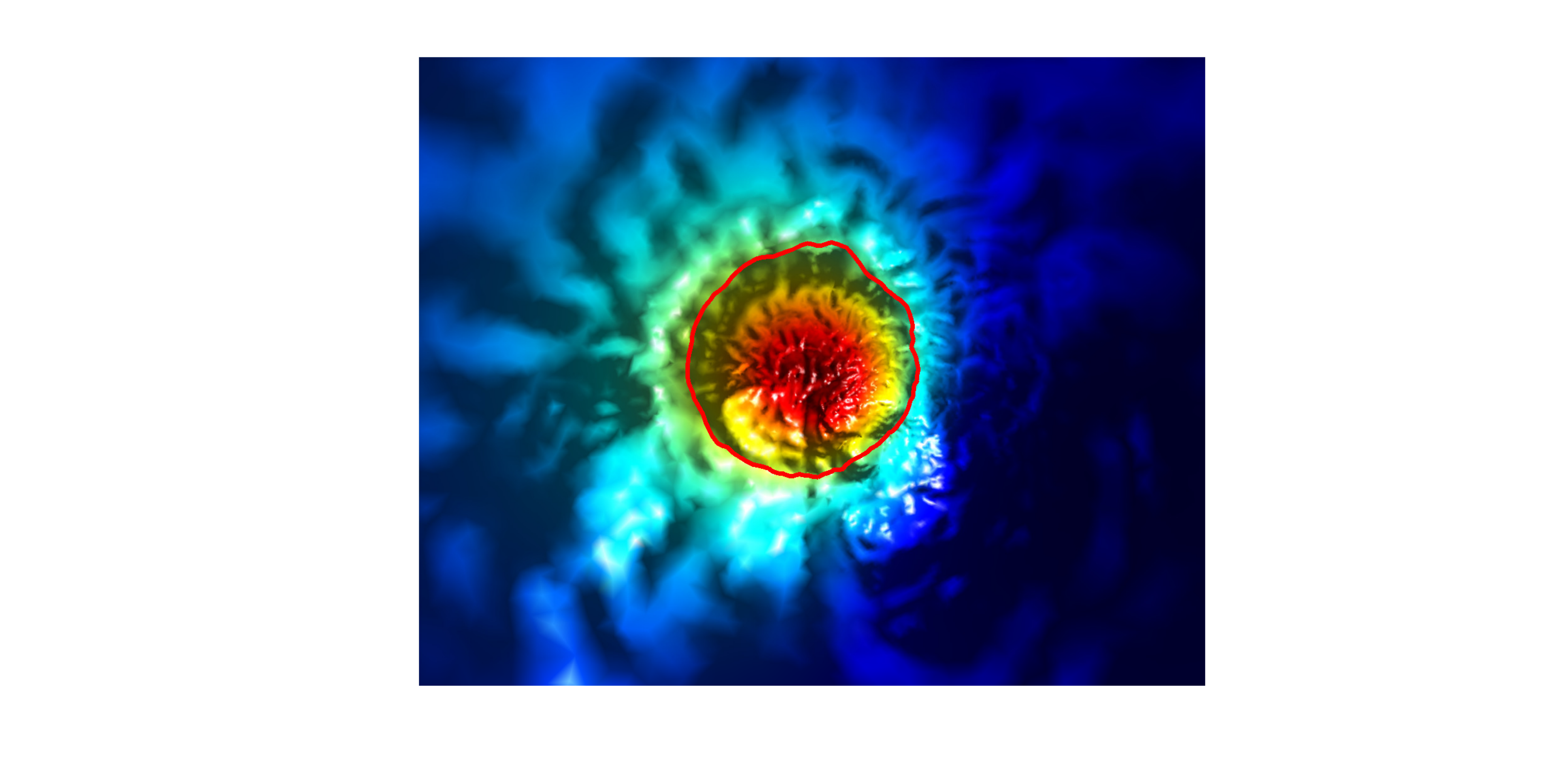}}\\
\multicolumn{2}{c}
{(c) Conformal welding result}\\
\includegraphics[height=0.235\textwidth]{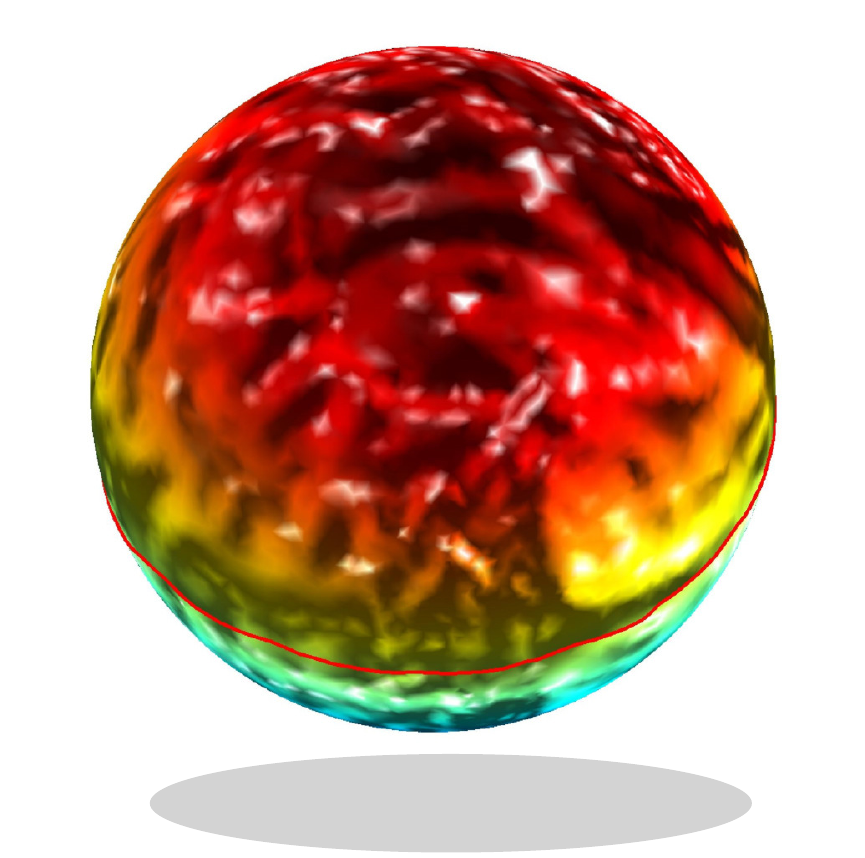}&
\includegraphics[height=0.24\textwidth]{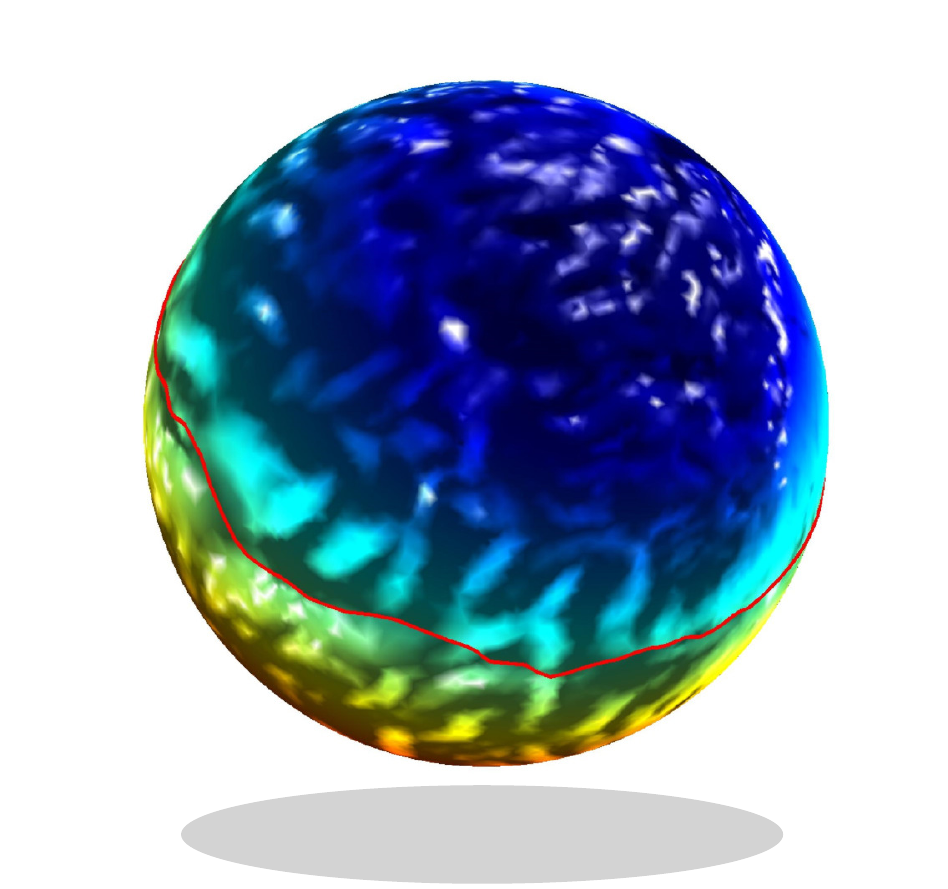}\\
(d) Front view & (e) Back view\\
spherical mapping & spherical mapping
\end{tabular}
\end{center}
\vspace{-2.5mm}
\caption{Conformal spherical mapping for the cortical surface.
\label{fig:brain_pipeline}}
\vspace{-5mm}
\end{figure}

\vspace{-3mm}
\subsection{Area-Preserving Spherical Mapping}
\label{sec:map_area}
This section introduces the optimal transportation map. The algorithmic details of optimal transport map can be found in our previous work \cite{zhao2013}. This is necessary for computing area-preserving spherical mapping. Area distortions can be completely eliminated by the optimal mass transport map. We recently
developed a variational principle for discrete optimal mass
transport map between domains in Euclidean space \cite{gu:2013:arXiv}.

\vspace{-2.5mm}
\subsubsection{Discrete Optimal Mass Transport Map}
Given the source and the target $(X,\mu)$ and $(Y,\nu)$, suppose
$\mu$ has compact support:
\[
    \Omega =supp~\mu:=\{x\in X| \mu(x)>0\},
\]
and $Y$ is discretized to $Y=\{y_1,y_2,\cdots,y_k\}$ with Dirac measure:
\[
    \nu=\sum_{j=1}^k \nu_j(y-y_j).
\]
Furthermore, the total mass is equal $\int_\Omega \mu(x)dx = \sum_j \nu_j$.

The optimal mass transport map $\varphi:(X,\mu)\to (Y,\nu)$ is
measure-preserving:
\[
    \int_{\varphi^{-1}(y_i)} \mu(x)dx = \nu_i,
\]
and minimizes the quadratic transport cost:
\[
    \varphi = argmin_{\tau_{\#}\mu = \nu} \int_\Omega \|x-\varphi(x)\|^2 \mu(x)dx.
\]
According to Brenier's theorem, there is a convex
function $u:\Omega\to \mathbb{R}$, such that the optimal map is
given by the gradient map of $u$, $\varphi: x\mapsto \nabla u(x)$. The
convex function $u$ can be approximated by a piecewise linear
function, constructed as follows. For each point $y_i \in Y$, one constructs a hyperplane $\pi_i:
\langle x,y_i\rangle + h_i = 0$, where the piecewise linear convex
function is defined by:
\begin{equation}
    u_{\mathbf{h}}(x) := \max_{i=1}^k \{ \langle x,y_i \rangle + h_i
    \}, \hspace{5mm} \mathbf{h}=(h_1,h_2,\cdots,h_n),
    \label{eqn:convex_function}
\end{equation}
where the heights $\mathbf{h}$ are unknowns. The gradient map of
$u_\mathbf{h}$ maps $X$ to discrete points $\{y_i\}$, the preimages
of all $y_i$'s partition $\Omega$ and each cell is denoted as
$W_i(\mathbf{h})$,
\begin{equation}
    \Omega = \bigcup_{i=1}^k W_i(\mathbf{h}) = \bigcup_{i=1}^k
    \{x\in \Omega | u_\mathbf{h}(x) = \langle x,y_i \rangle + h_i
    \}.
   \label{eqn:cell_decomposition}
\end{equation}
The total measure of each cell is denoted as $w_i(\mathbf{h})$. When
$\nabla u_\mathbf{h}$ is the optimal mass transport map,
$\mathbf{h}$ satisfies the following measure-preserving condition:
\begin{equation}
    w_i(\mathbf{h}) = \int_{W_i(\mathbf{h})} \mu(x) dx = \nu_i, \hspace{5mm} i =
    1, 2, \cdots, k.
    \label{eqn:area_constraint}
\end{equation}

The following theorem has been recently proved by the authors \cite{gu:2013:arXiv}, which lays down the algorithm foundation.

\begin{theorem}[Discrete Optimal Mass Transport Map] For any given measures $\mu$ and $\nu$ with equal total mass,
there must exist a height vector $\mathbf{h}$ unique up to adding a
constant vector $(c,c,\cdots, c)$. The convex function
(Eqn.~\ref{eqn:convex_function}) induces the cell decomposition of
$\Omega$ (Eqn. \ref{eqn:cell_decomposition}), such that the
\emph{area-preserving constraints} (Eqn.
\ref{eqn:area_constraint}) are satisfied. The gradient map
$grad~u_\mathbf{h}$ is the optimal mass transport map.
Furthermore, the height vector $\mathbf{h}$ is the unique global
optima of the convex energy:
\begin{equation}
    E(\mathbf{h}) = \int_{\Omega} u_\mathbf{h}(x) \mu(x) dx - \sum_{i=1}^k \nu_i
    h_i.
    \label{eqn:energy_3}
\end{equation}
\end{theorem}

The existence and uniqueness was first proven by Alexandrov
\cite{Alexandor} using a topological method; the existence was also
proven by Aurenhammer \cite{Aurenhammer:1987:PDP}; and the uniqueness and
optimality was proven by Brenier \cite{Brenier}. Gu et al.
\cite{gu:2013:arXiv} have provided a novel proof for the existence and
uniqueness based on the variational principle. The deep insight of
the variational framework provides us excellent opportunities for
designing the computational algorithm.

The optimal transport map algorithm is for optimizing the convex
energy using Newton's method in the admissible space of height
vectors:
\[
H_0:=\{ \mathbf{h} | \sum_{j=1}^k h_j = 0~and~
w_i(\mathbf{h}) > 0, \forall i=1,\cdots,k, \}.
\]
The gradient of the energy is given by:
\[
    \nabla E(\mathbf{h}) = (w_1(\mathbf{h})-\nu_1, w_2(\mathbf{h})-\nu_2, \cdots, w_k(\mathbf{h}) - \nu_k)^T.
\]
Suppose the cells $W_i(\mathbf{h})$ and $W_j(\mathbf{h})$ intersect at an edge $e_{ij} = W_i(\mathbf{h})\cap
W_j(\mathbf{h}) \cap \Omega$, then the Hessian of $E(\mathbf{h})$ is
given by:
\begin{equation}
    \frac{\partial^2 E(\mathbf{h})}{\partial h_i \partial h_j } = \left\{
    \begin{array}{ll}
    \int_{e_{ij}} \mu(x)dx /|y_j-y_i| & W_i(\mathbf{h}) \cap W_j(\mathbf{h})\cap \Omega \neq \emptyset\\
    0  & otherwise\\
    \end{array}
    \right.
    \label{eqn:hessian}
\end{equation}

In practice, the algorithm can be carried out using conventional
computational geometry algorithms. Computing the convex function (Eqn.~\ref{eqn:convex_function}) is equivalent to finding the upper envelope
of planes, computing the cell decomposition (Eqn.~\ref{eqn:cell_decomposition}) is equivalent to computing the power
Voronoi diagram.

Let $P={p_1,p_2,\cdots,p_n}$ be a set of sites on the plane, each $p_i$ is with a power $h_i$. The power distance between $p_i$ and $q$ is given by:
\[
    Pow(q,p_i):= \langle p_i - q,p_i - q \rangle + h_i,
\]
where $\langle, \rangle$ is the Euclidean inner product. The power Voronoi diagram of ${(p_i,h_i)}$ is a partition of the plane into cells:
\[
    \mathbb{R}^2 = \bigcup_{i=1}^n W_i,
\]
where each Voronoi cell is defined as:
\[
    W_i := {q \in \mathbb{R}^2 | Pow(q,p_i) \leq Pow(q,p_j),\forall_j}.
\]
The computation of power Voronoi is equivalent to computing the upper envelope of the planes:
\[
    {f_i (q) := \langle p_i, q \rangle - \frac{1}{2} (h_i + \langle p_i, p_i \rangle)}.
\]

\setlength{\tabcolsep}{2pt}
\begin{figure}[h]
\begin{center}
\begin{tabular}{cc}
\includegraphics[height=0.15\textwidth]{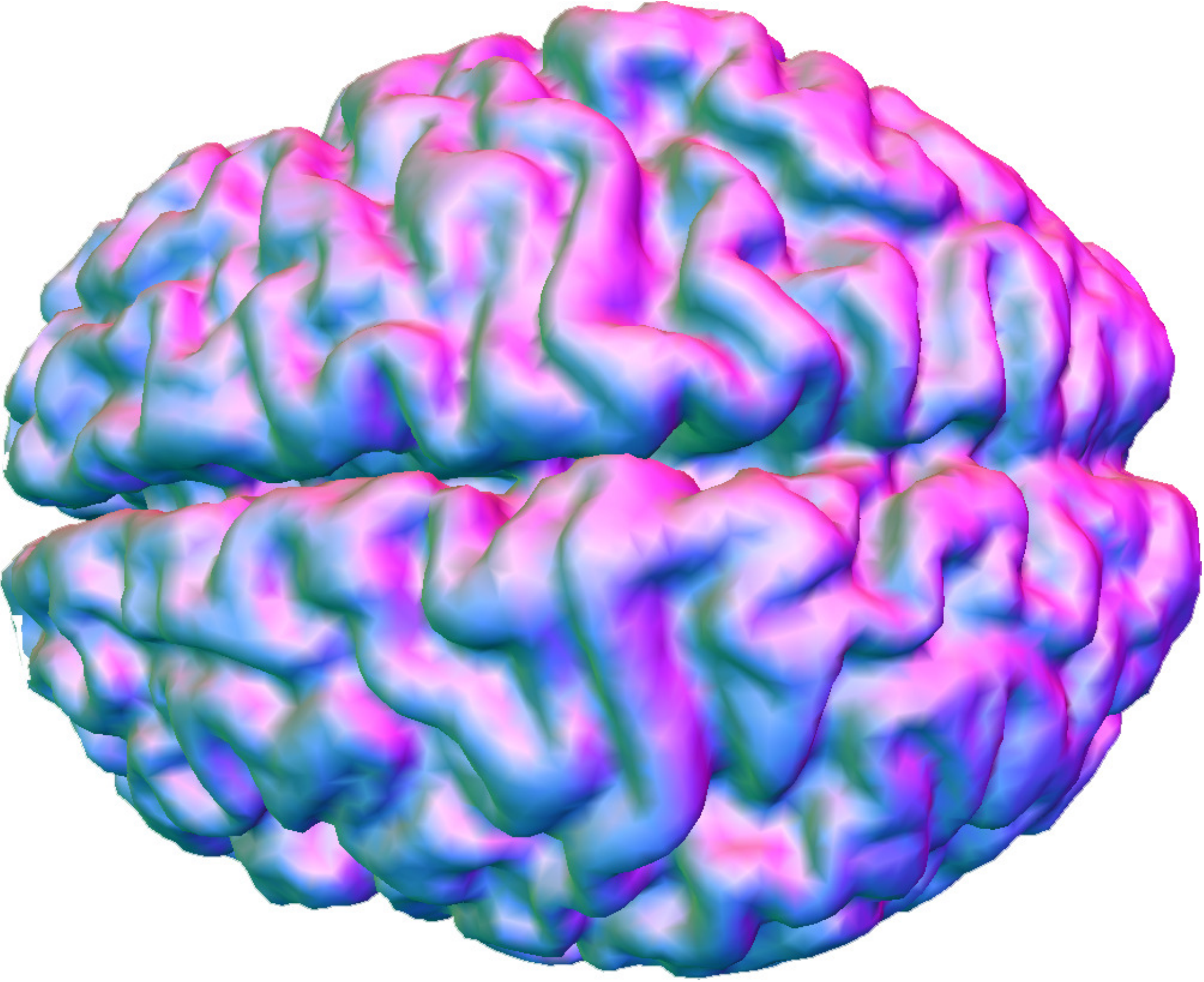}&
\includegraphics[height=0.15\textwidth]{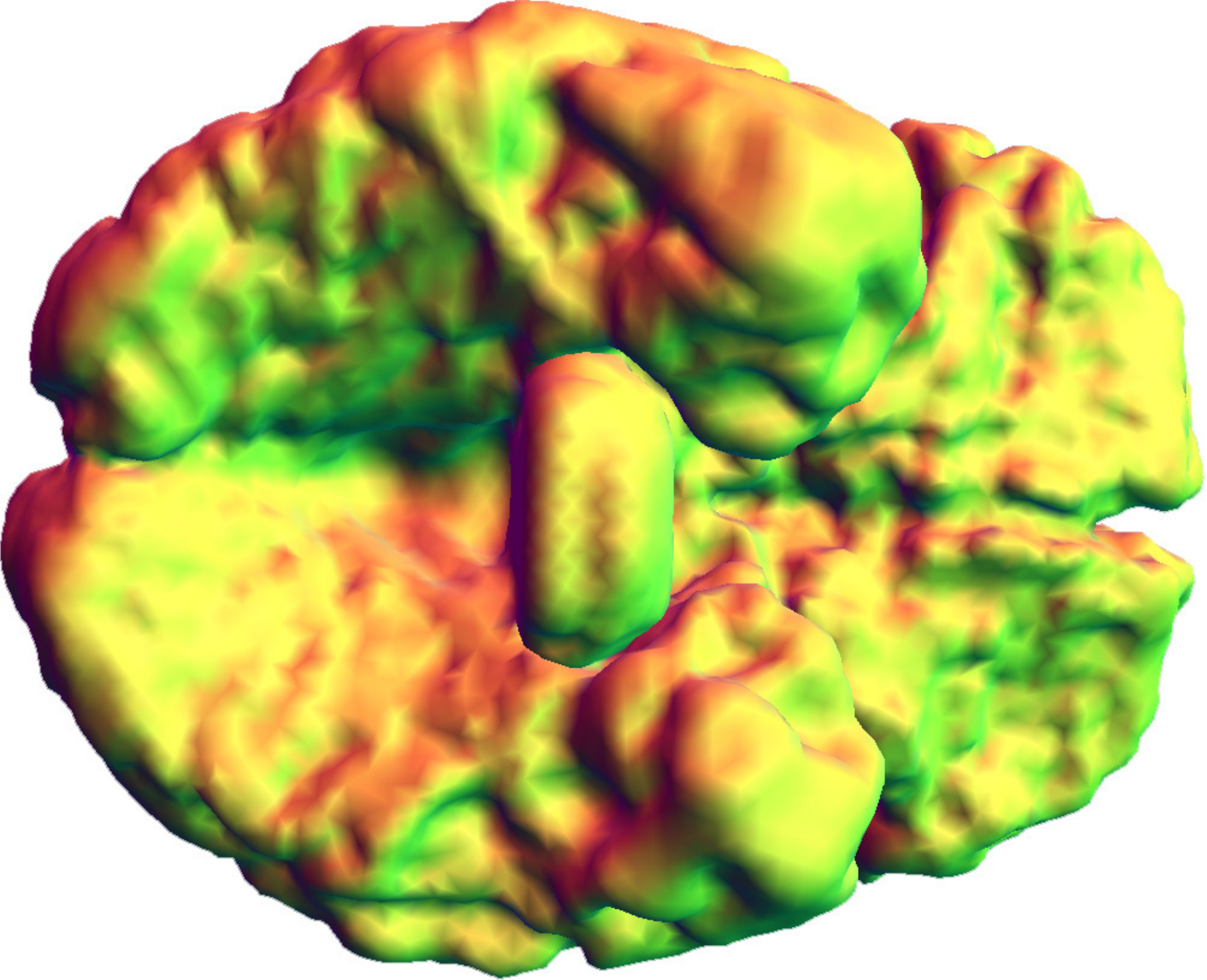}\\
(a) Brain top view& (b) Brain bottom view\\
\includegraphics[height=0.20\textwidth]{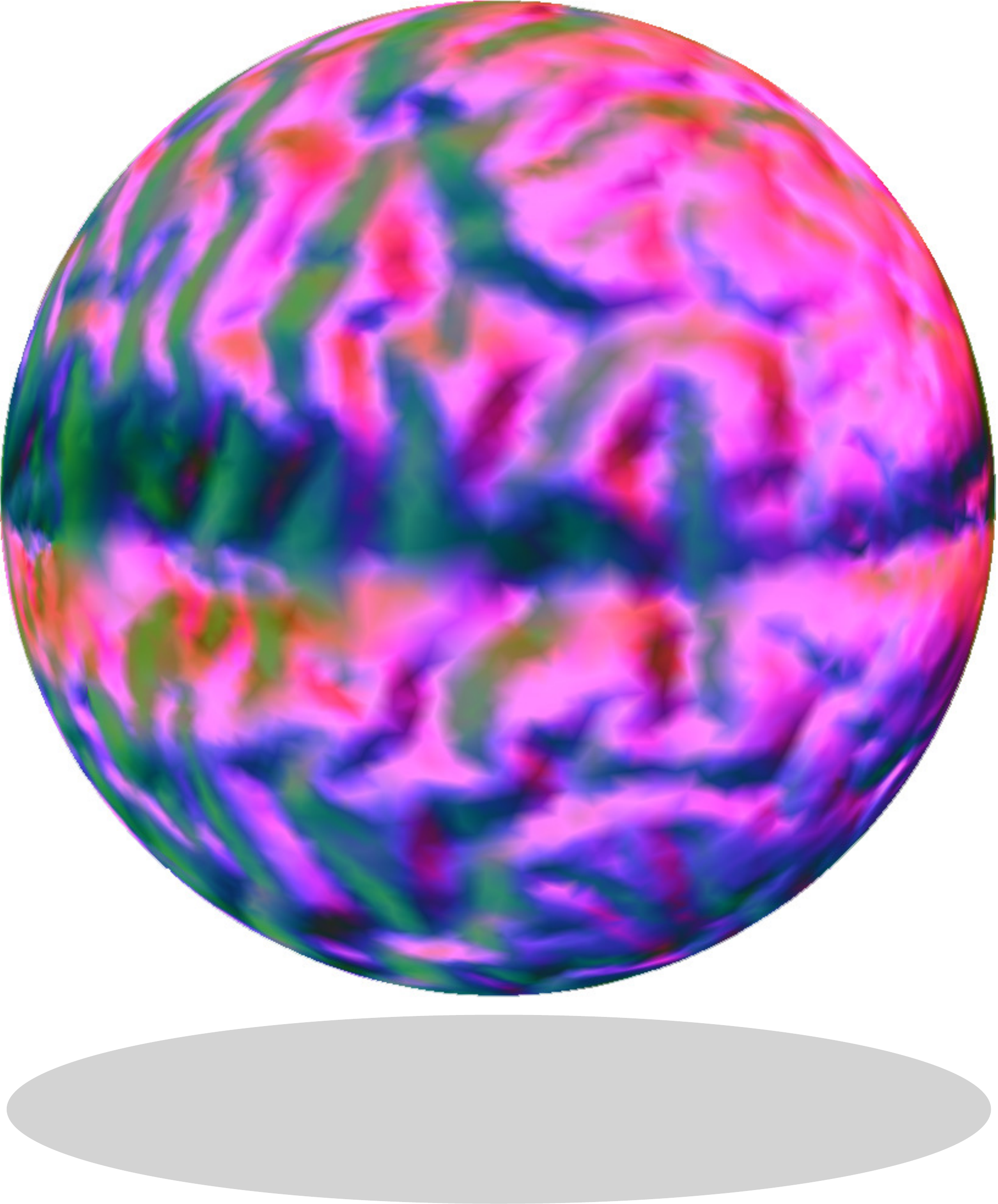}&
\includegraphics[height=0.20\textwidth]{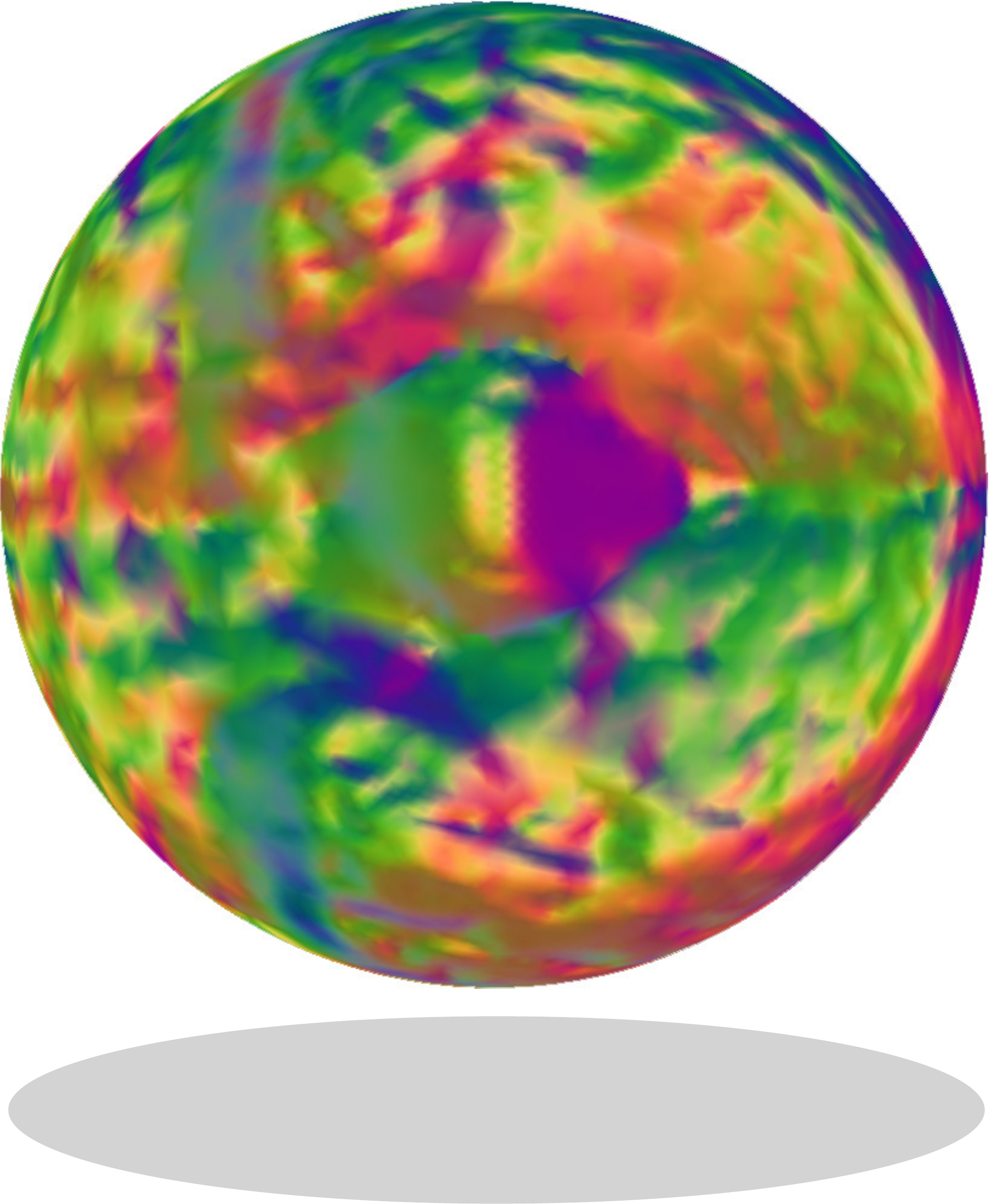}\\
(c) Conformal map & (d) Conformal map\\
 top view & bottom view\\
\includegraphics[height=0.20\textwidth]{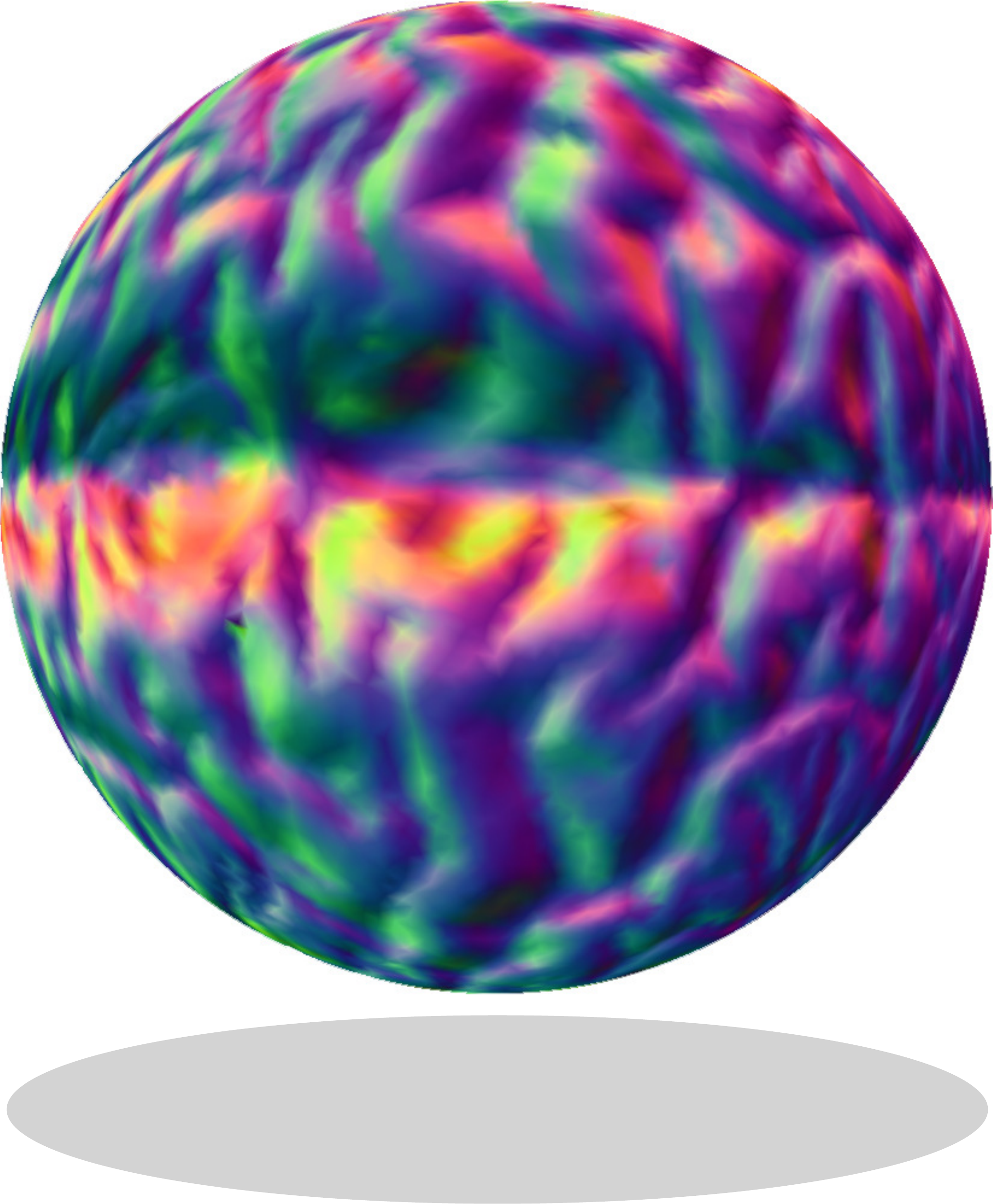}&
\includegraphics[height=0.20\textwidth]{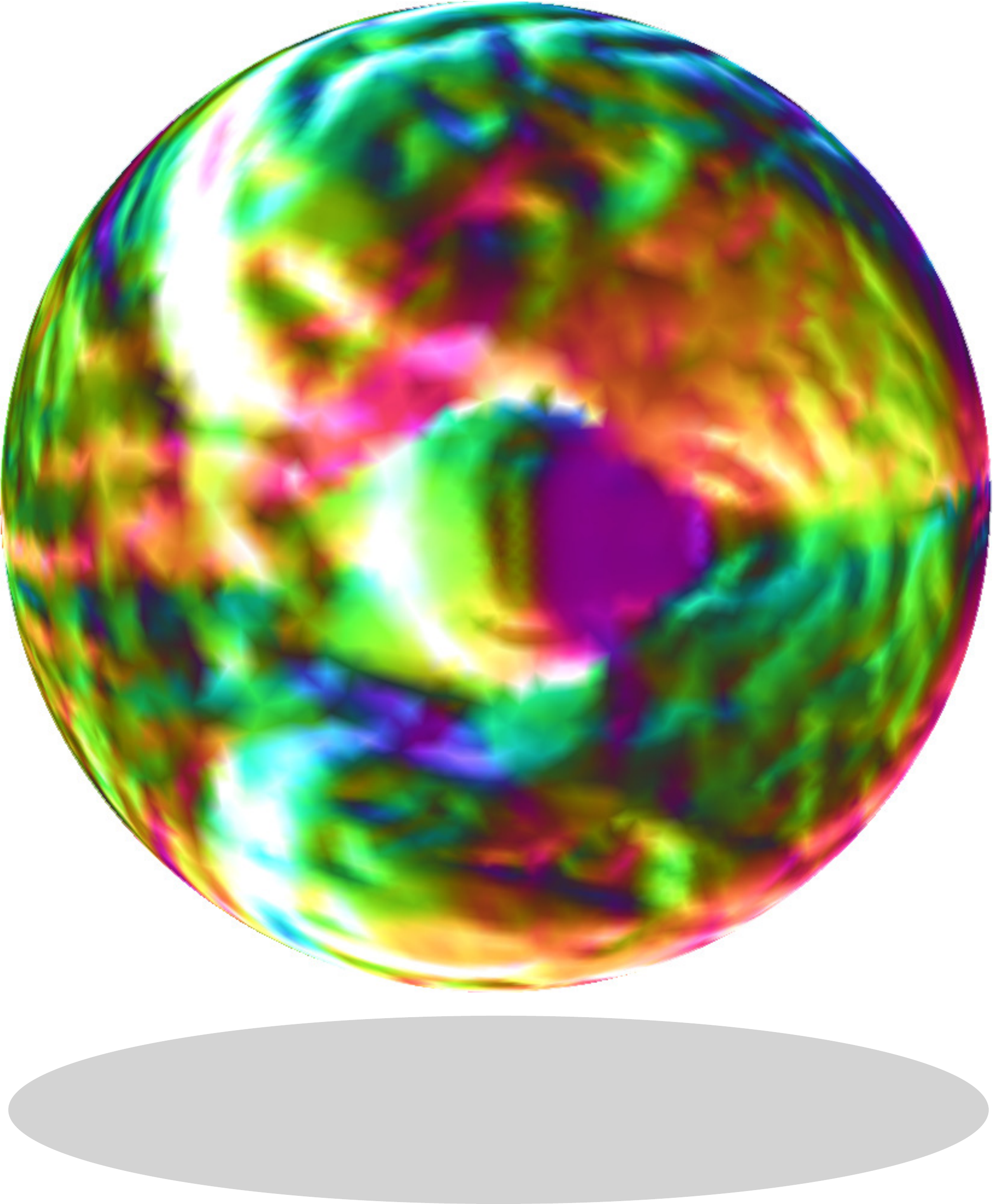}\\
(e) Area-preserving & (f) Area-preserving\\
map top view & map bottom view
\end{tabular}
\end{center}
\vspace{-3mm}
\caption{Conformal and area-preserving spherical mappings for the cortical surface.
\label{fig:brain_conformal_ap}}
\vspace{-5mm}
\end{figure}

\setlength{\tabcolsep}{0pt}
\begin{figure}[h]
\begin{center}
\begin{tabular}{cc}
\includegraphics[height=0.21\textwidth]{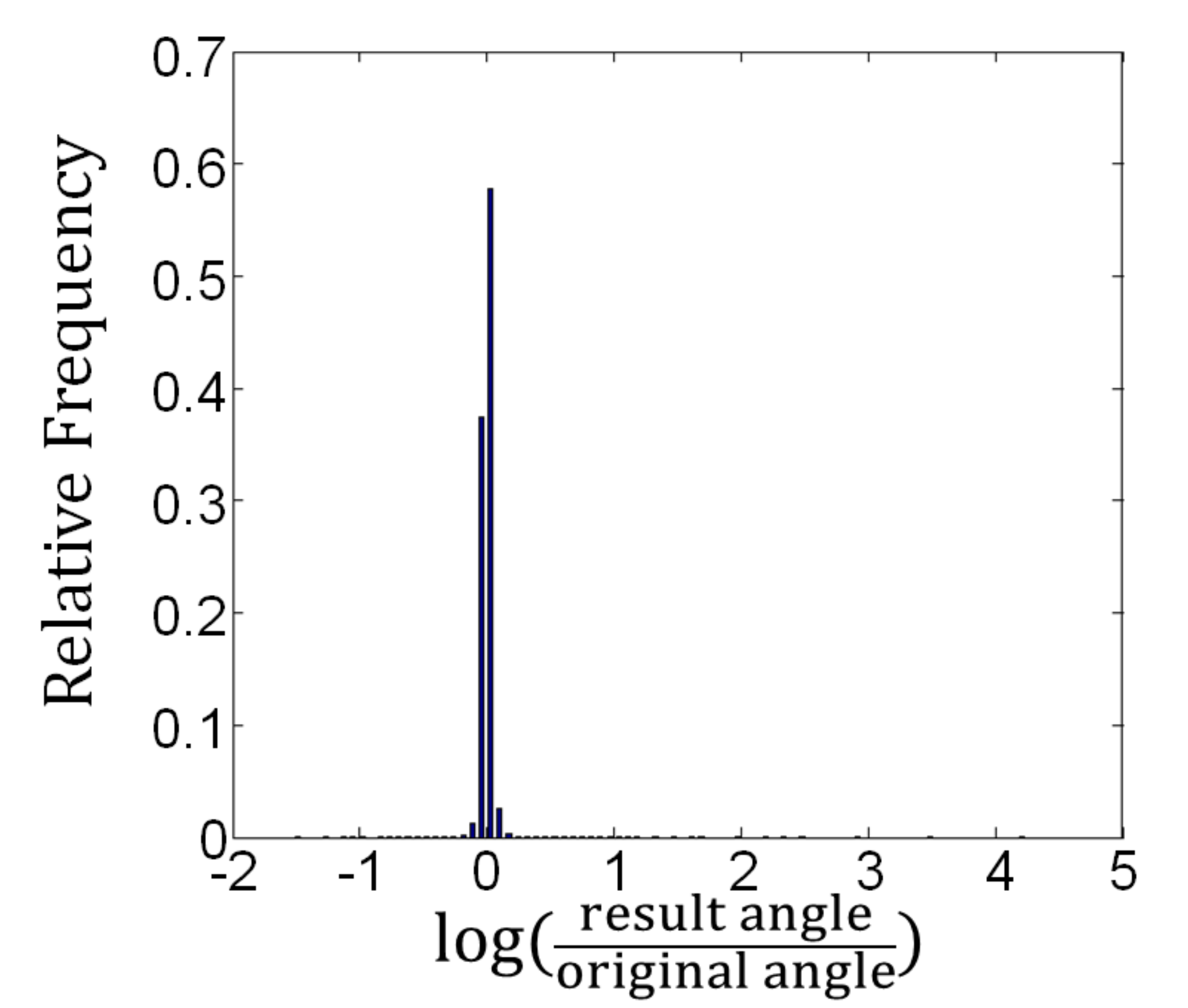}&
\includegraphics[height=0.21\textwidth]{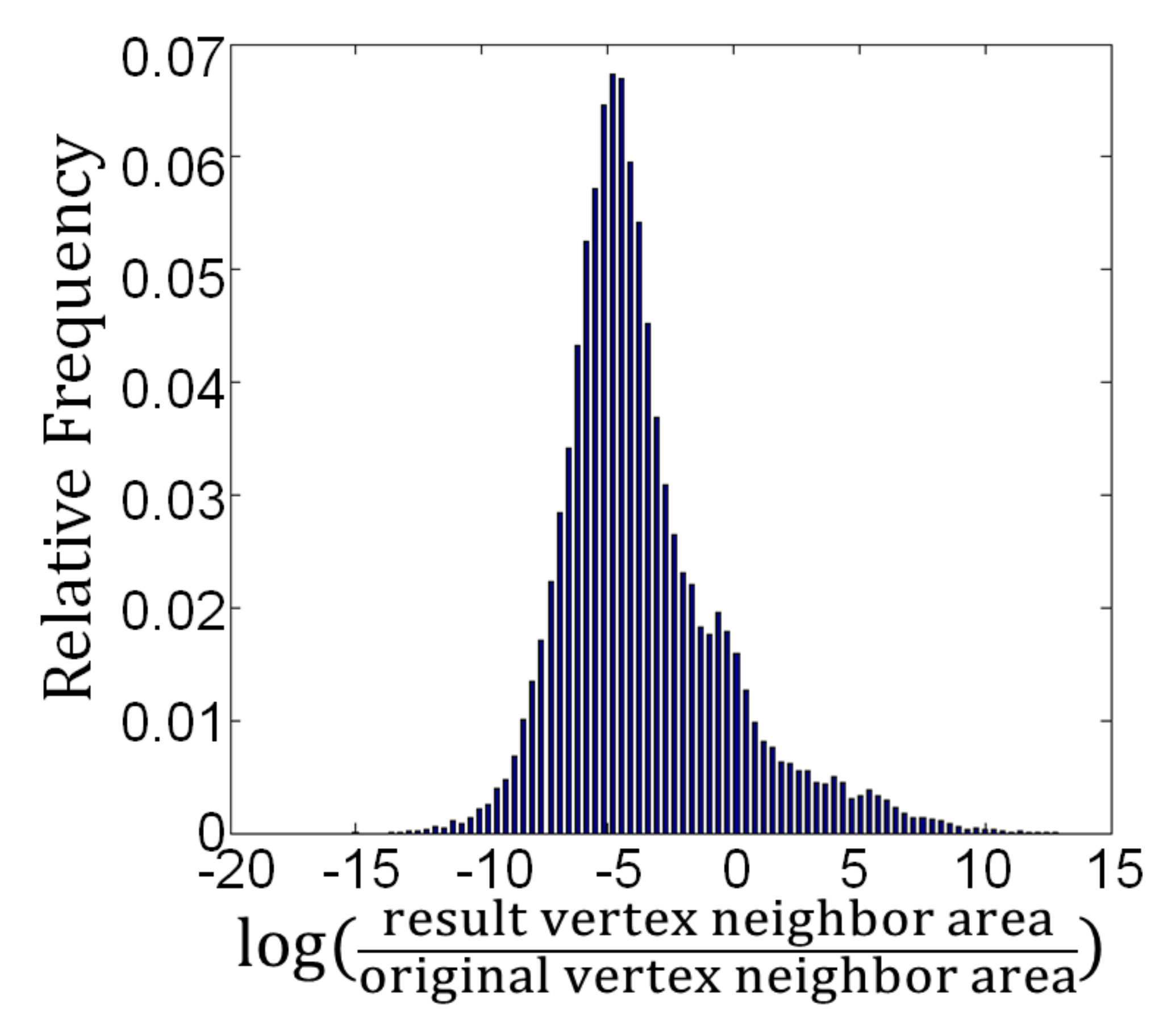}\\
(a) Angle distortion of & (b) Area distortion of\\
conformal mapping & conformal mapping\\
\includegraphics[height=0.21\textwidth]{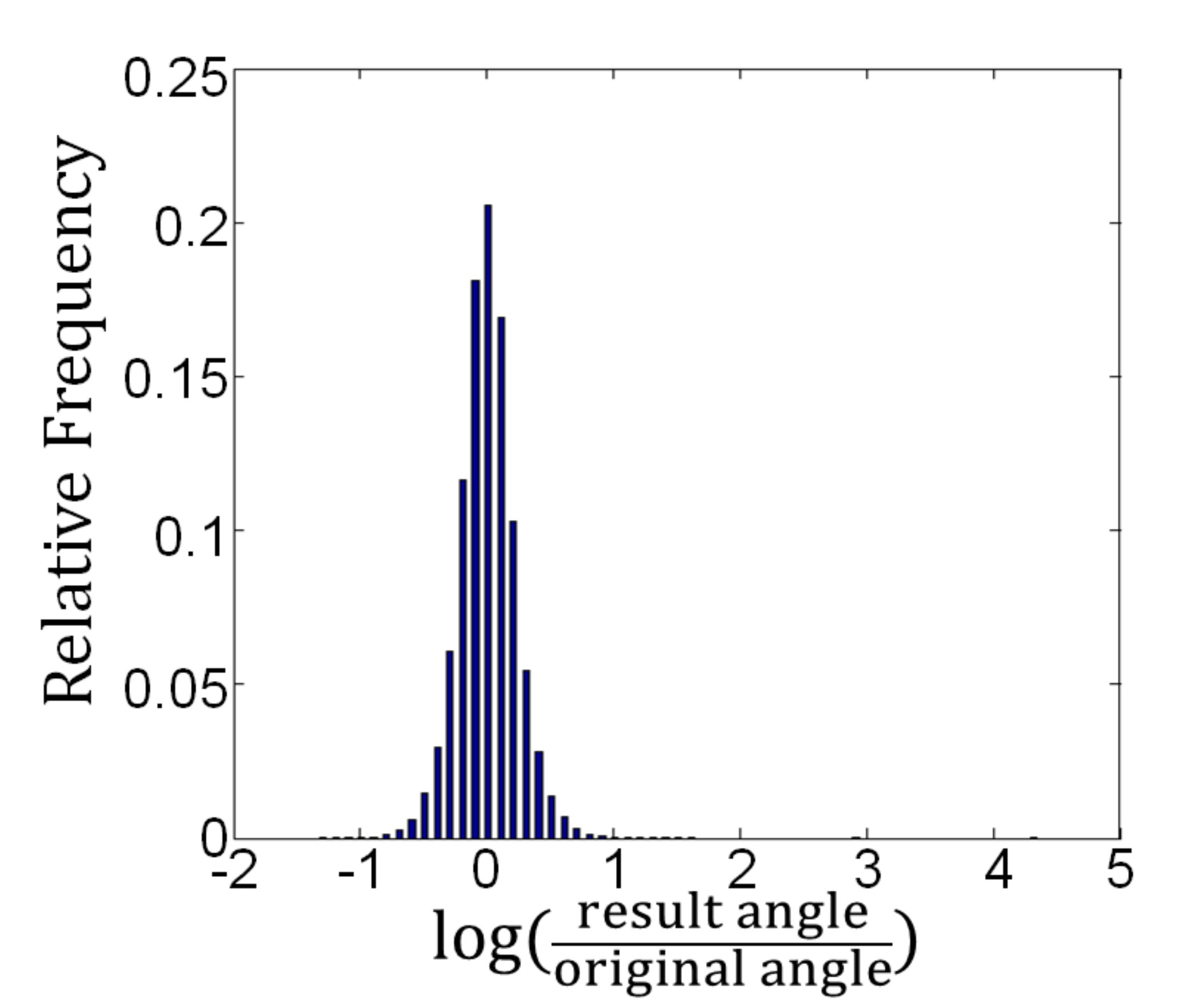}&
\includegraphics[height=0.21\textwidth]{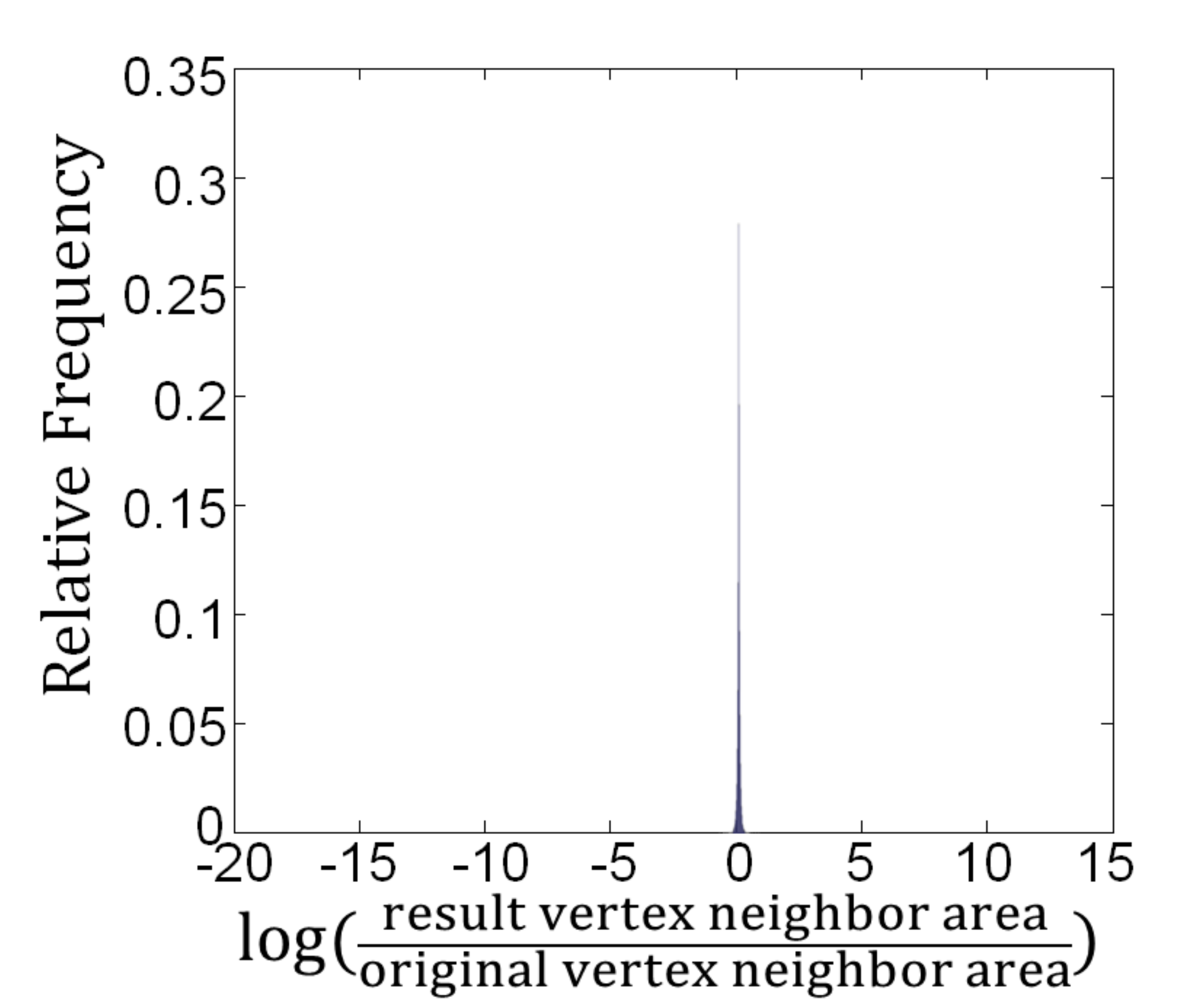}\\
(c) Angle distortion of & (d) Area distortion of\\
area-preserving mapping & area-preserving mapping\\
\end{tabular}
\end{center}
\vspace{-3mm}
\caption{Angle-distortion and area-distortion histograms for conformal mapping and area-preserving mapping of the cortical surface mapping in Fig.~\ref{fig:brain_conformal_ap}.\label{fig:brain_hist}}
\vspace{-4mm}
\end{figure}

\setlength{\tabcolsep}{5pt}
\begin{figure*}[ht!]
\begin{center}
\begin{tabular}{cccccc}
\includegraphics[height=0.17\textwidth]{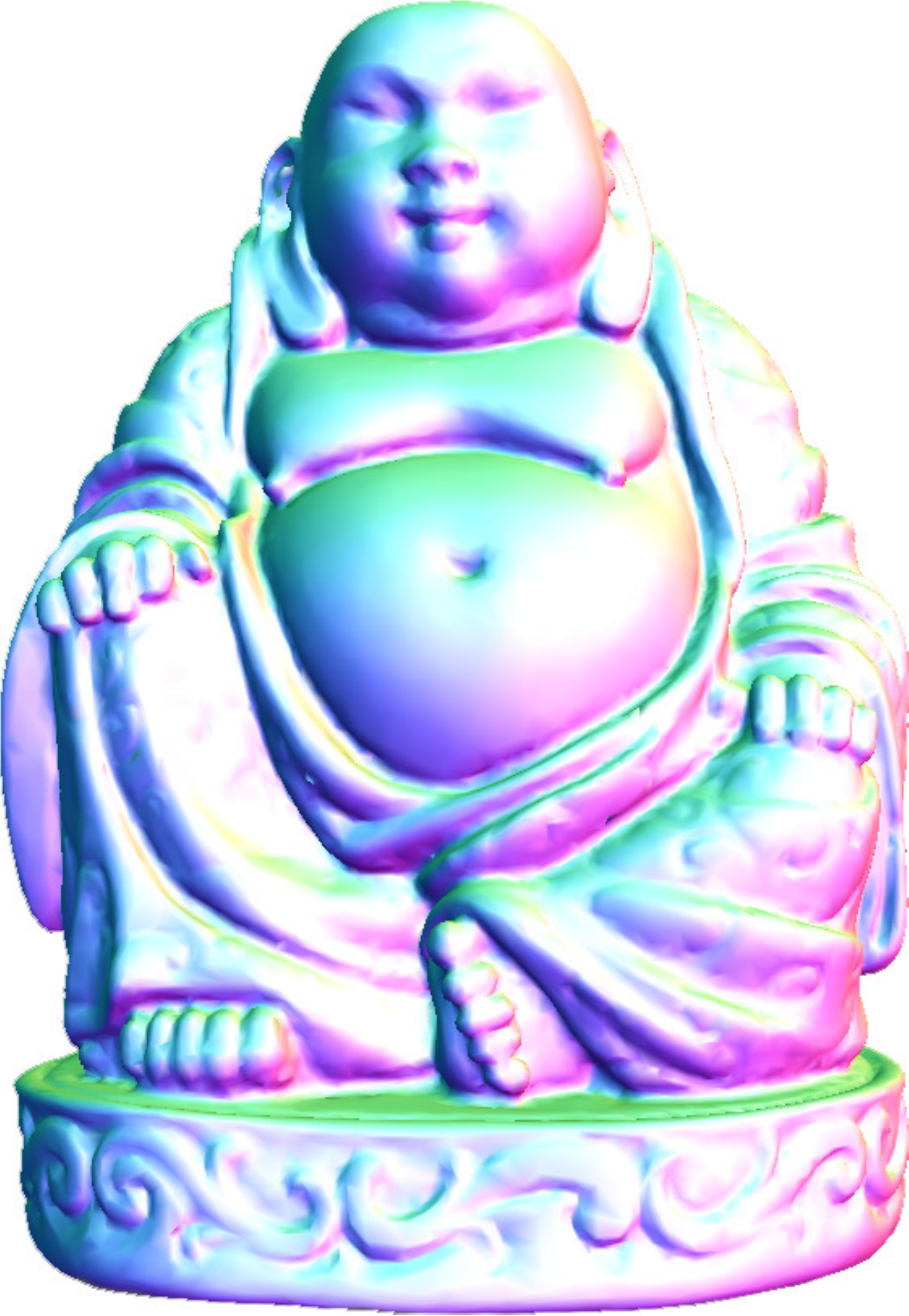}&
\includegraphics[height=0.17\textwidth]{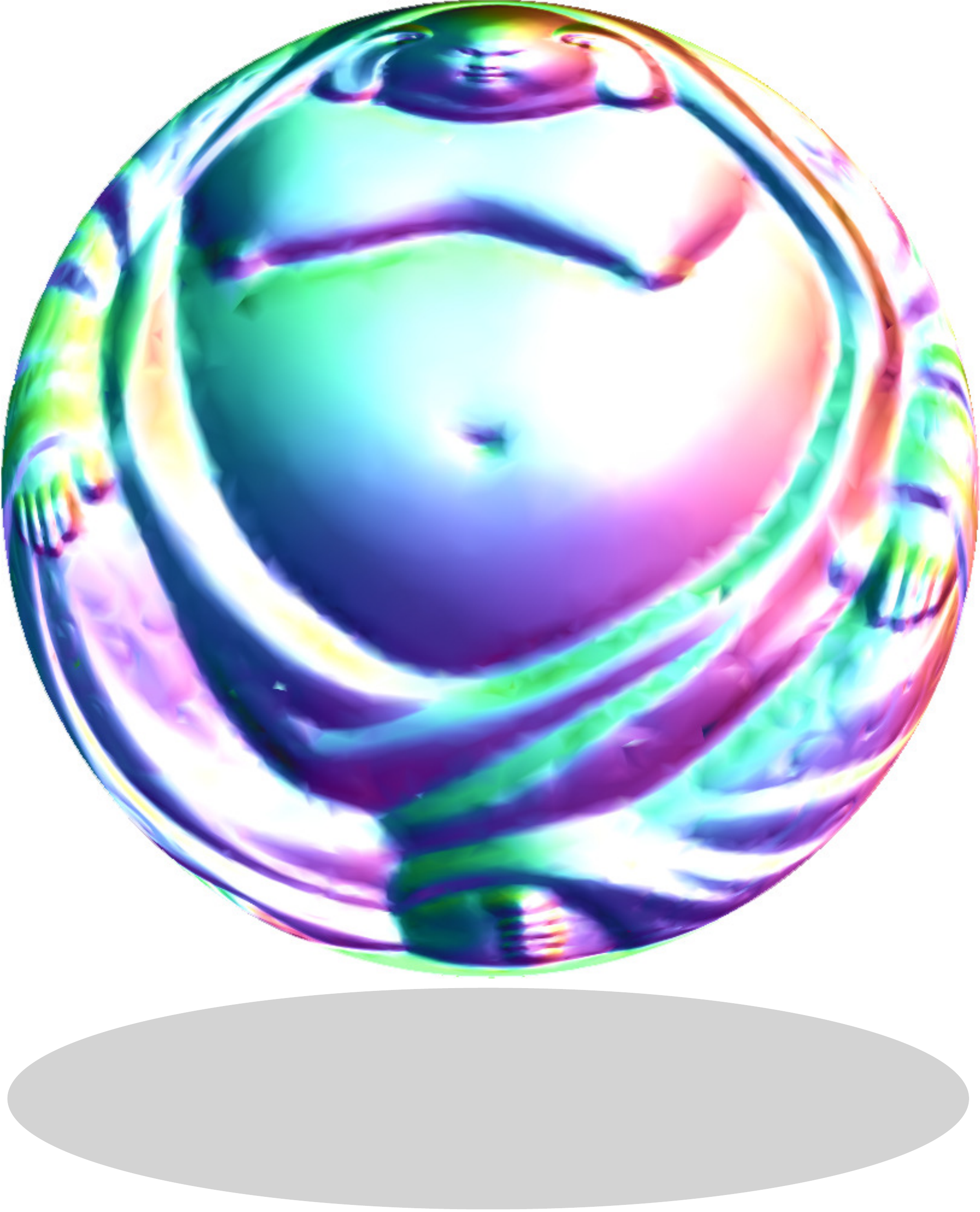}&
\includegraphics[height=0.17\textwidth]{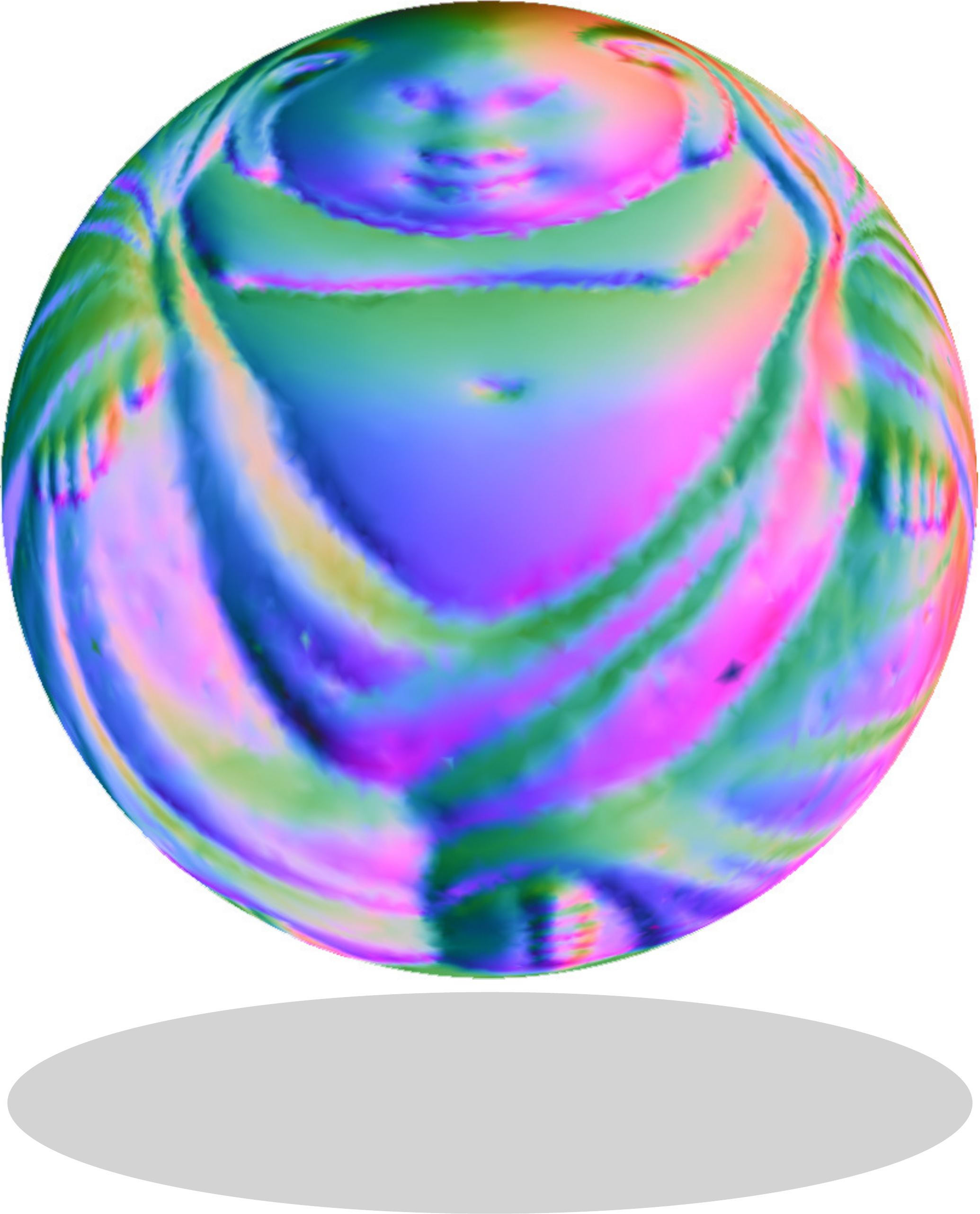}&
\includegraphics[height=0.17\textwidth]{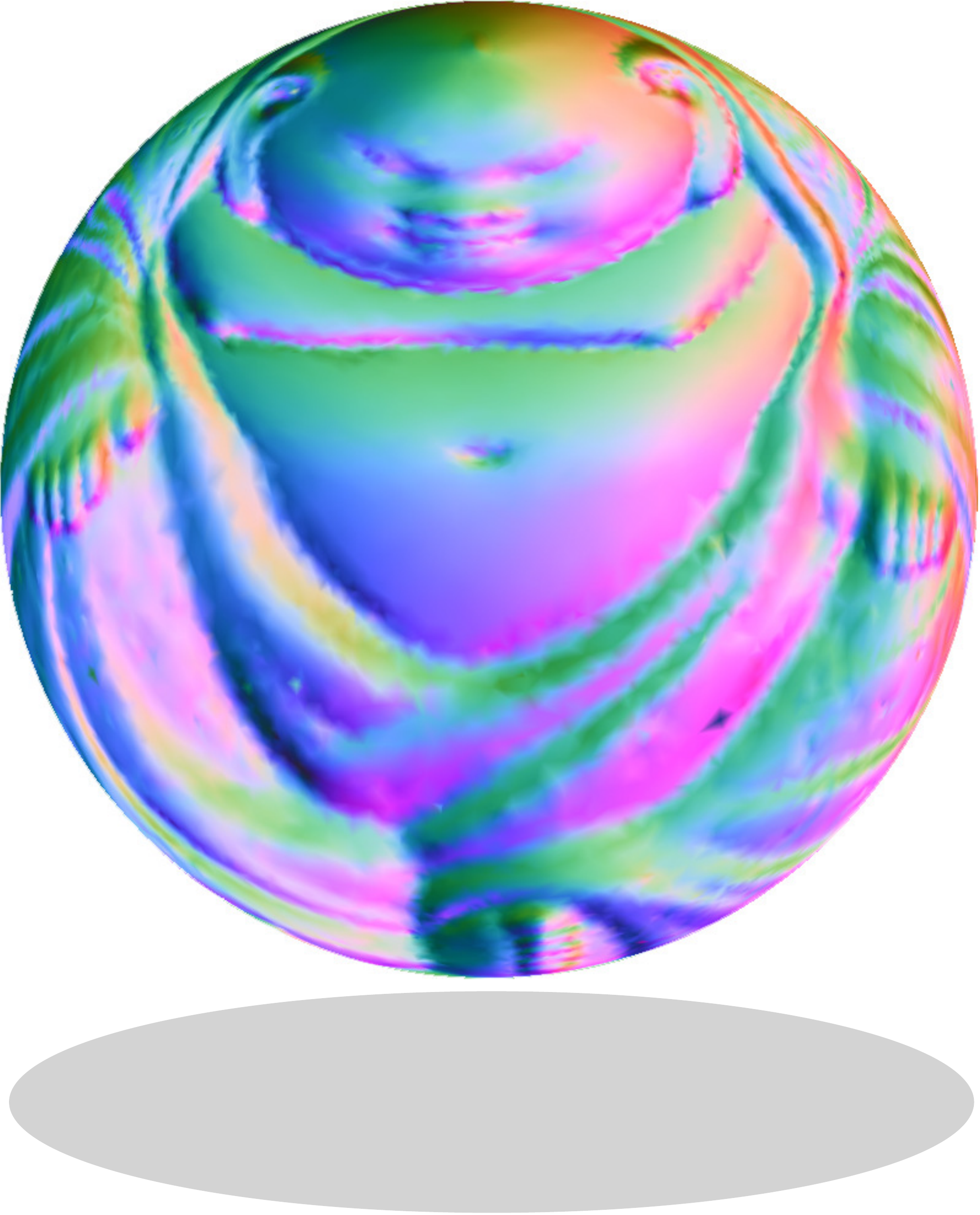}&
\includegraphics[height=0.17\textwidth]{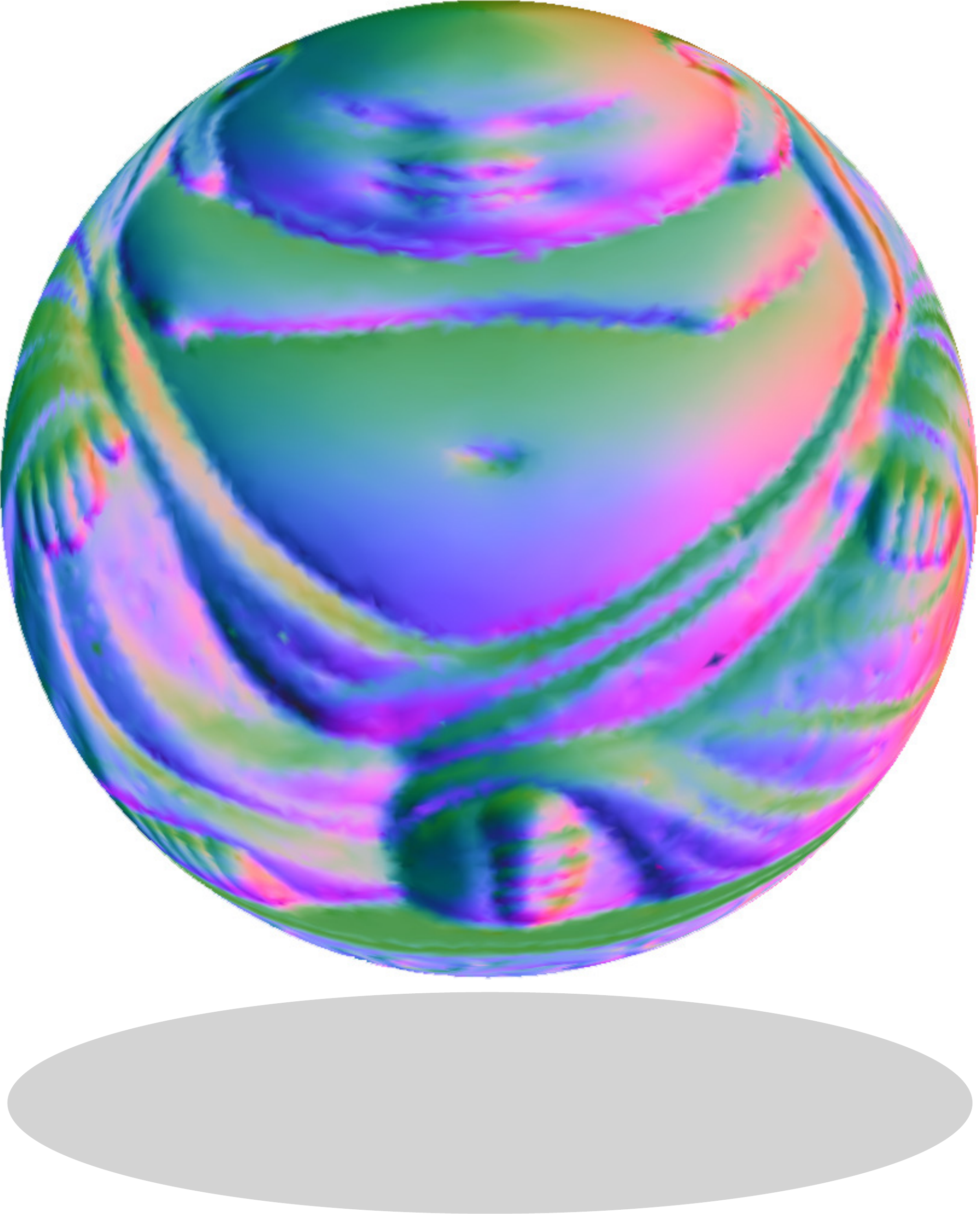}&
\includegraphics[height=0.17\textwidth]{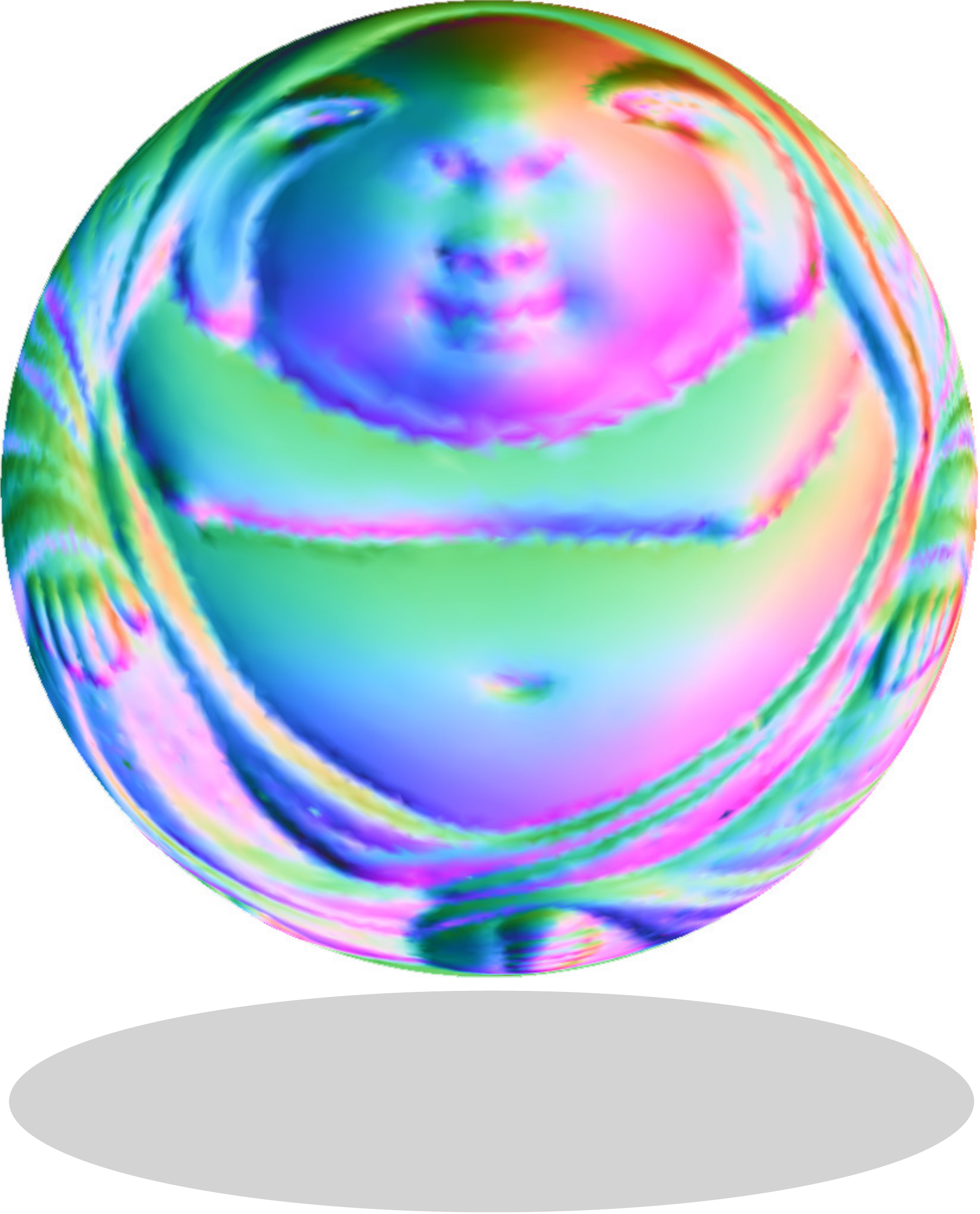}\\
(a) Buddha surface & (b) Angle-preserving & (c) $t=0.25$ & (d) Balanced map & (e) $t=0.75$ & (f) Area-preserving\\
topological sphere & map ($t=0$) & & $t=0.5$ & & map ($t=1$)\\
\end{tabular}
\end{center}
\vspace{-3mm}
\caption{ Angle-area distortion balancing maps: (a) Buddha surface and the corresponding spherical mappings with different trade-off values of $t$, between angle and area distortions. (b) $t=0$ (angle-preserving), (c) $t=0.25$, (d) $t=0.5$ (balanced mapping), (e) $t=0.75$ and (f) $t=1$ (area-preserving mapping).
\label{fig:buddaha_balanced_map}}
\vspace{-5mm}
\end{figure*}

\subsubsection{Area-Preserving Spherical Mapping}

Suppose the source surface is a topological sphere $(S,\mathbf{g})$ with a Riemannian metric $\mathbf{g}$, the area element induced by $\mathbf{g}$ is denoted as $dA_\mathbf{g}$ and by scaling, the total area of $S$ is $4\pi$. The target surface is the unit sphere $(\mathbb{S}^2,\mathbf{h})$, where $\mathbf{h}$ is the spherical Riemannian metric and the corresponding area element is $dA_{\mathbf{h}}$.
The conformal mapping is denoted as $\varphi:(S,\mathbf{g})\to (\mathbb{S}^2,\mathbf{h})$, therefore:
\[
    \mathbf{g} = e^{2\lambda} \mathbf{h}, \hspace{5mm}\varphi_{\#} dA_{\mathbf{g}} = e^{2\lambda\circ \varphi^{-1}} dA_{\mathbf{h}}.
\]
The stereographic projection maps the unit sphere onto the extended complex plane $\hat{\mathbb{C}}=\mathbb{C}\cup\{\infty\}$,
\[
\tau(x,y,z) = \left(\frac{x}{1-z},\frac{y}{1-z}\right).
\]
The push-forward measure of $dA_{\mathbf{h}}$ induced by the stereographic projection is:
\[
        \tau_{\#}dA_{\mathbf{h}} = \frac{4dxdy}{(1+x^2+y^2)^2} .
\]
Similarly, the push-forward measure of $\varphi_{\#}dA_{\mathbf{g}}$ induced by $\tau$ is:
\[
      (\tau\circ\varphi)_{\#} dA_{\mathbf{g}} = e^{2\lambda\circ\varphi^{-1}\circ\tau^{-1}(x,y)}\frac{4dxdy}{(1+x^2+y^2)^2} .
\]
Then, we compute the optimal mass transport map by finding a convex function $u:\hat{\mathbb{C}}\to \mathbb{R}$, such that:
\[
    \nabla u: \left(\hat{\mathbb{C}},  (\tau\circ\varphi)_{\#} dA_{\mathbf{g}} \right) \to \left (\hat{\mathbb{C}},  \tau_{\#} dA_{\mathbf{h}} \right ).
\]
By stereographic projection, the optimal mass transport map $\nabla u: \hat{\mathbb{C}} \to \hat{\mathbb{C}}$ induces the spherical automorphism $\tau^{-1}\circ \nabla u \circ \tau: \mathbb{S}^2\to \mathbb{S}^2$, as shown in the following diagram:
\begin{diagram}
(\mathbb{S}^2, \varphi_{\#} dA_{\mathbf{g}}) & \rTo^{\tau^{-1}\circ \nabla u\circ \tau} &(\mathbb{S}^2, dA_{\mathbf{h}})\\
\dTo^{\tau} && \dTo_{\tau}\\
(\hat{\mathbb{C}}, (\tau\circ\varphi)_{\#}dA_{\mathbf{g}} ) &\rTo^{\nabla u} & (\hat{\mathbb{C}},\tau_{\#}dA_{\mathbf{h}}) \\
\end{diagram}
The composition of $\varphi$ and the inverse of $\tau^{-1}\circ\nabla u\circ \tau$ is area-preserving, as shown in the following diagram:
\begin{equation}
   \eta := (\tau^{-1}\circ\nabla u\circ \tau)^{-1} \circ \varphi : (S,dA_{\mathbf{g}}) \to (\mathbb{S}^2, \varphi_{\#}dA_{\mathbf{g}})
    \label{eqn:spherical_area_map}
\end{equation}
\begin{diagram}
(S,dA_\mathbf{g}) &\rTo^\varphi & (\mathbb{S}^2,dA_\mathbf{h})\\
&\rdTo_\eta &\uTo_{\tau^{-1}\circ \nabla u \circ \tau} \\
& &(\mathbb{S}^2, \varphi_{\#} dA_{\mathbf{g}} )
\end{diagram}

Algorithmically, the push-foward measure is represented as a weight function defined on the vertex. The weight of a vertex equals to one third of the total areas
of triangles adjacent to the vertex on the original triangle mesh. Moreover, it's based on Newton's method, which is quadratically convergent and at each step the main task is to compute the convex hull with the complexity $O(n \log n)$.

In summary, we first conformally map the surface onto the unit sphere and then use stereographic projection to map the unit sphere onto the extended complex plane. The resultant mapping from the original surface onto the extended complex plane is angle-preserving and the area distortion is encoded as the conformal factor. The Riemannian metric of the original surface is equal to the product of the conformal factor and the planar Euclidean metric. The product of the conformal factor and the Euclidean area element is treated as the source measure, the Euclidean area element is the target measure, and the optimal transportation map is between these two measures. Hence, the optimal transport map is based on the geometry of the original surface.

In terms of stability, since the optimal transportation map continuously depends on the source and the target measures, the source measure continuously depends on the Riemannian metric of the input mesh. Therefore, smooth perturbations of the input mesh will change the Riemannian metric smoothly and in turn change the optimal transportation map smoothly. Therefore, this method is stable to smooth perturbations of the input.

\begin{algorithm}[!h]
\caption{Area-Preserving Spherical Mapping}
    \textbf { Input:}  Closed genus zero surface mesh $(M,\mathbf{g})$ with total area $4\pi$.

    \textbf { Output:} An area-preserving mapping $\eta: M \rightarrow \mathbb{S}^2$.

    \textbf{(1)} Compute a conformal spherical map $\varphi: (M,\mathbf{g})\to (\mathbb{S}^2,\mathbf{h})$, using Alg.~\ref{alg:spherical_conformal_map}. Compute the push forward measure induced by $\varphi$, $\varphi_{\#}dA_{\mathbf{g}}$. \\

    \textbf{(2)} Use stereographic projection $\tau: \mathbb{S}^2\to \hat{\mathbb{C}}$ to map the sphere to the extended plane.
    Compute the push-forward measures, $\tau_{\#}dA_{\mathbf{h}}$, $(\tau\circ\varphi)_{\#}dA_{\mathbf{g}}$. \\

    \textbf{(3)} Construct the optimal mass transport map $\nabla u: (\hat{\mathbb{C}}, (\tau\circ\varphi)_{\#}dA_{\mathbf{g}})\to (\hat{\mathbb{C}}, \tau_{\#}dA_{\mathbf{h}})$\\

    \textbf{(4)} Lift the optimal mass transport map to the sphere, $\tau^{-1}\circ\nabla u\circ \tau: (\mathbb{S}^2, \varphi_{\#}dA_{\mathbf{g}})\to (\mathbb{S}^2, dA_{\mathbf{h}})$. \\

    \textbf{(5)} Compose the inverse of $\tau^{-1}\circ\nabla u\circ \tau$ with the conformal map $\varphi$ to obtain the area-preserving map (Eqn.\ref{eqn:spherical_area_map}).\\
\label{alg:spherical_area_map}
\end{algorithm}

\vspace{-3.5mm}
\subsection{Angle-Area Distortion Balancing Maps}
\label{sec:map_blend}

In practice, it is highly desirable to achieve a good balance between angle distortion and area distortion. We use polar decomposition to accomplish this goal. In this section, we introduce the polar decomposition method which is necessary for interpolating between the conformal mapping in Section \ref{sec:conf_spherical_mapping} and the area-preserving mapping in Section \ref{sec:map_area}. In the previous discussion, the spherical conformal map is decomposed in the form $\varphi = (\tau^{-1}\circ \nabla u \circ \tau)^{-1} \circ s $, where $s$ is the area-preserving mapping. We construct a one-parameter family of measures \cite{mccann1997} on the extended complex plane:
\[
  \mu_t :=  (1-t) \tau_{\#} dA_{\mathbf{h}} + t (\tau\circ\varphi)_{\#}dA_{\mathbf{g}}, \hspace{3mm} 0\le t\le 1,
\]
and construct the corresponding optimal mass transport maps:
\[
    \nabla u_t : \left(\mathbb{C},(\tau\circ\varphi)_{\#} dA_{\mathbf{g}}\right)\to \left(\mathbb{C},\mu_t\right).
\]
The one-parameter family of mappings, $ \eta_t: (S,\mathbf{g})\to (\mathbb{S}^2, \mathbf{h})$:
\[
    \eta_t:=\left(\tau^{-1}\circ\nabla u_t \circ \tau \right)^{-1}\circ \varphi,\hspace{3mm} 0\le t \le 1,
\]
then $\eta_0$ is area-preserving, $\eta_1$ is angle-preserving. For $t$ between $0$ and $1$, the mapping $\eta_t$ is between angle-preserving and area-preserving. By choosing an appropriate value for $t$, we can select a good balance between them. By designing the measures, \emph{this method can be carried out on partial regions on the surface, therefore, we can achieve angle-area distortion balance locally instead of globally}. Fig.~\ref{fig:buddaha_balanced_map} shows the balancing maps at different values of $t$.

In effect, first we construct a conformal map: $\varphi:(S,\mathbf{g})\to (\mathbb{S}^2,\mathbf{h})$. Then, we define one parameter family of area elements (measures),
\[
    \mu_t = (1-t) dA_{\mathbf{h}} + t \varphi_{\#} dA_{\mathbf{g}}.
\]
$\mu_t$ connects the original spherical area element and the conformal image area element, when $t=0$, $\mu_0$ is the area element induced by the conformal mapping, when $t = 1$ , $\mu_1$ is the original spherical area element. Then, we construct OMT map $\tau_t: \mu_t \to \mu_1$, then $\tau_1 = id$, $\tau_0 = \eta^{-1}$. The one-parameter family of mappings
$\eta_t:(S,\mathbf{g})\to (\mathbb{S}^2,\mathbf{h})$ is given by the composition, $\eta_t := \tau_t \circ \varphi$. So $\eta_0 = \tau_0 \circ \eta \circ \sigma = \eta^{-1}\circ \eta \circ \sigma = \sigma$ is area-preserving; $\eta_1 = \tau_1 \circ \varphi = id \circ \varphi = \varphi$ is conformal.

\setlength{\tabcolsep}{0pt}
\begin{figure}[t!]
\begin{center}
\begin{tabular}{cccc}
\includegraphics[height=0.21\textwidth]{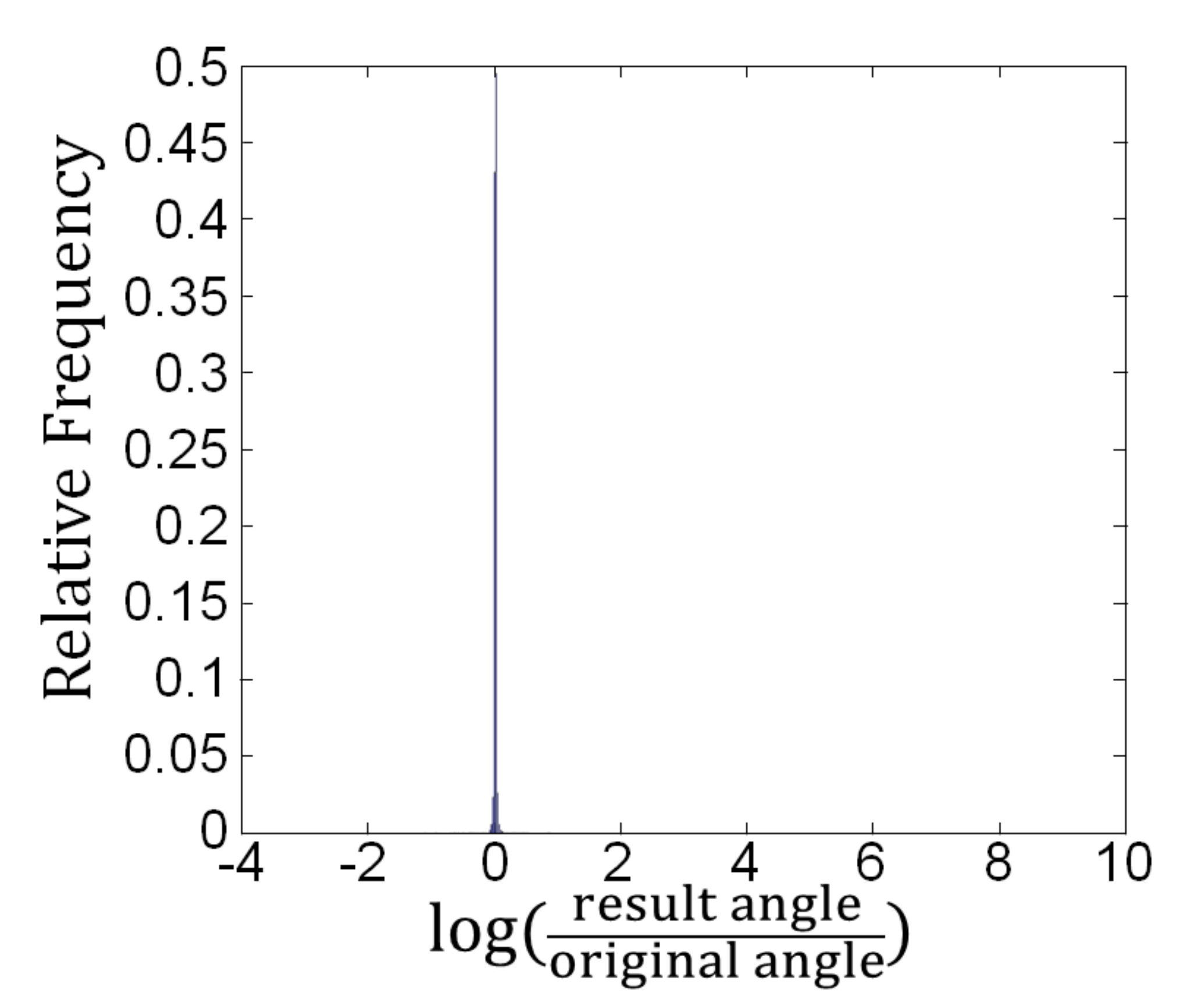}&
\includegraphics[height=0.21\textwidth]{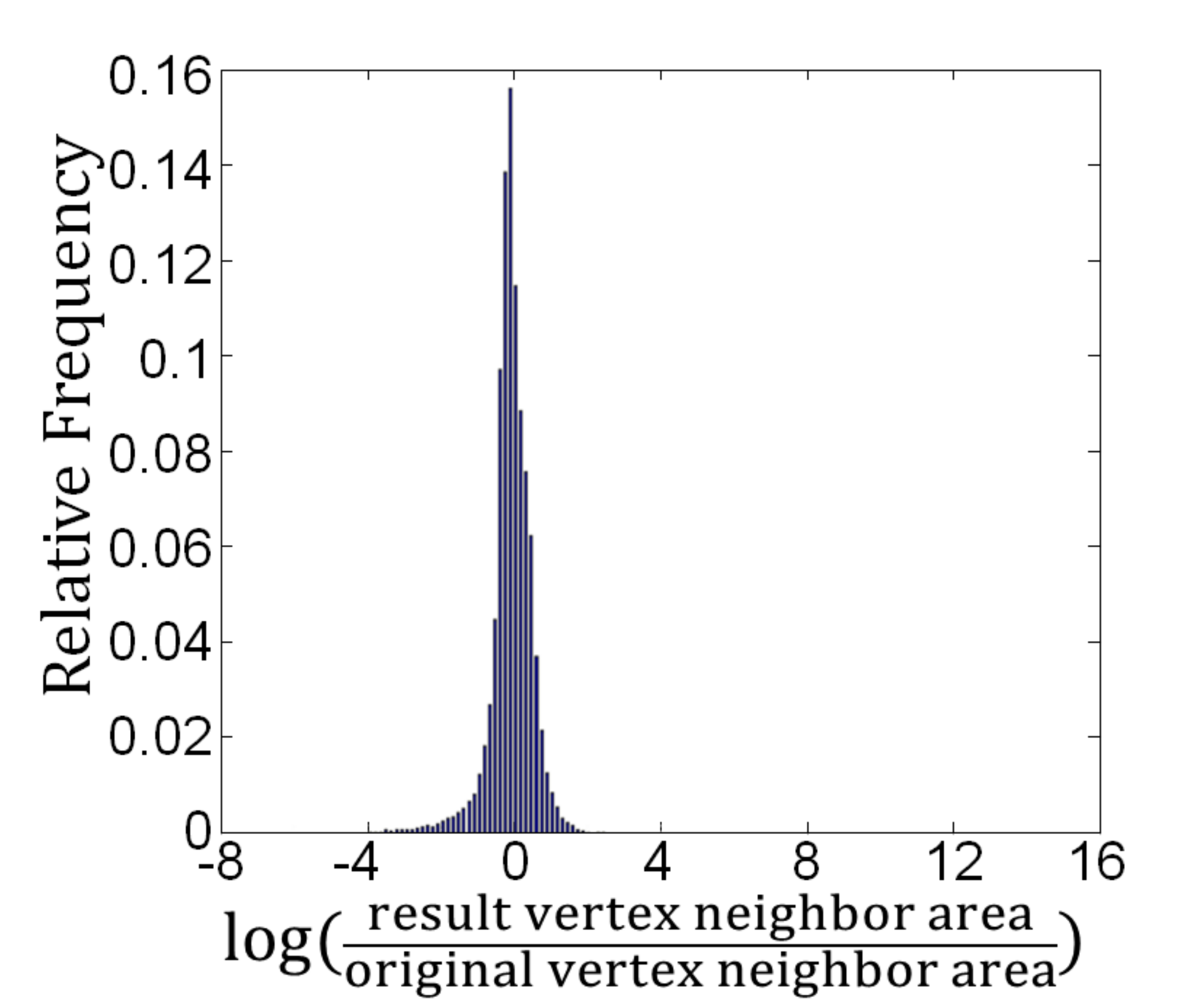}\\
(a) Angle distortion of & (b) Area distortion of\\
conformal map & conformal map\\
\includegraphics[height=0.21\textwidth]{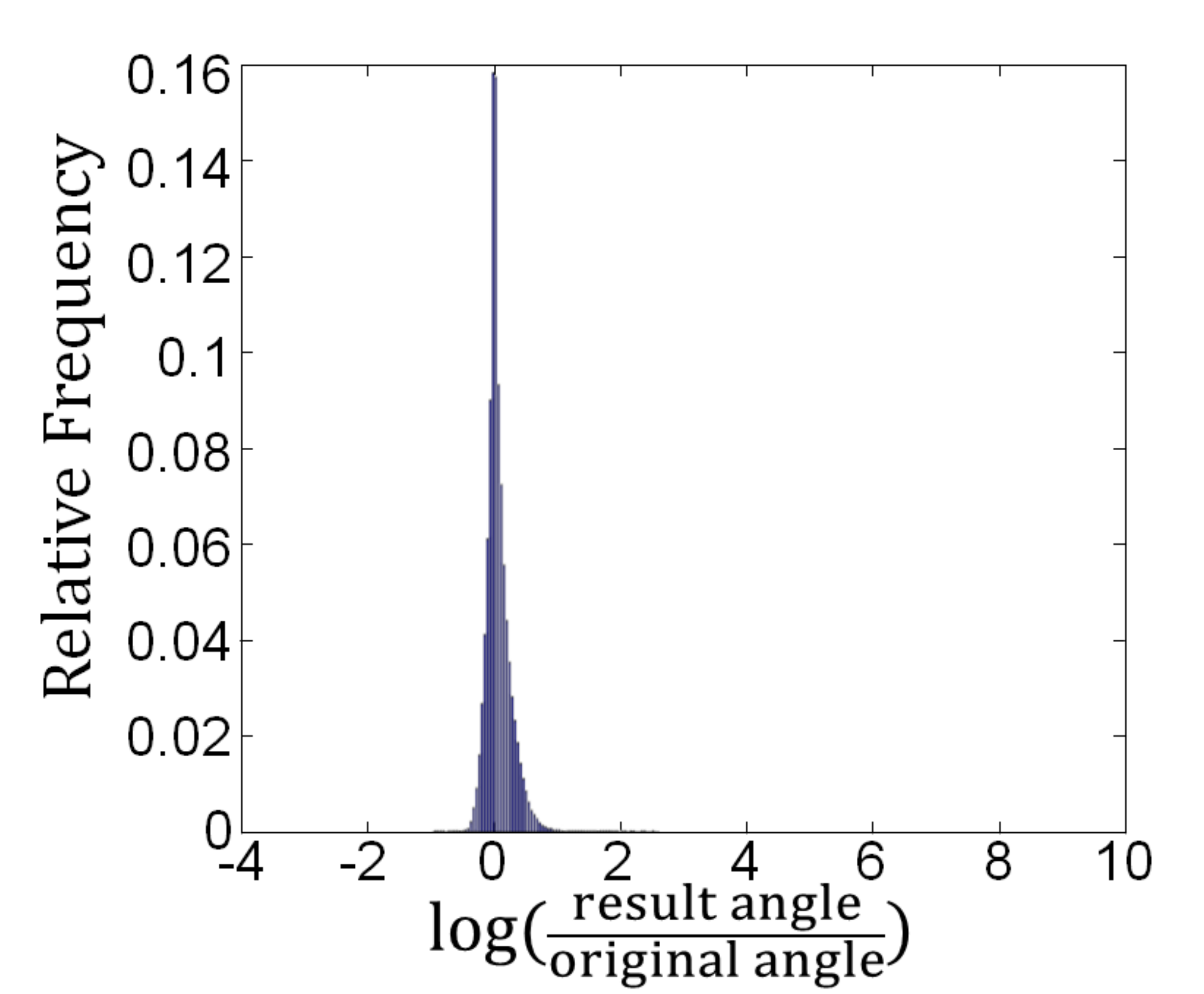}&
\includegraphics[height=0.21\textwidth]{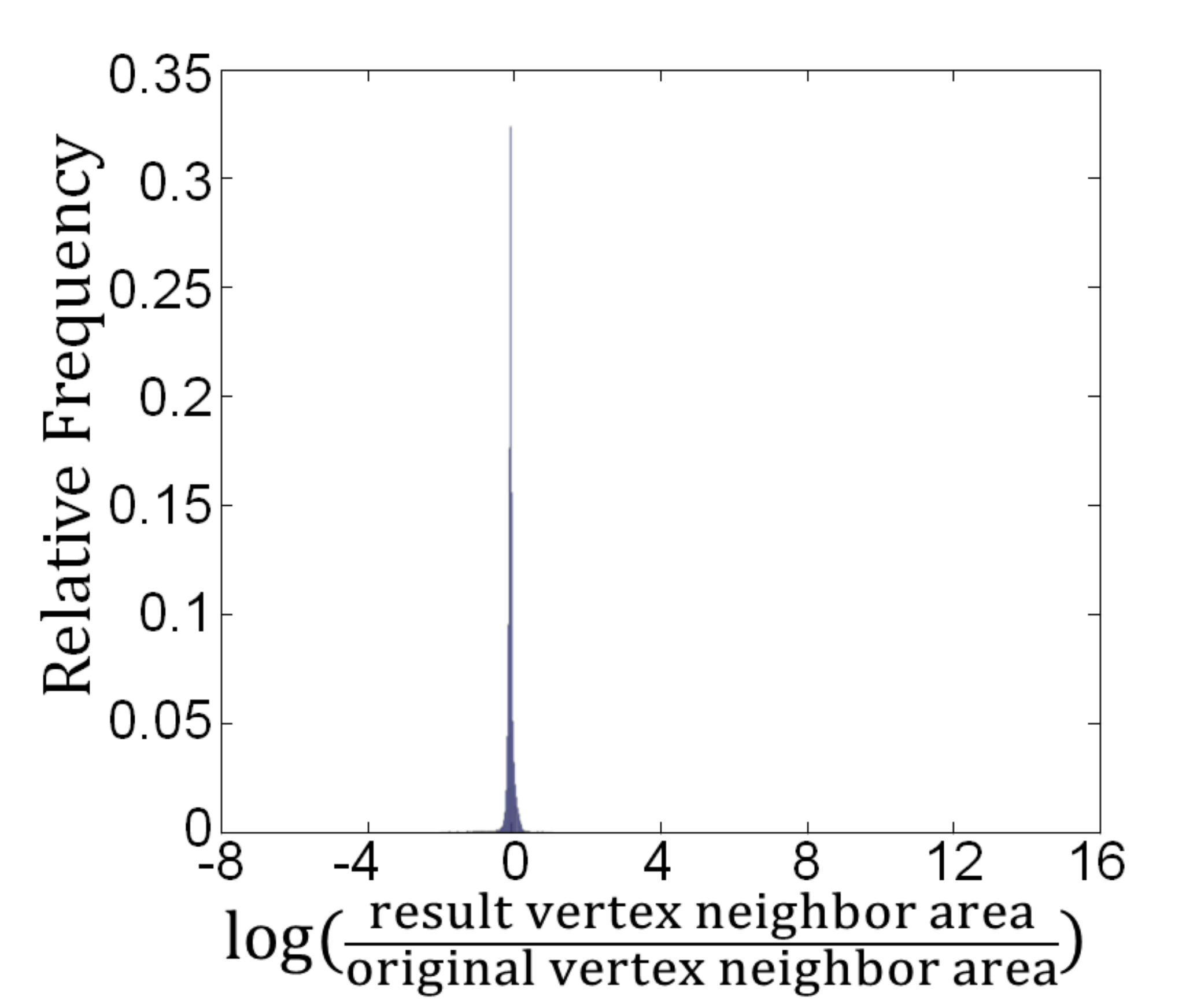}\\
(c) Angle distortion of & (d) Area distortion of\\
balanced map & balanced map\\
\includegraphics[height=0.21\textwidth]{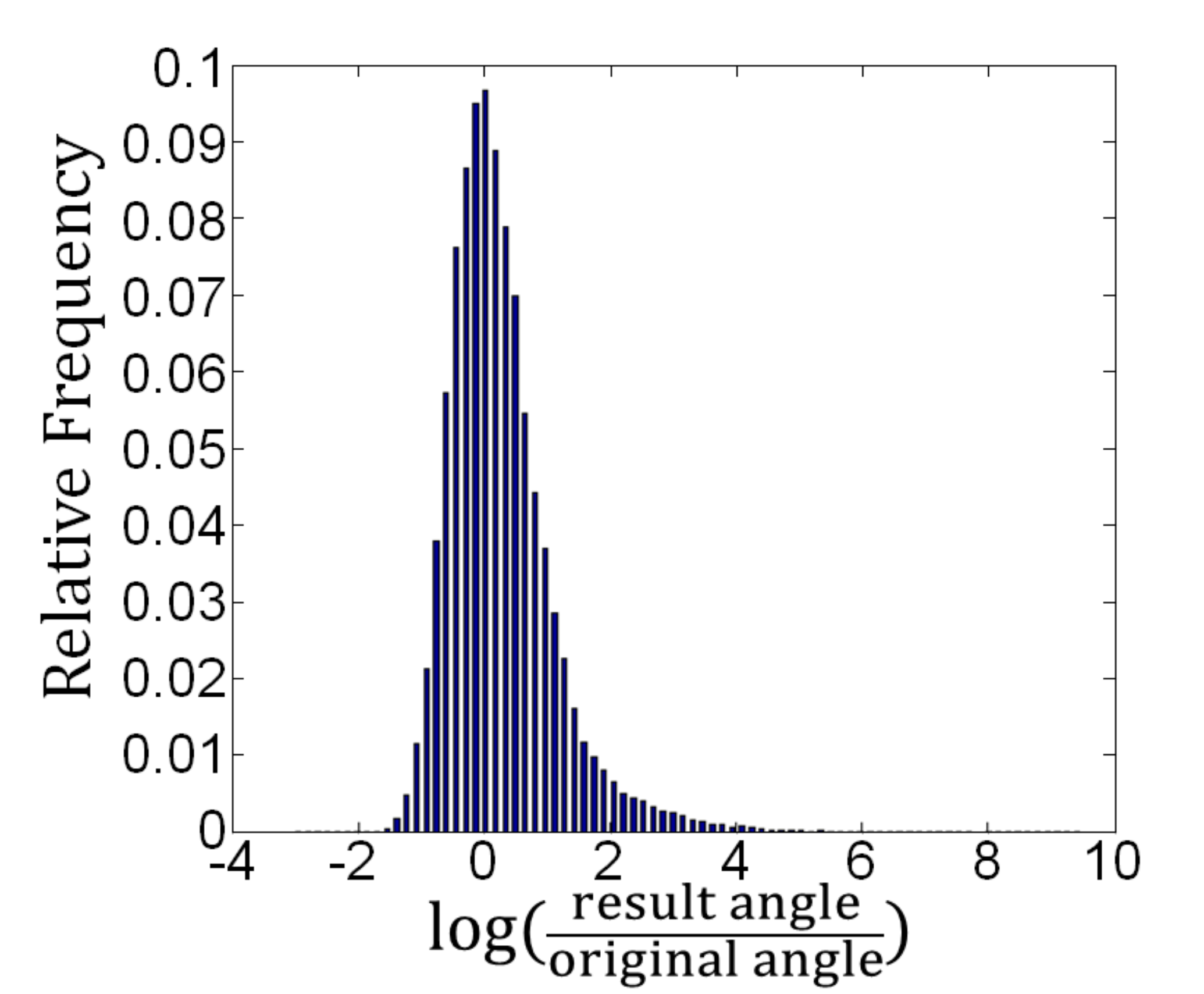}&
\includegraphics[height=0.21\textwidth]{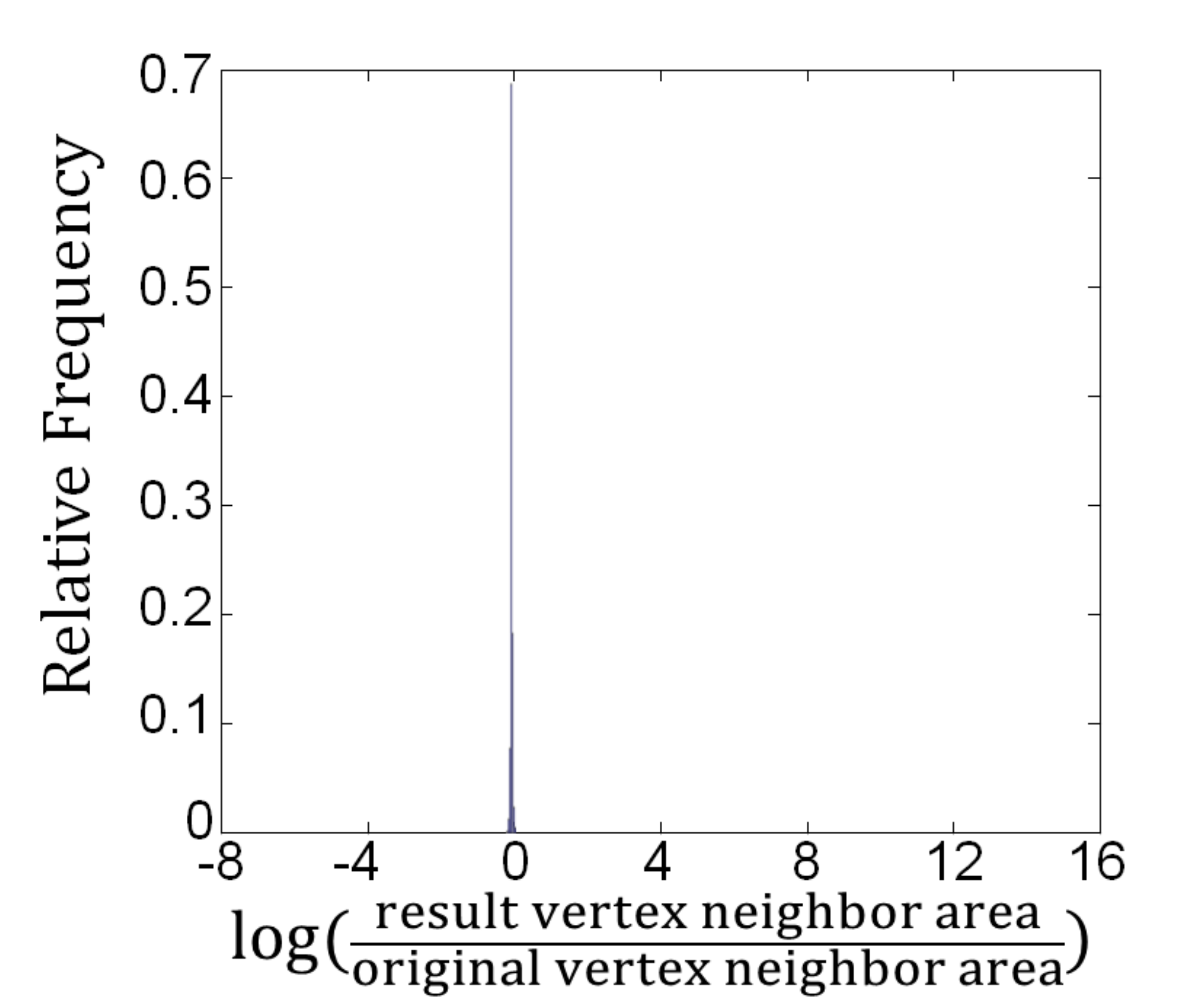}\\
(e) Angle distortion of & (f) Area distortion of\\
area-preserving map & area-preserving map\\
\end{tabular}
\end{center}
\vspace{-1.5mm}
\caption{Angle and area distortion of the conformal, balanced and the area-preserving map for the Buddha surface in Fig.\ref{fig:buddaha_balanced_map}(a).
\label{fig:buddha_balanced_map_hist}}
\vspace{-5mm}
\end{figure}

\vspace{-2.5mm}
\section{Experimental Results}

We implemented the dynamic Yamabe flow, conformal welding and optimal mass transport map algorithms, and applied them to various shape models. All the experiments (in the following sections) have been done on a laptop computer with Intel Core i7 CPU, M620 2.67GHz with 4GB memory. All the algorithms have been implemented using generic C++ on the Windows 7 operating system. All 3D shape models are represented as triangular meshes. The Bimba (Fig.~\ref{fig:bimba_pipeline}), Gargoyle (Fig.~\ref{fig:gargoyle2_pipeline}) and the Buddha (Fig.~\ref{fig:buddaha_balanced_map}) models are from public 3D geometry repositories \cite{aimatshape}. The human brain cortical surface is reconstructed from MRI data, using the Freesurfer pipeline \cite{freesurfer:2012} and the spherical mapping is shown in Fig.~\ref{fig:brain_conformal_ap}.

\setlength{\tabcolsep}{0pt}
\begin{figure}[t!]
\begin{center}
\begin{tabular}{cccc}
\includegraphics[height=0.21\textwidth]{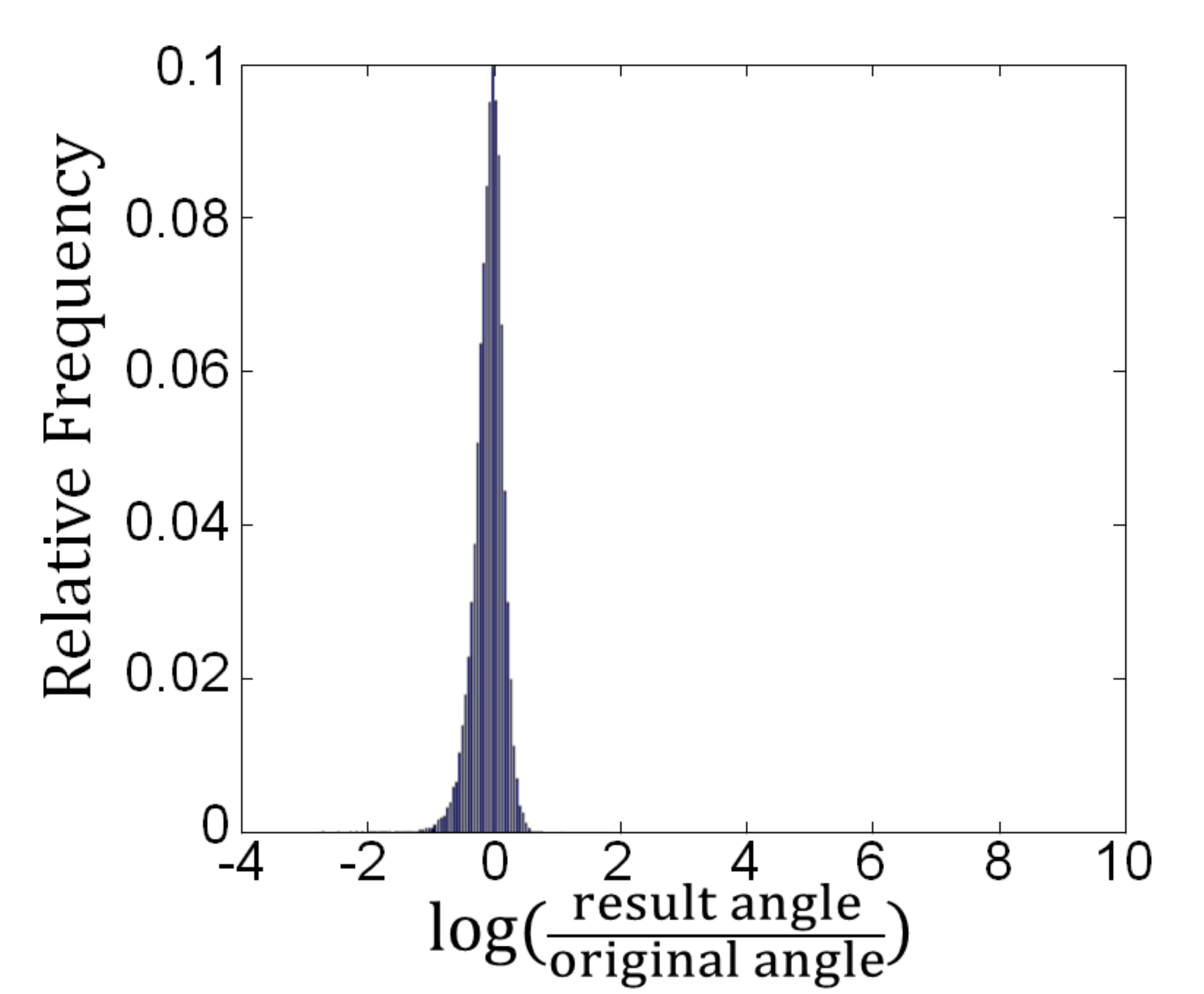}&
\includegraphics[height=0.21\textwidth]{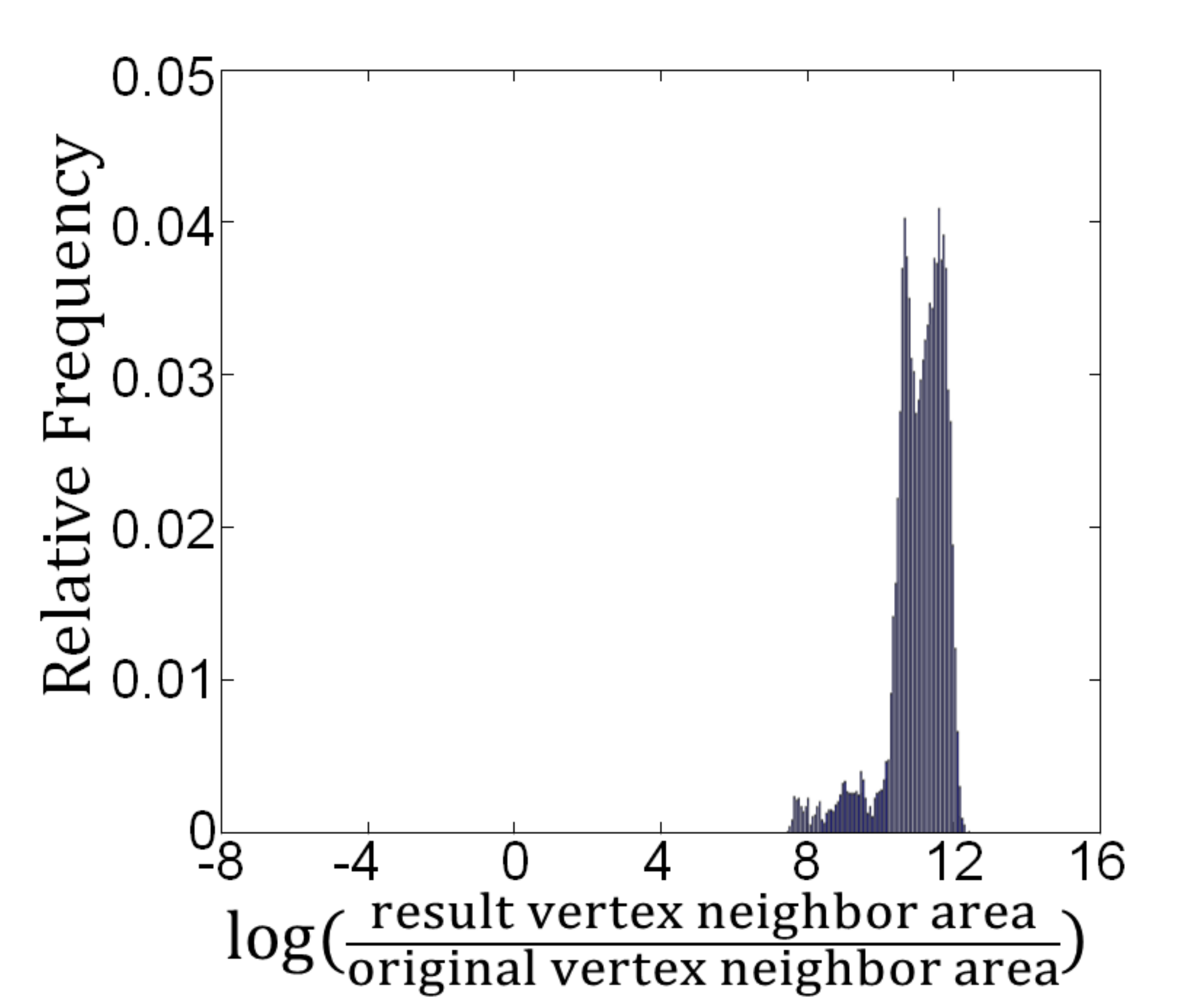}\\
(a) Angle distortion & (b) Area distortion\\
Haker et al. \cite{haker:2004} & Haker et al. \cite{haker:2004}\\
\includegraphics[height=0.21\textwidth]{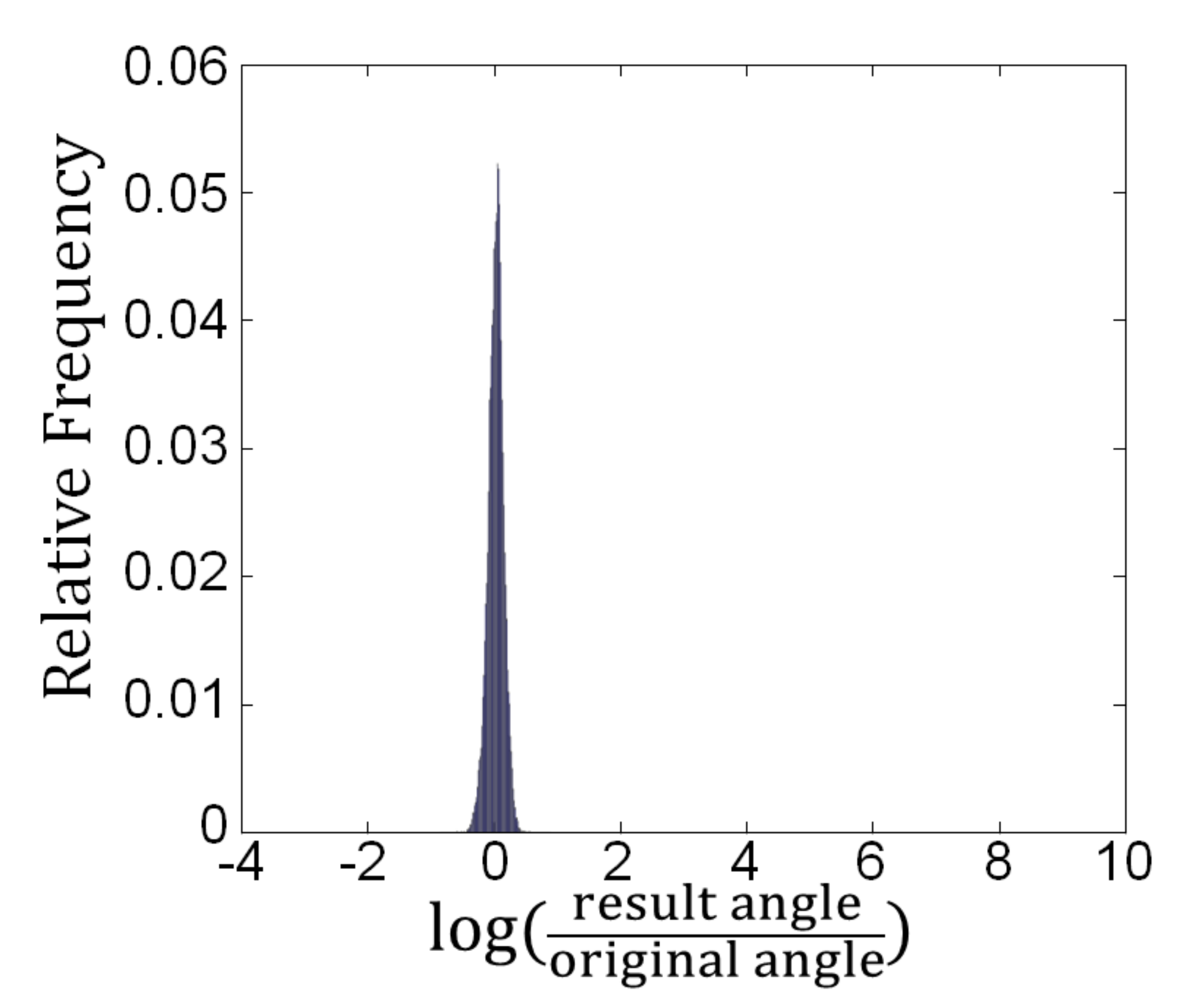}&
\includegraphics[height=0.21\textwidth]{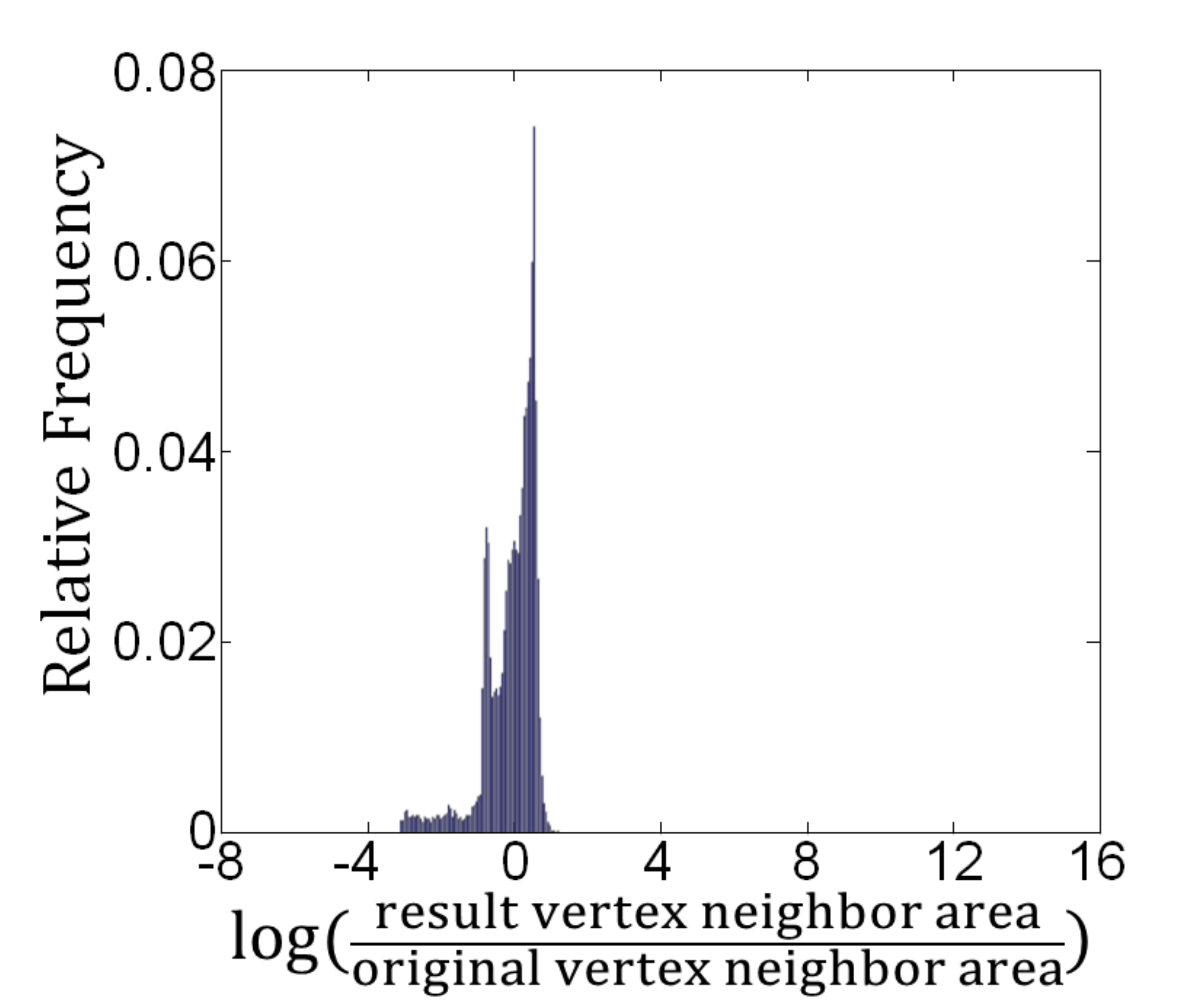}\\
(c) Angle distortion & (d) Area distortion\\
Gu et al. \cite{Gu04} & Gu et al. \cite{Gu04}\\
\includegraphics[height=0.21\textwidth]{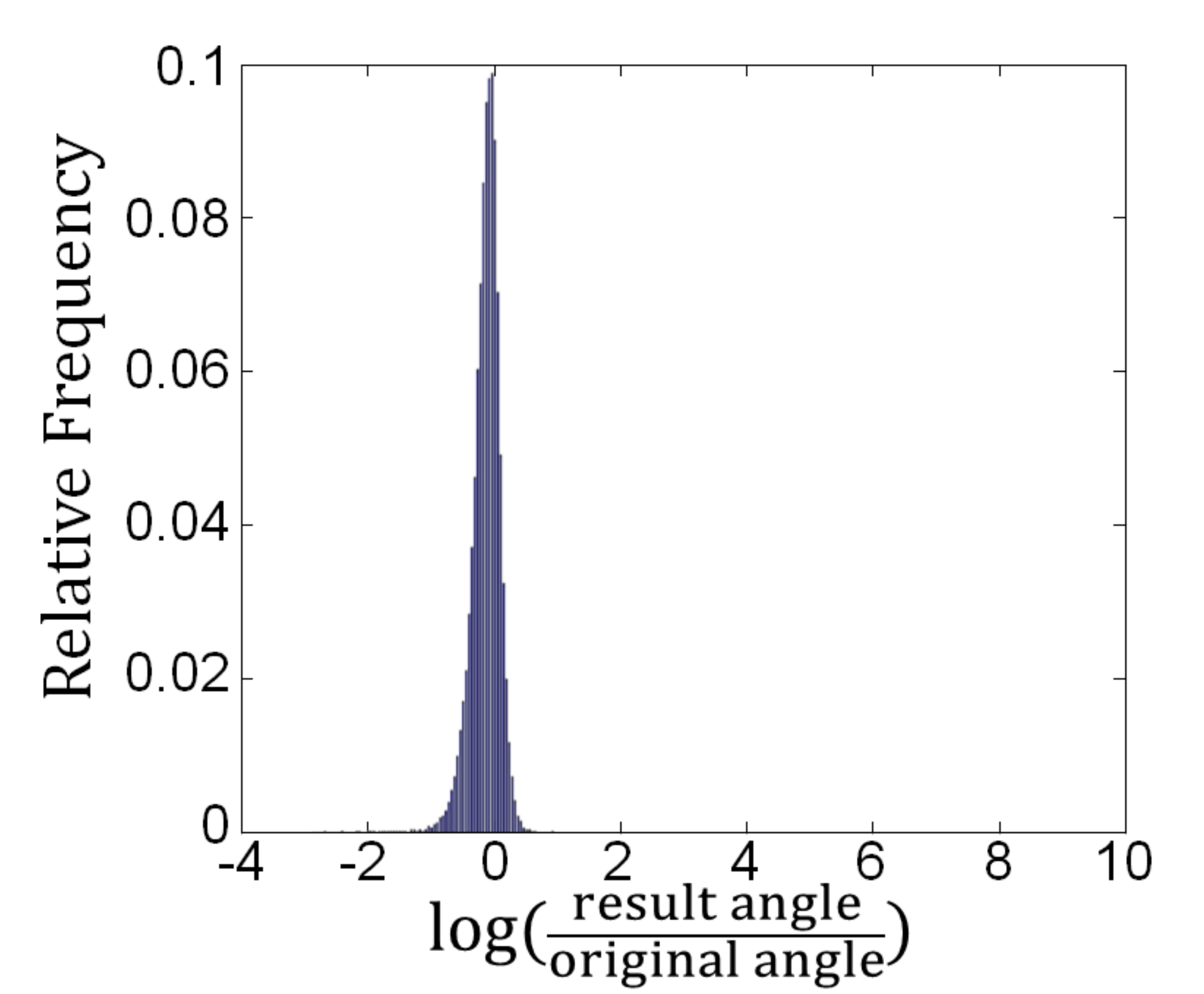}&
\includegraphics[height=0.21\textwidth]{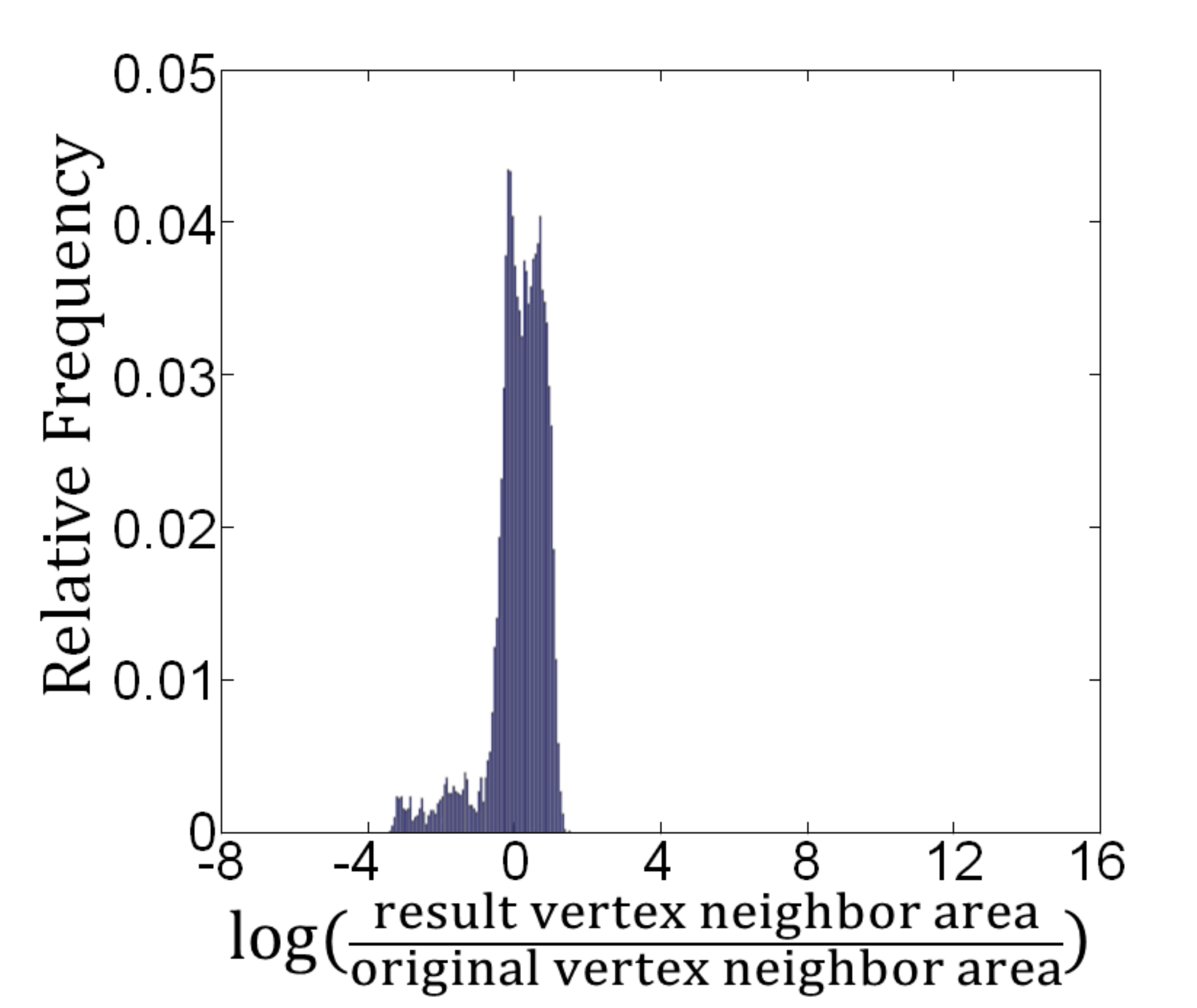}\\
(e) Angle distortion & (f) Area distortion\\
Kazhdan et al. \cite{kazhdan2012} & Kazhdan et al. \cite{kazhdan2012}\\
\includegraphics[height=0.21\textwidth]{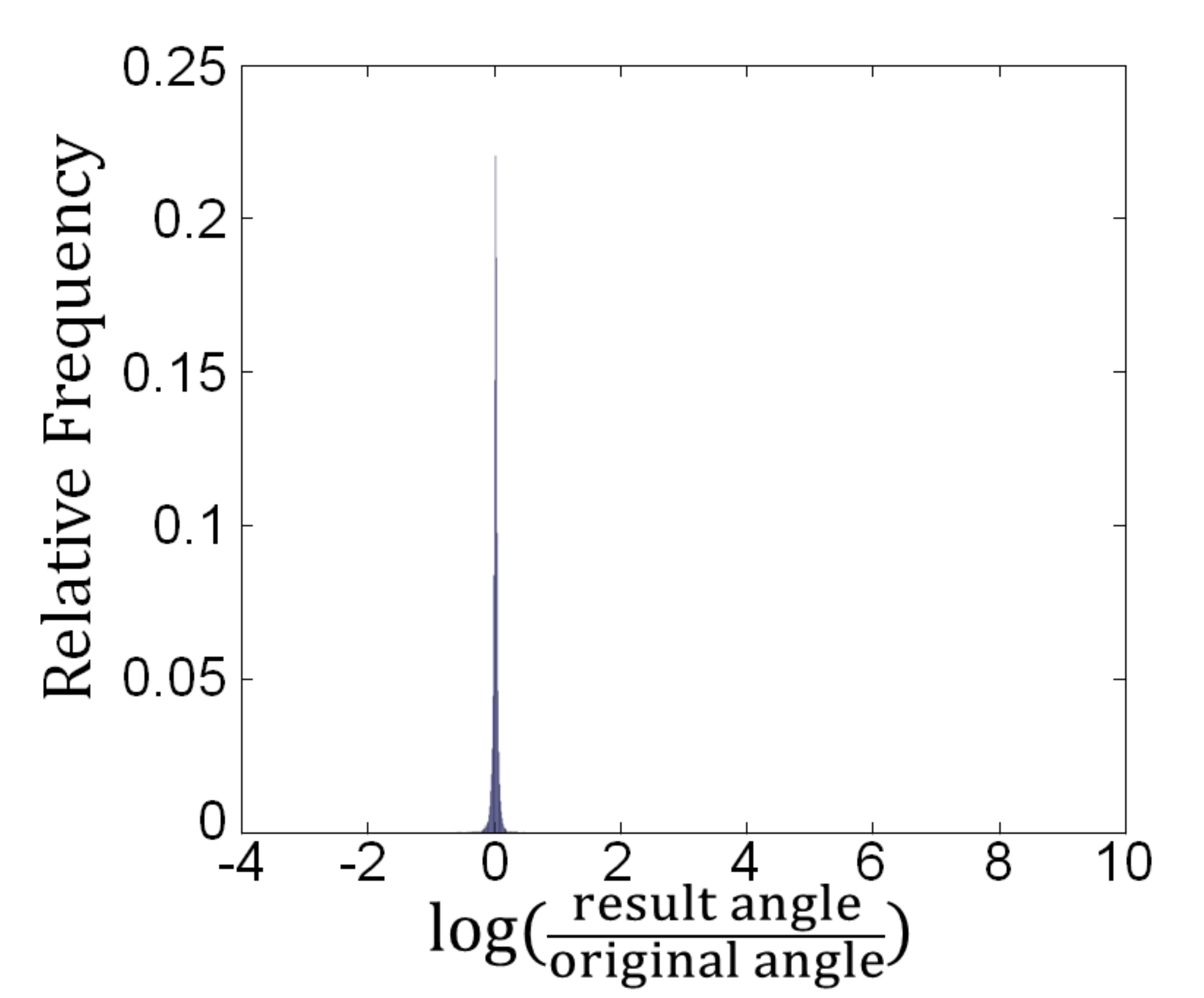}&
\includegraphics[height=0.21\textwidth]{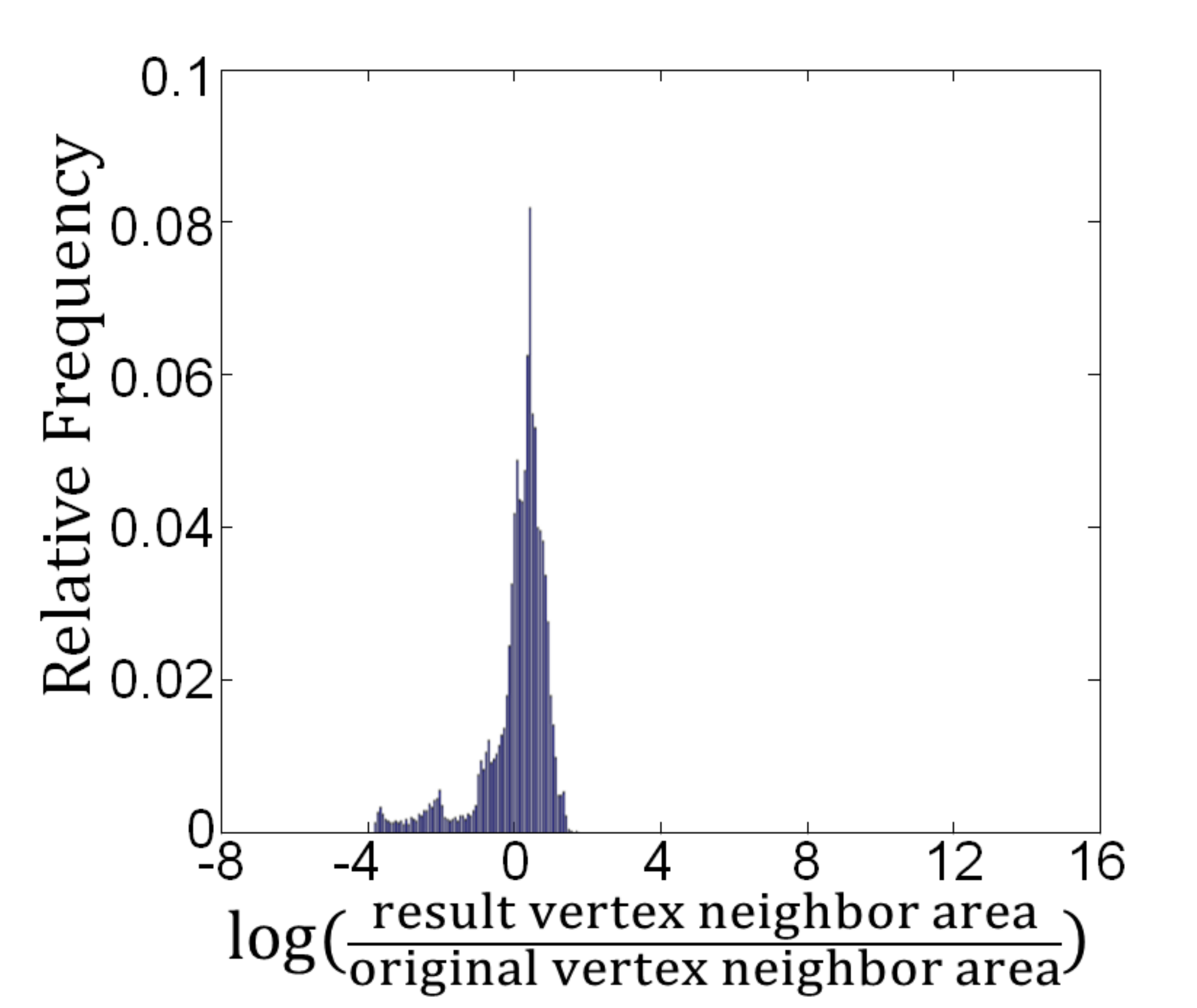}\\
(g) Angle distortion & (h) Area distortion\\
Crane et al. \cite{crane2013} & Crane et al. \cite{crane2013}\\
\end{tabular}
\end{center}
\vspace{-1.5mm}
\caption{Angle and area distortion of spherical mapping computed using: ((a) \& (b)) Haker et al. \cite{haker:2004}, ((c) \& (d)) Gu et al. \cite{Gu04}, ((e) \& (f)) Kazhdan et al. \cite{kazhdan2012}, and ((g) \& (h)) Crane et al. \cite{crane2013} for the Buddha surface in Fig.\ref{fig:buddaha_balanced_map}(a).
\label{fig:buddha_balanced_map_hist_comp}}
\vspace{-5mm}
\end{figure}

\vspace{-3.5mm}
\subsection{Angle-Area Distortion Statistics}

We compute the statistics of angle distortion and area distortion of all the parameterization methods including Haker et al. \cite{haker:2004}, Gu et al. \cite{Gu04}, Crane et al. \cite{crane2013} and Kazhdan et al. \cite{kazhdan2012}. The angle distortion is measured in the following way. For each triangular face, we measure three corner angles of the original input mesh, and the image mesh, then, compute the logarithm of the ratios between the two values. The histogram of the logarithms of all corner angles is plotted out, which visualizes the distribution of angle distortions. Similarly, in order to measure area distortions, for each vertex we compute the area of its neighboring faces, and compute the logarithm of the ratio between the image area and the original area. The histogram shows the area distortion distribution. The details of these measures can be found in the Appendix.

The histograms for the mappings of the cortical surface are shown in Fig.~\ref{fig:brain_hist}, those for the Buddha surface are demonstrated in Figs.~\ref{fig:buddha_balanced_map_hist} (computed using our algorithm) and \ref{fig:buddha_balanced_map_hist_comp} (computed using other methods). From the Buddha example, it can be seen that our conformal spherical parameterization algorithm produces very low angle distortion, and area-preserving parameterization obtains very low area-distortions and therefore the balanced map obtains a good balance between angle distortion and area distortion.

\vspace{-3.5mm}
\subsection{Comparison with Other Methods}
We quantify the area and angle distortion metrics of the spherical parameterization by using the signed singular values of the Jacobian of the transformation for each triangle \cite{degener:2003,hormann:2000,liu:2008}. Small angular and area distortions are indicated by a distortion value approaching 2. The details can be found in the Appendix. We ran our algorithm (SP) with different $t$ values, on a variety of inputs, in order to evaluate the computation time and convergence rate; $t=0$ indicates angle-preserving mapping, $t=0.5$ indicates balanced mapping and $t=1$ indicates area-preserving mapping. We compared the results of the SP algorithm with the results obtained after running the algorithm of Haker et al. \cite{haker:2004}, Gu et al. \cite{Gu04}, Crane et al. \cite{crane2013} and Kazhdan et al. \cite{kazhdan2012}. The values of the distortion measures obtained by the various algorithms are summarized in Tables \ref{tab:ang_comparison} and \ref{tab:area_comparison}. The running time of our algorithm for meshes of varying sizes is summarized in Table~\ref{tab:time_comparison}.

\setlength{\tabcolsep}{0pt}
\begin{table}[!t]
\centering
\captionsetup{labelformat=empty}
\caption{Table 1: Angular Distortion Statistics}
\label{tab:ang_comparison}
\vspace{-1.5mm}
\begin{tabular*}{0.48\textwidth}{@{\extracolsep{\fill}}ccccccccc}
\hline
\hline
Model & \# of & Gu & Haker & Kazhdan & Crane & SP & SP & SP\\
      & faces &    &       &         &       & ($t=0$) & ($t=0.5$) & ($t=1$)\\
\hline
Skull & 10K & 2.116 & 2.121 & 2.115 & 2.112 & 2.107 & 2.125 & 2.163\\
Bimba & 20K & 2.152 & 2.165 & 2.153 & 2.149 & 2.141 & 2.161 & 2.185\\
Armadillo & 30K & 2.324 & 2.582 & 2.327 & 2.319 & 2.311 & 2.578 & 2.675\\
Brain & 50K & 2.106 & 2.109 & 2.104 & 2.098 & 2.091 & 2.111 & 2.240\\
Gargoyle & 70K & 2.781 & 2.812 & 2.779 & 2.767 & 2.756 & 2.833 & 3.001\\
Bunny & 90K & 2.693 & 2.741 & 2.694 & 2.691 & 2.687 & 2.746 & 2.832\\
\hline
\end{tabular*}
\vspace{-2mm}
\end{table}

\begin{table}[!t]
\centering
\captionsetup{labelformat=empty}
\caption{Table 2: Area Distortion Statistics}
\label{tab:area_comparison}
\vspace{-1.5mm}
\begin{tabular*}{0.48\textwidth}{@{\extracolsep{\fill}}ccccccccc}
\hline
\hline
Model & \# of & Gu & Haker & Kazhdan & Crane & SP & SP & SP\\
      & faces &    &       &         &       & ($t=0$) & ($t=0.5$) & ($t=1$)\\
\hline
Skull & 10K & 2.527 & 2.558 & 2.531 & 2.529 & 2.519 & 2.109 & 2.102\\
Bimba & 20K & 2.671 & 2.715 & 2.670 & 2.668 & 2.667 & 2.321 & 2.238\\
Armadillo & 30K & 4.123 & 4.287 & 4.128 & 4.120 & 4.119 & 3.778 & 3.761\\
Brain & 50K & 2.745 & 2.886 & 2.751 & 2.742 & 2.738 & 2.624 & 2.613\\
Gargoyle & 70K & 4.362 & 4.424 & 4.365 & 4.358 & 4.358 & 3.841 & 3.836\\
Bunny & 90K & 3.512 & 3.635 & 3.517 & 3.510 & 3.509 & 2.691 & 2.685\\
\hline
\end{tabular*}
\vspace{-2mm}
\end{table}

\begin{table}[!t]
\centering
\captionsetup{labelformat=empty}
\caption{Table 3: Performance Statistics}
\label{tab:time_comparison}
\vspace{-1.5mm}
\begin{tabular*}{0.4\textwidth}{@{\extracolsep{\fill}}ccccc}
\hline
\hline
Model & \# of & SP ($t=0$) & SP ($t=0.5$) & SP ($t=1$)\\
      & faces & time(sec) & time(sec) & time(sec)\\
\hline
Skull & 10K & 54 & 437 & 430\\
Bimba & 20K & 67 & 576 & 568\\
Armadillo & 30K & 92 & 871 & 865\\
Brain & 50K & 136 & 1258 & 1254\\
Gargoyle & 70K & 245 & 1641& 1638\\
Bunny & 90K & 378 & 2112 & 2106\\
\hline
\end{tabular*}
\vspace{-2mm}
\end{table} 
\vspace{-2.5mm}
\section{Conclusion And Future Work}

Spherical mesh parameterization emphasizes the balance between angle and area distortion. The current work introduces a general framework with solid theoretic foundations. For angle-preserving parameterization, we combine the dynamic Yamabe flow method and conformal welding, and for area-preserving mapping, we have developed the discrete mass transport method. The balance between the two mappings can be achieved by prescribing the target area measure. The proposed framework is grounded in sound theoretic foundations, is more efficient compared to conventional algorithms, is capable of controlling the balance with high precision, and can be extended to general surfaces. Our experimental results demonstrate the efficiency and efficacy of our methods.

In the future, we will generalize the current method to surfaces with more complicated topologies. Furthermore, we would like to generalize optimal mass transport maps to higher dimensions. The theoretic foundation for optimal mass transport in higher dimensions has been fully established; the major difficulty is the space and time complexity of computing power Voronoi diagrams in higher dimensional spaces. Moreover, we will extend our current approach to the medical imaging field for volumetric human organ registration and comparison. 

\vspace{-3mm}
\ifCLASSOPTIONcompsoc
  \section*{Acknowledgments}
\else
  \section*{Acknowledgment}
\fi
This paper has been supported by NSF grants IIS0916235, CCF-0702699, CNS0959979, and CCF1544267 and NIH grant R01EB7530.

\vspace{-3mm}

\begin{thebibliography}{10}

\bibitem{aimatshape}
{AIM} at shape repository.
\newblock {\em http://shapes.aimatshape.net}.

\bibitem{alexa:2002}
M.~Alexa.
\newblock Recent advances in mesh morphing.
\newblock {\em Computer Graphics Forum}, 21(2):173--198, 2002.

\bibitem{Alexandor}
A.~D. Alexandrov.
\newblock {\em Convex Polyhedra}.
\newblock Springer, 2005.

\bibitem{Aurenhammer:1987:PDP}
F.~Aurenhammer.
\newblock Power diagrams: properties, algorithms and applications.
\newblock {\em SIAM J. Comput.}, 16(1):78--96, Feb. 1987.

\bibitem{bobenko2010}
A.~Bobenko, U.~Pinkall, and B.~Springborn.
\newblock Discrete conformal maps and ideal hyperbolic polyhedra.
\newblock {\em arXiv preprint arXiv:1005.2698}, 2010.

\bibitem{Brenier}
Y.~Brenier.
\newblock Polar factorization and monotone rearrangement of vector-valued
  functions.
\newblock {\em Com. Pure Appl. Math.}, 64:375--417, 1991.

\bibitem{crane2013}
K.~Crane, U.~Pinkall, and P.~Schr{\"o}der.
\newblock Robust fairing via conformal curvature flow.
\newblock {\em ACM Transactions on Graphics (TOG)}, 32(4):61, 2013.

\bibitem{deGoes:2012:BNOT}
F.~de~Goes, K.~Breeden, V.~Ostromoukhov, and M.~Desbrun.
\newblock Blue noise through optimal transport.
\newblock {\em ACM Transactions on Graphics}, 31(6):171, 2012.

\bibitem{deGoes:2011:AOTARRSS}
F.~de~Goes, D.Cohen-Steiner, P.~Alliez, and M.~Desbrun.
\newblock An optimal transport approach to robust reconstruction and
  simplification of 2{D} shapes.
\newblock {\em Eurographics Sym. on Geometry Processing}, 30(5):1593--1602,
  2011.

\bibitem{degener:2003}
P.~Degener, J.~Meseth, and R.~Klein.
\newblock An adaptable surface parameterization method.
\newblock {\em Proceedings of 12th Int'l Meshing Roundtable}, 3:201--213, 2003.

\bibitem{Dominitz10}
A.~Dominitz and A.~Tannenbaum.
\newblock Texture mapping via optimal mass transport.
\newblock {\em IEEE Transactions on Visualization and Computer Graphics},
  16(13):419--432, 2010.

\bibitem{freesurfer:2012}
B.~Fischl.
\newblock Freesurfer.
\newblock {\em Neuroimage}, 62(2):774--781, 2012.

\bibitem{FH05}
M.~S. Floater and K.~Hormann.
\newblock Surface parameterization: a tutorial and survey.
\newblock In N.~A. Dodgson, M.~S. Floater, and M.~A. Sabin, editors, {\em
  Advances in multiresolution for geometric modelling}, pages 157--186.
  Springer Verlag, 2005.

\bibitem{gotsman:2003}
C.~Gotsman, X.~Gu, and A.~Sheffer.
\newblock Fundamentals of spherical parameterization for 3{D} meshes.
\newblock {\em ACM Transactions on Graphics}, 22(3):358--363, 2003.

\bibitem{gu:2013_2:arXiv}
X.~Gu, F.~Luo, J.~Sun, and T.~Wu.
\newblock A discrete uniformization theorem for polyhedral surfaces.
\newblock {\em arXiv:1309.4175}, 2013.

\bibitem{gu:2013:arXiv}
X.~Gu, F.~Luo, J.~Sun, and S.-T. Yau.
\newblock Variational principles for {M}inkowski type problems, discrete
  optimal transport, and discrete {M}onge-{A}mp{\`e}re equations.
\newblock {\em arXiv:1302.5472}, 2013.

\bibitem{Gu04}
X.~Gu, Y.~Wang, T.~F. Chan, P.~M. Thompson, and S.-T. Yau.
\newblock Genus zero surface conformal mapping and its application to brain
  surface mapping.
\newblock {\em IEEE Transactions on Medical Imaging}, 23(8):949--958, 2004.

\bibitem{haker:2000}
S.~Haker, S.~Angenent, A.~Tannenbaum, R.~Kikinis, G.~Sapiro, and M.~Halle.
\newblock Conformal surface parameterization for texture mapping.
\newblock {\em IEEE Transactions on Visualization and Computer Graphics},
  6(2):181--189, 2000.

\bibitem{haker:2004}
S.~Haker, L.~Zhu, A.~Tannenbaum, and S.~Angenent.
\newblock Optimal mass transport for registration and warping.
\newblock {\em International Journal of Computer Vision}, 60(3):225--240, 2004.

\bibitem{hormann:2000}
K.~Hormann and G.~Greiner.
\newblock {MIPS}: An efficient global parametrization method.
\newblock {\em Curve and Surface Design.}, pages 153--162, 2000.

\bibitem{Kantorovich48}
L.~V. Kantorovich.
\newblock On a problem of {M}onge.
\newblock {\em Uspekhi Mat. Nauk.}, 3:225--226, 1948.

\bibitem{kazhdan2012}
M.~Kazhdan, J.~Solomon, and M.~Ben-Chen.
\newblock Can mean-curvature flow be made non-singular?
\newblock {\em Eurographics Symposium on Geometry Processing}, 2012.

\bibitem{kharevych:2006}
L.~Kharevych, B.~Springborn, and P.~Schr{\"o}der.
\newblock Discrete conformal mappings via circle patterns.
\newblock {\em ACM Transactions on Graphics}, 25(2):412--438, 2006.

\bibitem{kobbelt:1999}
L.~P. Kobbelt, J.~Vorsatz, and U.~Labsik.
\newblock A shrink wrapping approach to remeshing polygonal surfaces.
\newblock {\em Computer Graphics Forum}, 18(3):119--130, 1999.

\bibitem{lablee2015spectral}
O.~Labl{\'e}e.
\newblock {\em Spectral Theory in Riemannian Geometry}.
\newblock European Mathematical Society Publishing House, 2015.

\bibitem{liu:2008}
L.~Liu, L.~Zhang, Y.~Xu, C.~Gotsman, and S.~J. Gortler.
\newblock A local/global approach to mesh parameterization.
\newblock In {\em Computer Graphics Forum}, volume~27, pages 1495--1504. Wiley
  Online Library, 2008.

\bibitem{Marshall2007}
D.~E. Marshall and S.~Rohde.
\newblock Convergence of a variant of the zipper algorithm for conformal
  mapping.
\newblock {\em SIAM Journal on Numerical Analysis}, 45(6):2577--2609, 2007.

\bibitem{mccann1997}
R.~J. McCann.
\newblock A convexity principle for interacting gases.
\newblock {\em Advances in Mathematics}, 128(1):153--179, 1997.

\bibitem{Merigot}
Q.~M{\'{e}}rigot.
\newblock A multiscale approach to optimal transport.
\newblock {\em Computer Graphics Forum}, 30(5):1583--1592, 2011.

\bibitem{saba:2005}
S.~Saba, I.~Yavneh, C.~Gotsman, and A.~Sheffer.
\newblock Practical spherical embedding of manifold triangle meshes.
\newblock {\em International Conference Shape Modeling and Applications}, pages
  256--265, 2005.

\bibitem{shapiro:1998}
A.~Shapiro and A.~Tal.
\newblock Polyhedron realization for shape transformation.
\newblock {\em The Visual Computer}, 14(8):429--444, 1998.

\bibitem{sheffer:2007}
A.~Sheffer, K.~Hormann, B.~Levy, M.~Desbrun, K.~Zhou, E.~Praun, and H.~Hoppe.
\newblock Mesh parameterization: Theory and practice.
\newblock {\em ACM SIGGRAPPH, course notes}, 2007.

\bibitem{sheffer:2005}
A.~Sheffer, B.~L{\'e}vy, M.~Mogilnitsky, and A.~Bogomyakov.
\newblock Abf++: fast and robust angle based flattening.
\newblock {\em ACM Transactions on Graphics}, 24(2):311--330, 2005.

\bibitem{sheffer2006}
A.~Sheffer, E.~Praun, and K.~Rose.
\newblock Mesh parameterization methods and their applications.
\newblock {\em Foundations and Trends in Computer Graphics and Vision},
  2(2):105--171, 2006.

\bibitem{solomon2015}
J.~Solomon, F.~de~Goes, P.~A. Studios, G.~Peyr{\'e}, M.~Cuturi, A.~Butscher,
  A.~Nguyen, T.~Du, and L.~Guibas.
\newblock Convolutional wasserstein distances: Efficient optimal transportation
  on geometric domains.
\newblock {\em ACM Transactions on Graphics (Proc. SIGGRAPH 2015)}, 2015.

\bibitem{solomon2014}
J.~Solomon, R.~Rustamov, L.~Guibas, and A.~Butscher.
\newblock Earth mover's distances on discrete surfaces.
\newblock {\em ACM Transactions on Graphics (TOG)}, 33(4):67, 2014.

\bibitem{springborn:2008}
B.~Springborn, P.~Schr{\"o}der, and U.~Pinkall.
\newblock Conformal equivalence of triangle meshes.
\newblock {\em ACM Transactions on Graphics}, 27(3):77, 2008.

\bibitem{su:2013:CVPR}
Z.~Su, W.~Zeng, R.~Shi, Y.~Wang, J.~Sun, and X.~Gu.
\newblock Area-preserving brain mapping.
\newblock {\em IEEE Computer Vision and Pattern Recognition (CVPR)}, pages
  2235--2242, 2013.

\bibitem{Rehmana09}
T.Rehman, E.Haber, G.Pryor, J.Melonakos, and A.Tannenbaum.
\newblock 3{D} nonrigid registration via optimal mass transport on the {GPU}.
\newblock {\em Medical Image Analysis}, 13:931--40, 2009.

\bibitem{zayer:2006}
R.~Zayer, C.~Rossl, and H.-P. Seidel.
\newblock Curvilinear spherical parameterization.
\newblock {\em IEEE International Conference on Shape Modeling and
  Applications}, pages 57--64, 2006.

\bibitem{Zengbook:2013}
W.~Zeng and X.~D. Gu.
\newblock {\em Ricci Flow for Shape Analysis and Surface Registration:
  Theories, Algorithms and Applications}.
\newblock Springer Publishing Company, Incorporated, 2013.

\bibitem{zhao2013}
X.~Zhao, Z.~Su, X.~D. Gu, A.~Kaufman, J.~Sun, J.~Gao, and F.~Luo.
\newblock Area-preservation mapping using optimal mass transport.
\newblock {\em IEEE Transactions on Visualization and Computer Graphics},
  19(12):2838--2847, 2013.

\bibitem{zhu:03:APVMS}
L.~Zhu, S.~Haker, and A.~Tannenbaum.
\newblock Area-preserving mappings for the visualization of medical structures.
\newblock {\em Medical Image Computing and Computer-Assisted Intervention
  (MICCAI)}, 2879:277--284, 2003.

\end{thebibliography}

\vspace{-10mm}
\begin{IEEEbiography}[{\includegraphics[width=1in,height=1.25in]{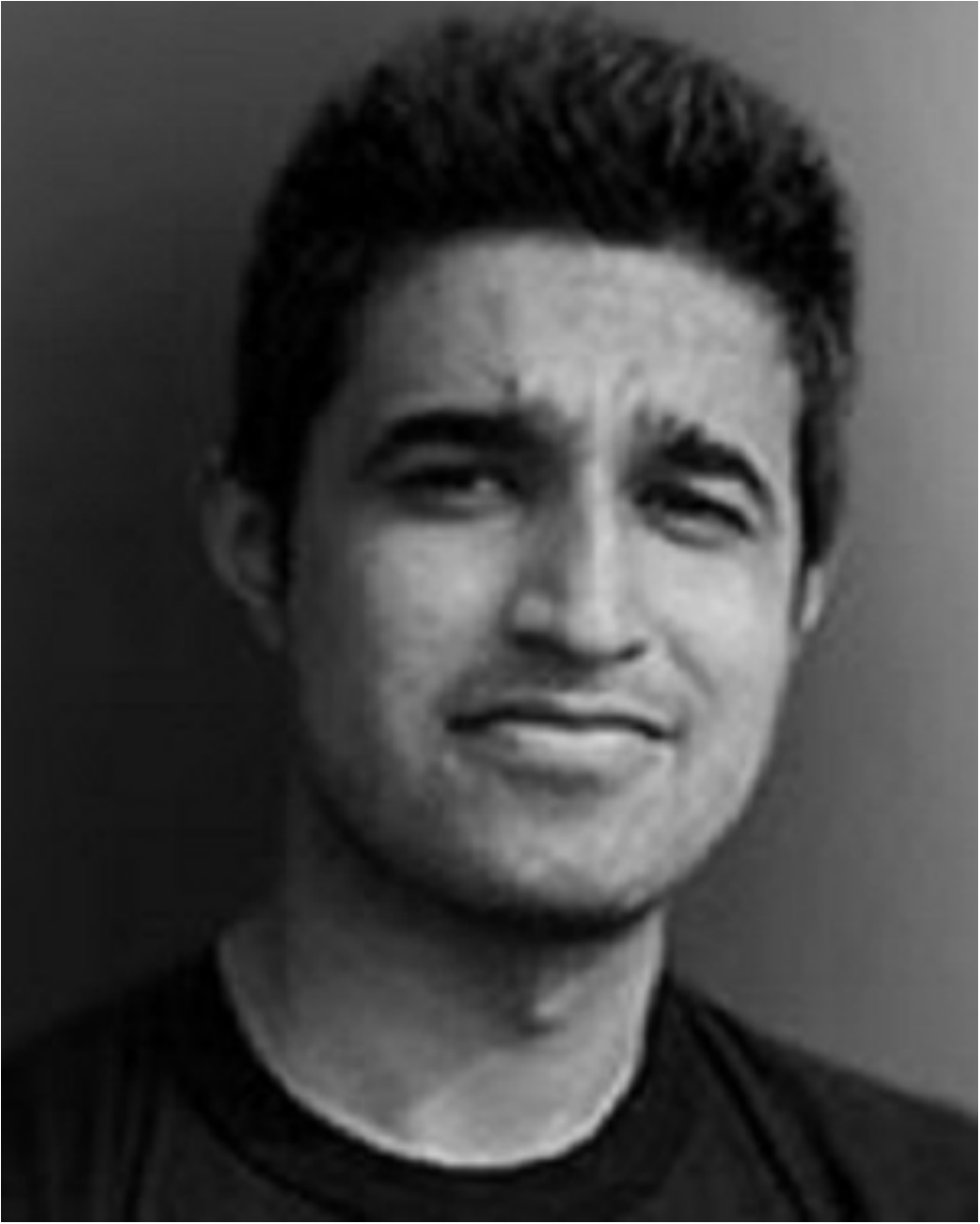}}]{Saad Nadeem}
is a PhD student at Computer Science department, Stony Brook University. His research interests include computer vision, computer graphics and visualization.
\end{IEEEbiography}
\vspace{-11mm}
\begin{IEEEbiography}[{\includegraphics[width=1in,height=1.25in,clip,keepaspectratio]{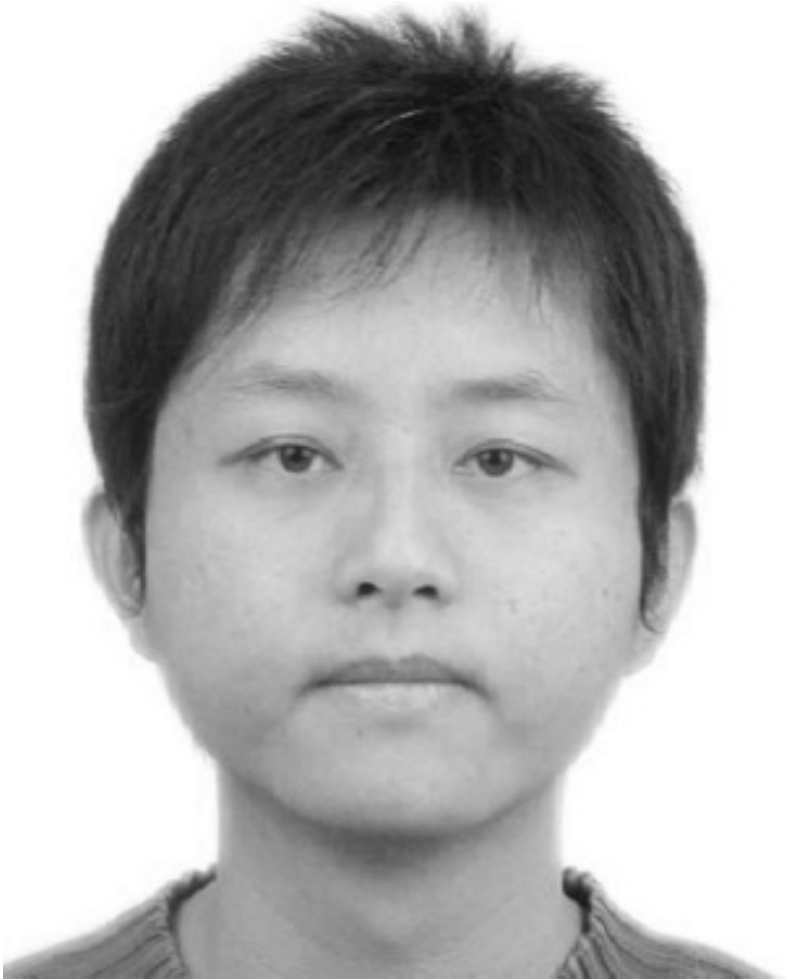}}]{Zhengyu Su}
is a PhD student at Computer Science department, Stony Brook University. His research interests are computational geometry, computer graphics and computer vision.
\end{IEEEbiography}
\vspace{-11mm}
\begin{IEEEbiography}[{\includegraphics[width=1in,height=1.25in,clip,keepaspectratio]{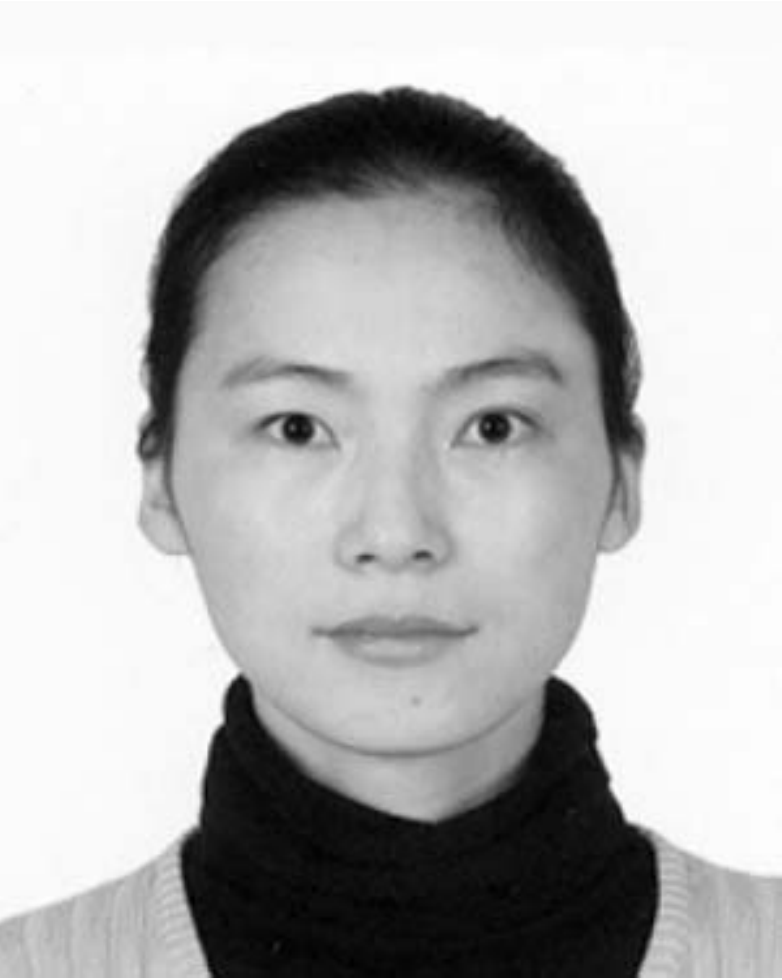}}]{Wei Zeng}
received her Ph.D. degree from the Institute of Computing Technology, Chinese Academy of Sciences in 2008. She is an assistant professor in the School of Computing and Information Sciences, Florida International University, Miami, Florida. Her research interests include computational conformal geometry, discrete Ricci flow, and surface matching, registration, tracking, recognition and shape analysis.
\end{IEEEbiography}
\vspace{-13mm}
\begin{IEEEbiography}[{\includegraphics[width=1in,height=1.25in,clip,keepaspectratio]{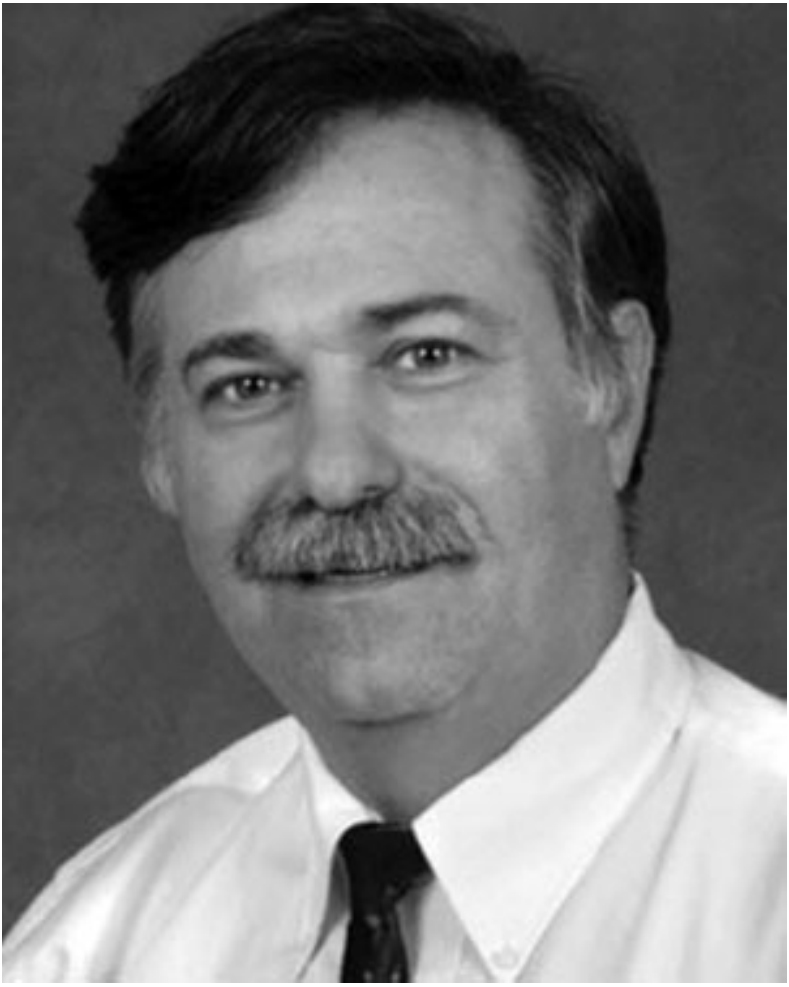}}]{Arie Kaufman}
is a Distinguished Professor and Chair of the Computer Science Department, the Director of the Center for Visual Computing (CVC), and the Chief Scientist of the Center of Excellence in Wireless and Information Technology (CEWIT) at Stony Brook University. He received his PhD in Computer Science from Ben-Gurion University, Israel, in 1977. He is internationally recognized for his pioneering and seminal contributions to visualization, graphics, virtual reality, and their applications, especially in biomedicine. He is Fellow of IEEE, Fellow of ACM, member of the European Academy of Sciences, recipient of the IEEE Visualization Career Award, and was inducted into the LI Technology Hall of Fame. He was the founding Editor-in-Chief of IEEE Transaction on Visualization and Computer Graphics (TVCG), 1995-1998.
\end{IEEEbiography}
\vspace{-11mm}
\begin{IEEEbiography}[{\includegraphics[width=1in,height=1.25in,clip,keepaspectratio]{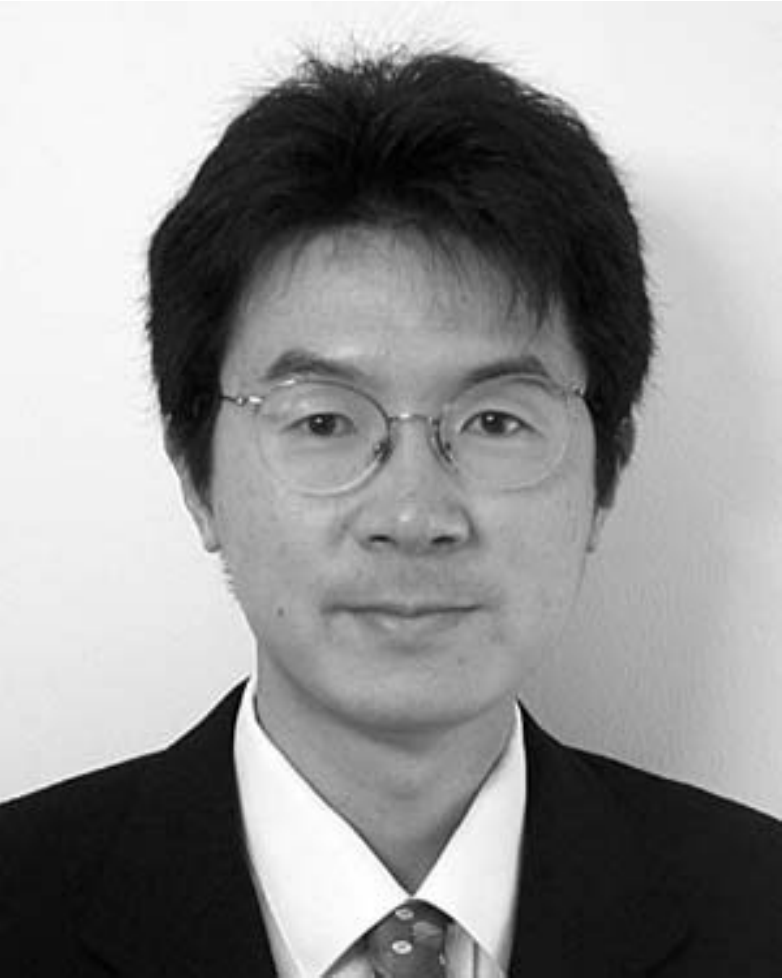}}]{Xianfeng Gu}
received the Ph.D. degree in computer science from Harvard University, Cambridge, MA, USA, in 2003. He is an associate professor of Computer Science and the Director of the 3D Scanning Laboratory with the Department of Computer Science at Stony Brook University, Stony Brook, NY, USA. His current research interests include computer vision, graphics, geometric modeling, and medical imaging. His major works include global conformal surface parameterization in graphics, tracking and analysis of facial expression in vision, manifold splines in modeling, brain mapping and virtual colonoscopy in medical imaging, and computational conformal geometry. He won the U.S. National Science Foundation CAREER Award in 2004.
\end{IEEEbiography}

\section*{Appendix}
\vspace{-2.5mm}
\subsection*{A.1 Conformal Mapping and Uniformization}

\textbf{Conformal Mapping: } A conformal mapping between two surfaces preserves angles.
 \begin{definition}[Conformal Mapping]
 Suppose $(S_1,\mathbf{g}_1)$ and $(S_2,\mathbf{g}_2)$ are two surfaces with Riemannian metrics, $\mathbf{g}_1$ and $\mathbf{g}_2$, respectively. A mapping $\phi: S_1 \to S_2$ is called \emph{conformal} if the pullback metric of $\mathbf{g}_2$
induced by $\phi$ on $S_1$ differs from $\mathbf{g}_1$ by a positive
scalar function: $\phi^{*}\mathbf{g}_2 = e^{2\lambda}\mathbf{g}_1$, where $\lambda: S_1 \to \mathbb{R}$ is a scalar function, called the
\emph{conformal factor}.
\end{definition}

By conformal mapping, surfaces can be classified according to the conformal equivalence relation.
\begin{definition}[Conformal Equivalence]
 Suppose $(S_1,\mathbf{g}_1)$ and $(S_2,\mathbf{g}_2)$ are two Riemannian surfaces. If there is a conformal diffeomorphism between them $\varphi:S_1\to S_2$, then the two surfaces are conformally equivalent.
\end{definition}

\begin{figure}[h]
\begin{center}
\begin{tabular}{ccc}
\includegraphics[width=0.15\textwidth]{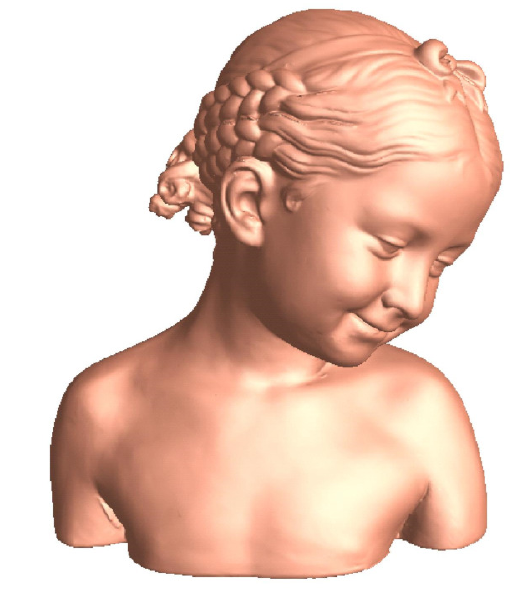}&
\includegraphics[width=0.135\textwidth]{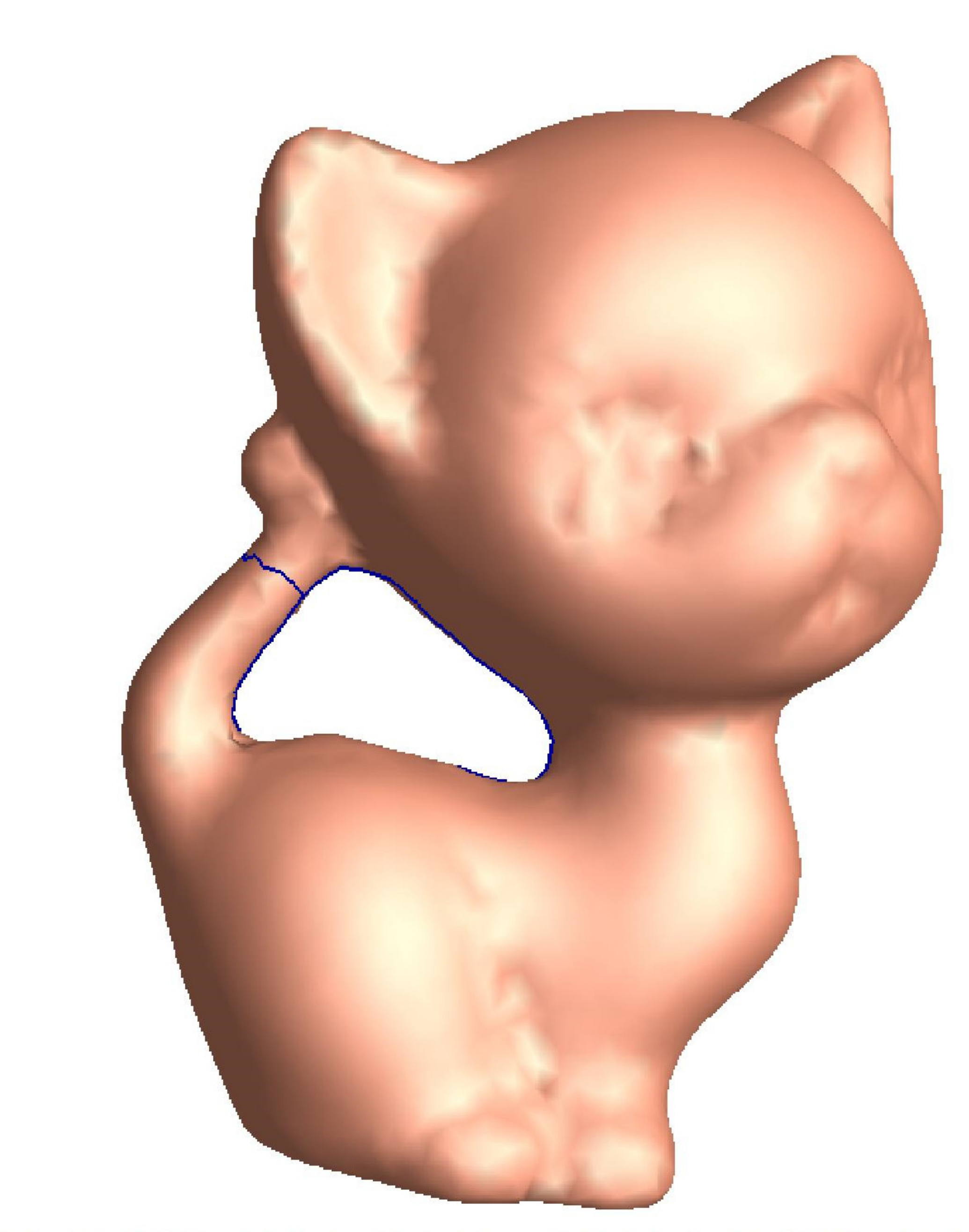}&
\includegraphics[width=0.155\textwidth]{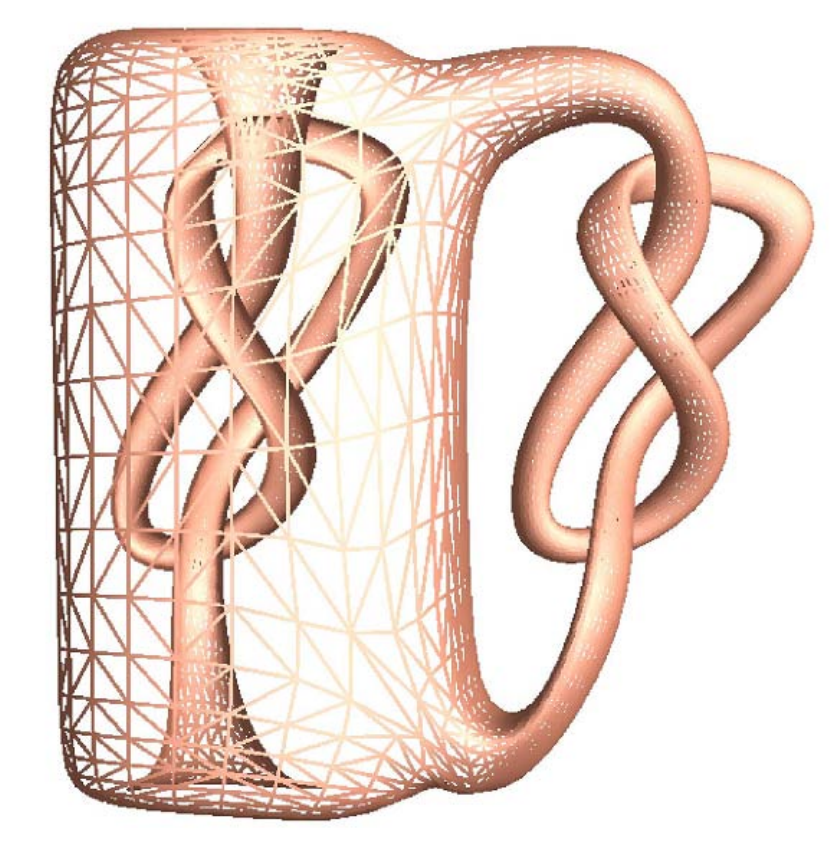}\\
(a) & (b) & (c)\\
\includegraphics[width=0.15\textwidth]{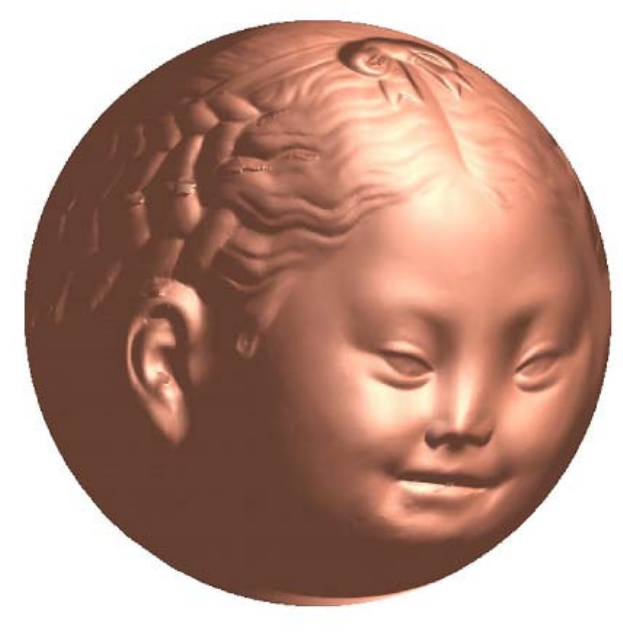}&
\includegraphics[width=0.135\textwidth]{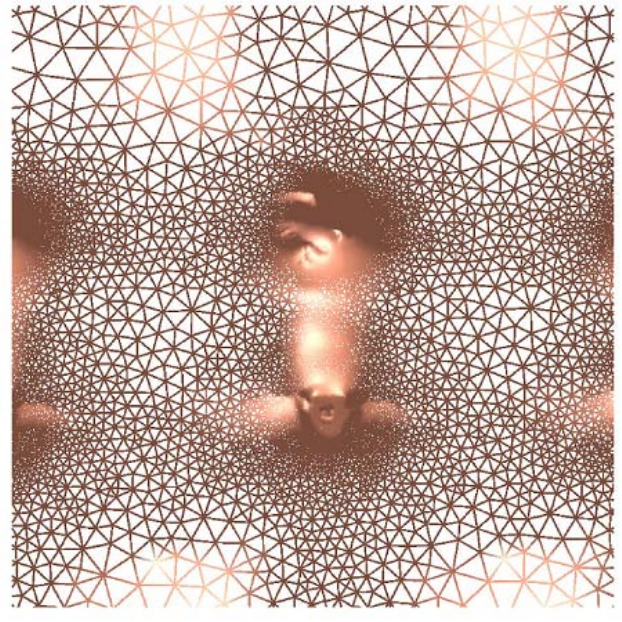}&
\includegraphics[width=0.155\textwidth]{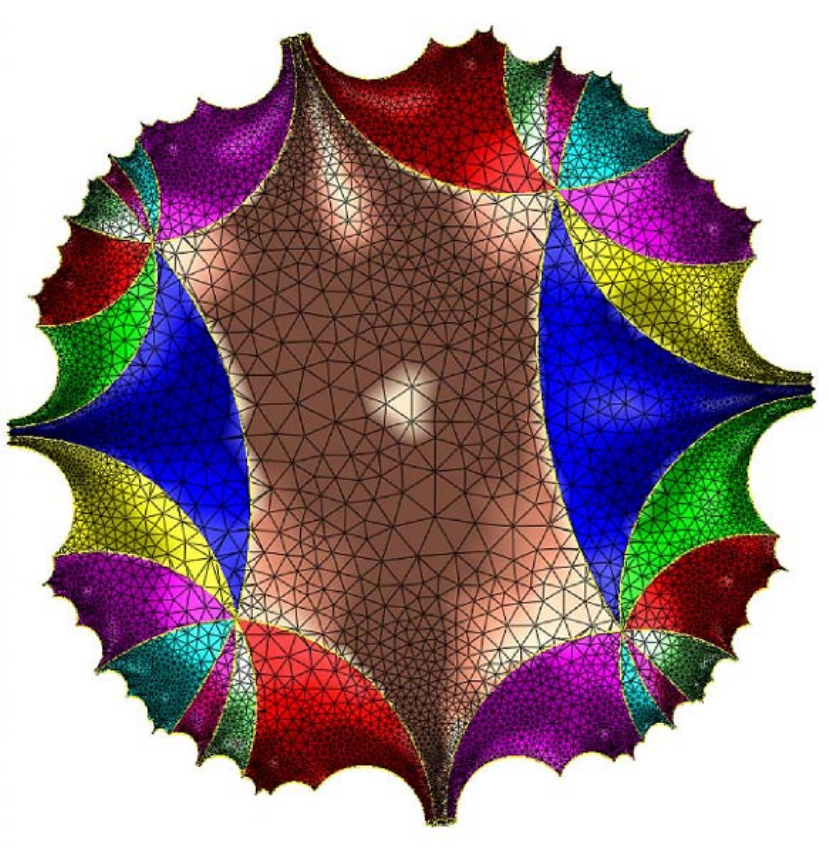}\\
(d) & (e) & (f)
\end{tabular}
\end{center}
\vspace{-3mm}
\caption{ Surface uniformization for closed surfaces \cite{Zengbook:2013}: (a) Genus 0 surface and the corresponding (d) spherical mapping. (b) Genus 1 surface and the corresponding (e) Euclidean plane. (c) Genus 2 surface and the corresponding (f) hyperbolic plane; in (f), different fundamental domains are color-coded, with each color representing one fundamental domain.\label{fig:closed_surface_uniformization}}
\vspace{-3mm}
\end{figure}

\noindent
\textbf{Uniformization: } The surfaces in each conformal equivalence class share the same complete conformal invariants, the so-called \emph{conformal module}. The most straightforward way to define conformal module is via the uniformization theorem, which states that all metric surfaces can be conformally mapped to one of three canonical spaces: the unit sphere $\mathbb{S}^2$, the Euclidean plane $\mathbb{E}^2$, or the hyperbolic plane $\mathbb{H}^2$.

In essence, two surfaces are conformally equivalent if and only if they share the same conformal modules. For example, if a topological annulus can be conformally mapped onto a planar annulus and the two boundaries are concentric circles with radii $R$ and $r$, then the conformal module can be formulated as $\frac{-1}{2} \pi$ ln $\frac{R}{r}$. If there exists a conformal mapping between two topological annuli, then they share the same conformal module, and vice versa.

\begin{theorem}[Uniformization]
Given a compact, closed surface $S$ with a Riemannian metric $g$, there exists a scalar function $\lambda: S\to\mathbb{R}$, such that the metric $e^{2\lambda}\mathbf{g}$ induces constant Gaussian curvature. If the Euler number of the surface $\chi(S)$ is positive, zero or negative, then the constant is $+1$, $0$ or $-1$, respectively.
\end{theorem}

\begin{figure}[h]
\begin{center}
\begin{tabular}{ccc}
\includegraphics[width=0.14\textwidth]{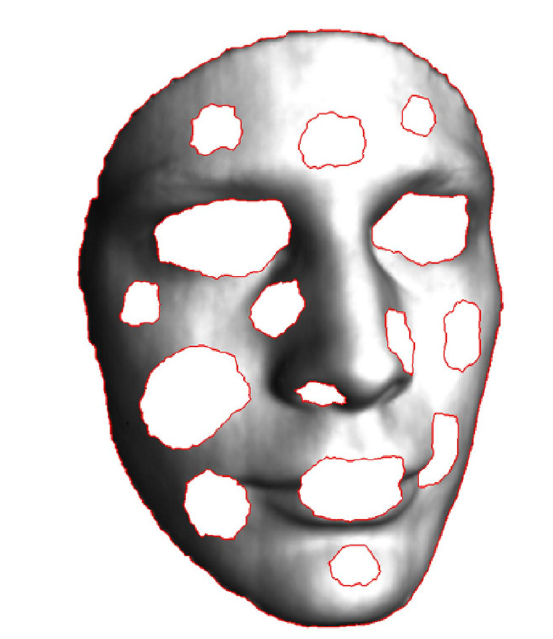}&
\includegraphics[width=0.15\textwidth]{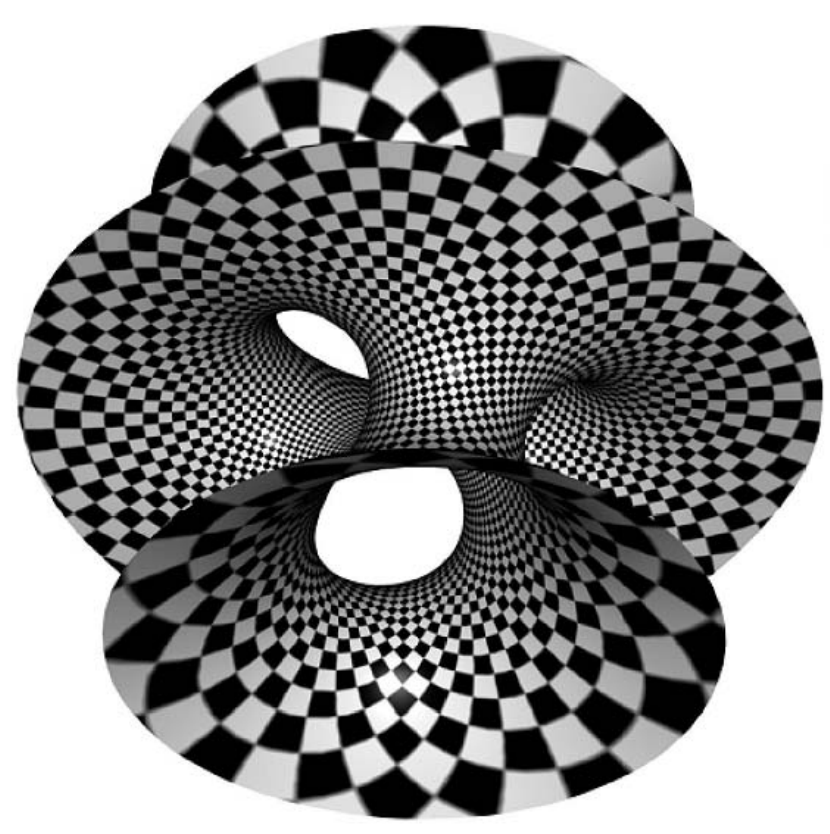}&
\includegraphics[width=0.15\textwidth]{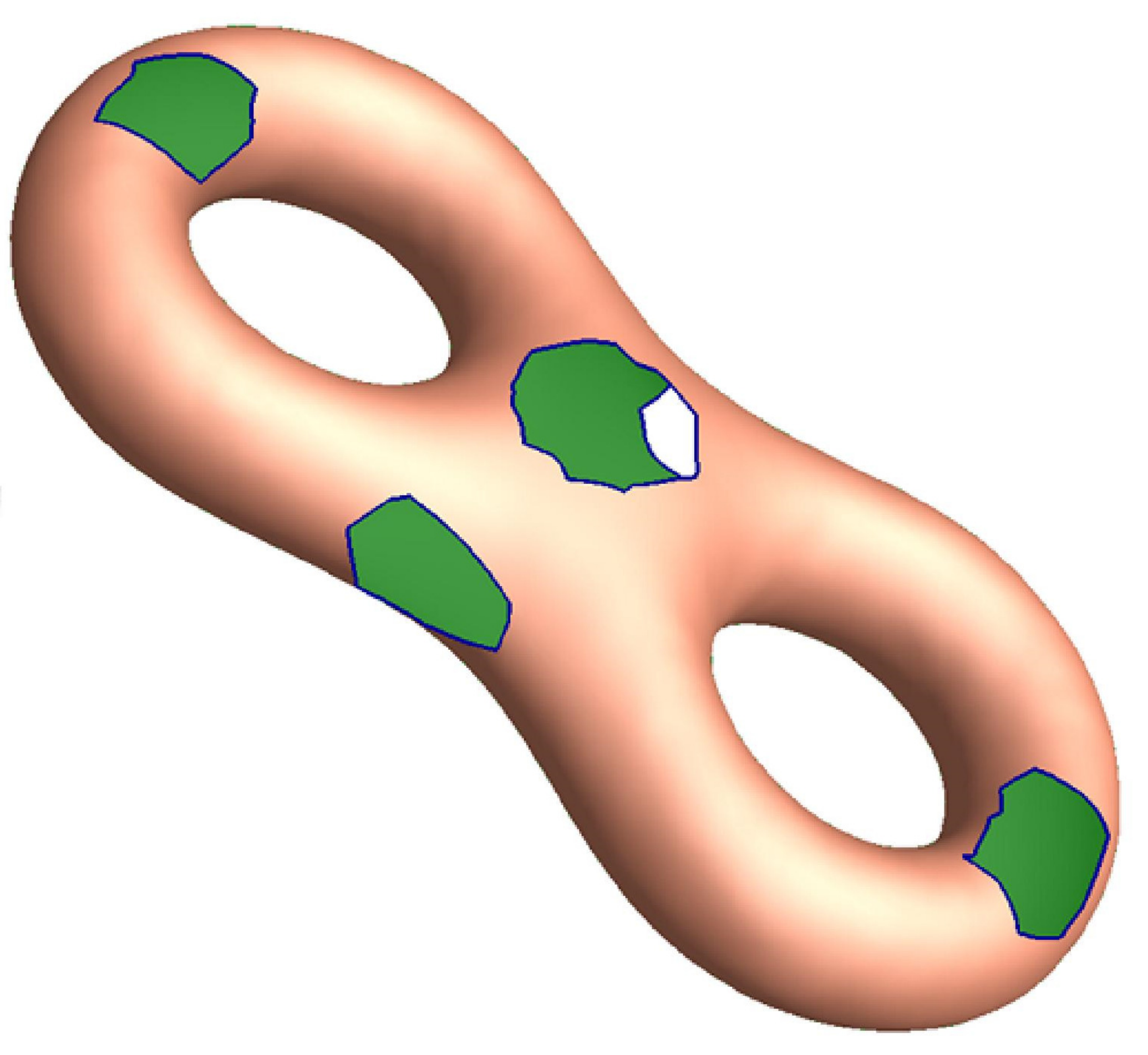}\\
(a) & (b) & (c)\\
\includegraphics[width=0.14\textwidth]{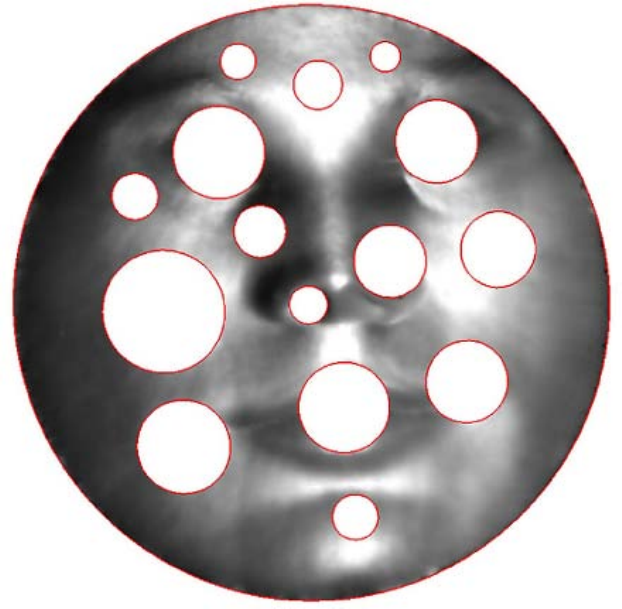}&
\includegraphics[width=0.145\textwidth]{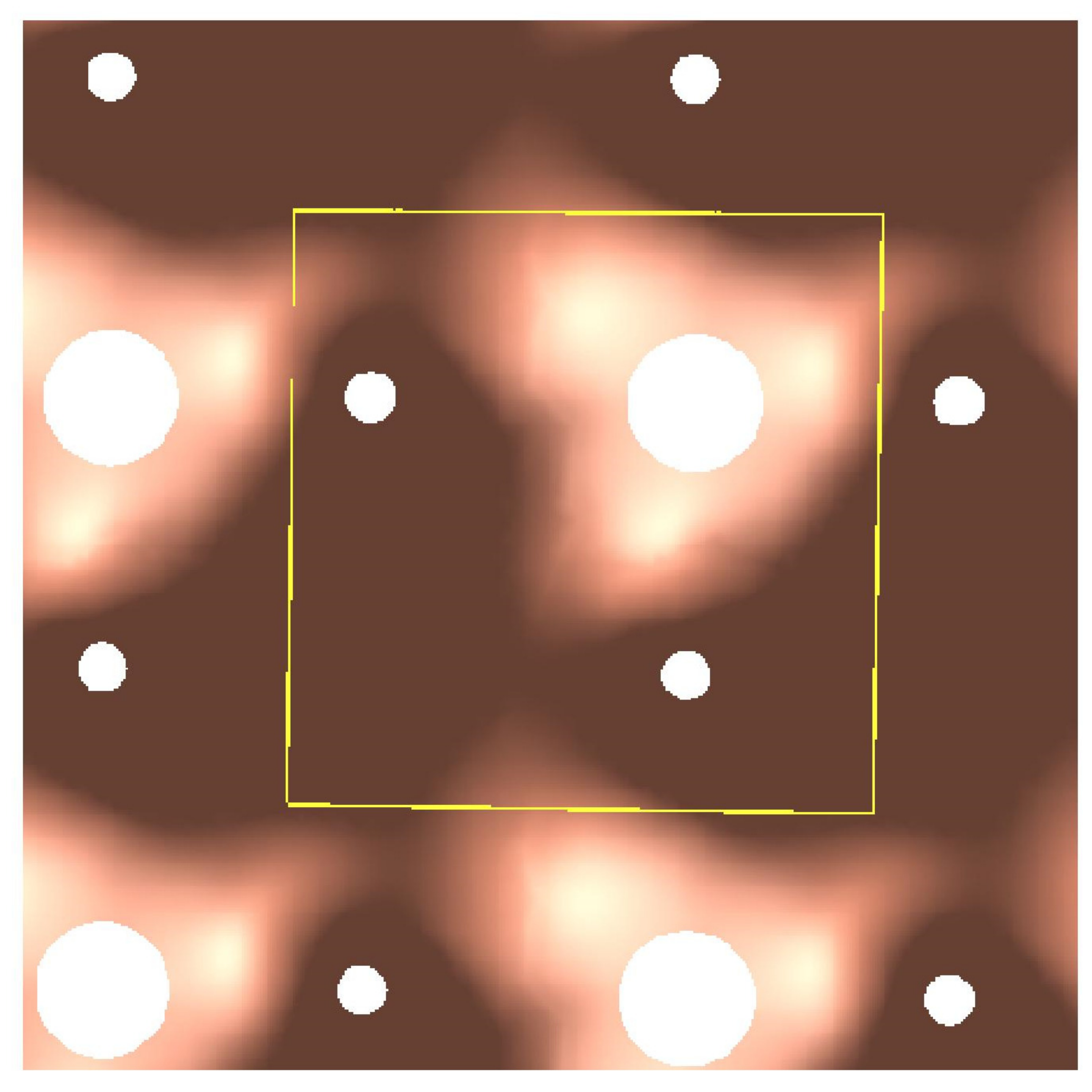}&
\includegraphics[width=0.15\textwidth]{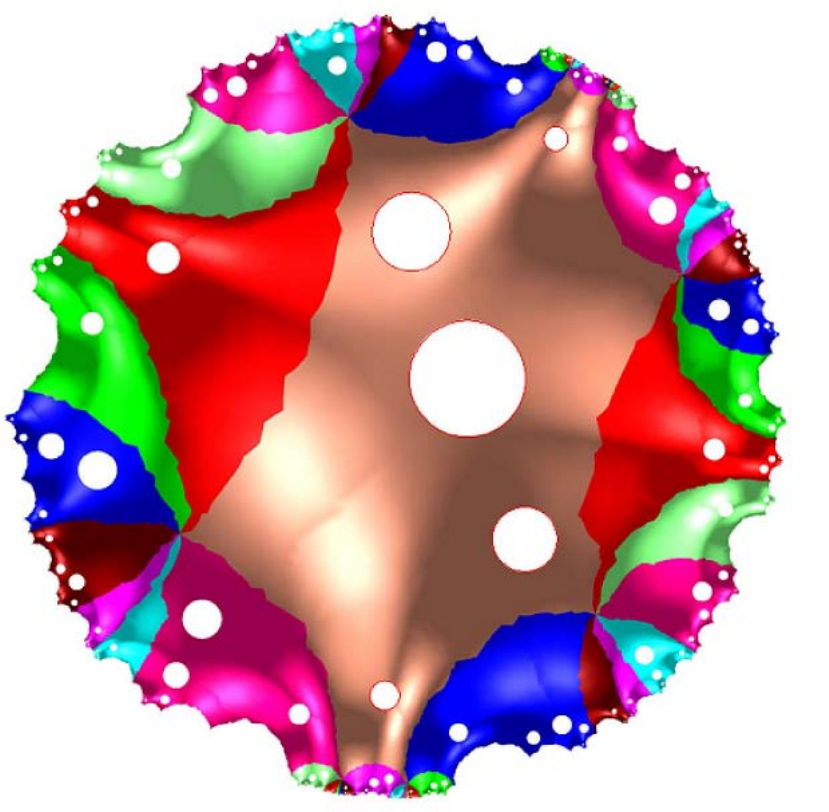}\\
(d) & (e) & (f)
\end{tabular}
\end{center}
\vspace{-3mm}
\caption{ Surface uniformization for surfaces with boundaries \cite{Zengbook:2013}: (a) Genus 0 surface with holes and the corresponding (d) planar circle domain. (b) Genus 1 surface with 3 holes and the corresponding (e) Euclidean plane; the yellow box in (e) specifies the 1-fundamental domain; the texture in (b) shows that the mapping to (e) is conformal. (c) Genus 2 surface with 5 holes and the corresponding (f) hyperbolic plane; in (f), different fundamental domains are color-coded, with each color representing one fundamental domain. \label{fig:open_surface_uniformization}}
\vspace{-5mm}
\end{figure}

If the Riemannian surfaces are with boundaries, then they can be conformally mapped to circle domains on the canonical spaces $\mathbb{S}^2, \mathbb{E}^2$ and $\mathbb{H}^2$, whose complements are spherical, Euclidean or hyperbolic disks. Figures \ref{fig:closed_surface_uniformization} and \ref{fig:open_surface_uniformization} show the uniformization for closed surfaces and surfaces with boundaries, respectively.

\noindent
\textbf{Ricci Flow: } Surface Ricci flow is a powerful tool for computing uniformization. \emph{Ricci flow} refers to the process of deforming the Riemannian metric $\mathbf{g}$ proportional to the curvature, such that the curvature $K$ evolves according to a heat diffusion process, eventually making the Gaussian curvature constant everywhere. Assuming that the metric $\mathbf{g}=(g_{ij})$ is in local coordinates, Hamilton \cite{ric88} introduces the normalized surface Ricci flow.
\begin{definition}[Ricci Flow]
The normalized surface Ricci flow is:
\[
    \frac{dg_{ij}}{dt} = \left(\frac{4\pi\chi(S)}{A(0)}-2K\right) g_{ij},
\]
\end{definition}
where $A(0)$ is the initial total area of the surface. Surface Ricci flow conformally deforms the Riemannian metric, and converges to a constant curvature metric, proved by Hamilton and Chow \cite{Chow06}. Furthermore, Ricci flow can be used to compute the unique conformal Riemannian metric with the prescribed curvature.
\begin{theorem}[Hamilton and Chow \cite{Chow06}]
Suppose $(S,\mathbf{g})$ is a closed surface with a Riemannian metric. The normalized surface Ricci flow will converge to a Riemannian metric of constant Gaussian curvature and the convergence is exponentially fast.
\end{theorem}

\vspace{-2.5mm}
\subsection*{A.2 Optimal Mass Transport Theory} 

\textbf{Optimal Mass Transport: } The problem of finding a map that minimizes the inter-domain transport cost while preserveing measure quantities was first studied by Monge \cite{Monge} in the 18th century.

Let $X$ and $Y$ be two metric spaces with probability measures $\mu$ and $\nu$ respectively. Assume $X$ and $Y$ have equal total measures $\int_X \mu = \int_Y \nu$.

\begin{definition}[Measure-Preserving Mapping]
A map $\varphi: X\to Y$ is \emph{measure preserving} if for any measurable set $B\subset Y$, $\mu(\varphi^{-1}(B)) = \nu(B)$.
\end{definition}
The mapping $\varphi$ induces a push-forward measure $\varphi_{\#}\mu$ on the target, for any measurable set $B\subset Y$,
\[
    \varphi_{\#}\mu(B) := \int_{\varphi^{-1}(B)} \mu(x) dx.
\]
$\varphi$ is measure-preserving if and only if $\varphi_{\#}\mu=\nu$. Let us denote the transport cost for sending $x\in X$ to $y\in Y$ by $c(x,y)$, then the total \emph{transport cost} is given by:
\begin{equation}
    \mathcal{C}(\varphi):=\int_X c(x,\varphi(x)) \mu(x)dx.
    \label{eqn:transport_cost}
\end{equation}

\begin{definition}[Optimal Mass Transport Map] Given metric spaces with
probabilities measures $(X,\mu)$, $(Y,\nu)$ and the transport
cost function $c: X\times Y\to \mathbb{R}$, the optimal mass transport map is a measure-preserving map $\varphi:X\to Y$, which minimizes the transport cost,
\[
    \varphi = argmin_{\tau_{\#}\mu=\nu} \int_X c(x,\varphi(x)) \mu(x)dx.
\]
\end{definition}

At the end of 1980's, Brenier \cite{Brenier} discovered the intrinsic connection between optimal mass transport map and convex geometry.

\begin{theorem}[Brenier] Suppose $X$ and $Y$ are in the Euclidean space $\mathbb{R}^n$, and the transport cost is the quadratic Euclidean distance $c(x,y) = |x-y|^2$. If $\mu$ is absolutely continuous and $\mu$ and $\nu$ have finite second order moments, then there exists a convex function $u: X\to \mathbb{R}$ and its gradient map $ x\to \nabla u(x)$ is the unique optimal mass transport map.
\end{theorem}

This theorem converts the problem of finding the optimal mass transport map to solving the following Monge-Amp\`{e}re partial differential equation:
\[
    \mu(x) det \left( \frac{\partial^2 u(x)}{\partial x_i \partial x_j}\right ) = \nu \circ \nabla~u(x).
\]

In the current work, the metric space is the Euclidean plane and the transport cost is the square of the Euclidean distance. If the source is a convex planar domain, then the solution to the optimal mass transport problem exists, and can be solved in a variational framework.

Fig. 1 in the paper shows one example of an optimal mass transport map. The brain cortical surface (a) is mapped onto the planar disk via a conformal mapping in (c). The conformal factor defines a probability on the disk $e^{2\lambda(x,y)}dxdy$. The optimal mass transport map $T: (\mathbb{D}, e^{2\lambda(x,y)}dxdy) \to (\mathbb{D}, dxdy )$ is shown in (d). The mapping from (a) to (d) is area-preserving. The optimal transport, in this example, is with respect to the Euclidean metric on the disk, not the geodesic distance, or the Euclidean distance in $\mathbb{R}^3$.

\noindent
\textbf{Polar Factorization: } The following polar factorization theorem plays a fundamental role
in the current project.

\begin{theorem}[Polar Factorization \cite{Brenier}]
Let $\Omega_0$ and $\Omega_1$ be two convex subdomains of
$\mathbb{R}^n$, with smooth boundaries, each with a positive density
function, $\mu_0$ and $\mu_1$ respectively, with the same total mass
$\int_{\Omega_0} \mu_0 = \int_{\Omega_1} \mu_1$. Let
$\varphi:(\Omega_0,\mu_0)\to(\Omega_1,\mu_1)$ be diffeomorphic
mapping, then $\varphi$ has a unique decomposition of the form
\begin{equation}
    \varphi = (\nabla u) \circ s,
\end{equation}
where $u:\Omega_0\to\mathbb{R}$ is a convex function,
$s:(\Omega_0,\mu_0)\to (\Omega_0,\mu_0)$ is a measure-preserving
mapping. This is called the polar factorization of $\varphi$ with
respect to $\mu_0$.
\end{theorem}

This means a general diffeomorphism $\varphi:(\Omega_0,\mu_0)\to
(\Omega_1,\mu_1)$, where $\mu_1 = \varphi_{\#}\mu_0$ can be
decomposed to the composition of a measure-preserving map
$s:(\Omega_0,\mu_0)\to (\Omega_0,\mu_0)$ and a $L^2$ optimal mass
transport map $\nabla u: (\Omega_0,\mu_0)\to (\Omega_1,\mu_1)$.
This decomposition is unique.

Furthermore, if $\Omega_0$ coincides with $\Omega_1$, then $s$ is
the unique $L^2$ projection of $\varphi$ in the space of all measure-preserving mappings of $(\Omega_0,\mu_0)$. Namely, $\tau$ minimizes
the $L^2$ distance among all measure-preserving mappings,
\[
s=argmin_{\tau} \int_{\Omega_0} \| \varphi(x) - \tau(x)\|^2 \mu_0(x)
dx, ~~\tau_{\#}\mu_0=\mu_0.
\]

\subsection*{A.3 Area and Angle Distortion Measurements}

The parameterization quality is measured by both angle and area distortions. We introduce two methods to measure these distortions.

\subsection*{A.3.1 Histograms for the Angle and Area Distortions (Section 4.1)}

The area distortion is computed as follows. Assume the parameterization is $\phi: M \rightarrow \mathbb{S}$. For each vertex $v_i$, the \emph{area distortion} is defined as
\[
    \epsilon_i := \log \frac{\sum_{j,k} A([\phi(v_i),\phi(v_j),\phi(v_k)])}{\sum_{j,k} A([v_i,v_j,v_k])}
\]
where $A(.)$ represents the area of a triangle, and $[v_i,v_j,v_k]$ is the triangle formed by ${v_i,v_j,v_k}$. We then plot the histograms of ${\epsilon_i}$. Similarly the \emph{angle distortion} at a corner angle is given by
\[
    \eta_{ijk} := \log \frac{\angle \phi(v_i)\phi(v_j)\phi(v_k)}{\angle v_i v_j v_k},
\]
we then plot the histograms of ${\eta_{ijk}}$.

The angle-preserving (conformal) mapping should ideally be close to zero angle distortions everywhere, whereas the area-preserving mapping should be close to zero area distortions everywhere.

\subsection*{A.3.2 Signed Singular Values of the Jacobian (Section 4.2)}
The parameterization is a piecewise linear mapping. On each face, the linear mapping is represented as a matrix $J$ and the singular values are the eigenvalues of $J^{TJ}$. The measurement in Section 4.2 is:
\[
\lambda_{max} / \lambda_{min} + \lambda_{min}/\lambda_{max}
\]
If the parameterization is angle-preserving, then $\lambda_{max} = \lambda_{min}$; if the parameterization is area-preserving, then $\lambda_{max} \lambda_{min} = 1$. If the parameterization is close to be isometric (both angle-preserving and area-preserving), then the measurement equals to 2.

\end{document}